\documentclass[%
 reprint,
superscriptaddress,
notitlepage,
nofootinbib,
 amsmath,amssymb,
 aps,
 onecolumn,
prd,
]{revtex4-1}

\usepackage{graphicx}
\usepackage{dcolumn}
\usepackage{bm}
\usepackage{hyperref}
\hypersetup{colorlinks}

\usepackage{color}
\usepackage[caption=false]{subfig}
\def\beq{\begin{eqnarray}}
\def\eeq{\end{eqnarray}}

\def\mnras{\rm{MNRAS}}
\def\aap{\rm{A\&A}}
\def\baas{\rm{BAAS}}
\def\physrep{\rm{Phys.~Rep.}}
\def\apj{\rm{ApJ}}                 
\def\apjl{\rm{ApJ}}                
\def\apjs{\rm{ApJS}}               
\def\jcap{\rm{JCAP}}

\newcommand{\av}[1]{\langle{#1\rangle}} 
\newcommand{\dpt}[1]{\delta^{(#1)}}
\newcommand{\dptR}[1]{\delta_R^{(#1)}}
\newcommand{\dptM}[1]{\delta_M^{(#1)}}
\let\vec\mathbf

\newcommand{\Mpch}{h^{-1}\mathrm{Mpc}}
\newcommand{\hMpc}{h\,\mathrm{Mpc}^{-1}}

\newcommand{\new}[1]{#1}
\newcommand{\resub}[1]{#1}

\begin{document}



\title{What does the Marked Power Spectrum Measure? \\ \normalsize Insights from Perturbation Theory}

\author{Oliver H.\,E. Philcox}
\email{ohep2@cantab.ac.uk}
\affiliation{Department of Astrophysical Sciences, Princeton University,\\ 8 Ivy Lane, Princeton, NJ 08540, USA}%

\author{Elena Massara}
\affiliation{Waterloo Centre for Astrophysics, University of Waterloo,\\
200 University Ave W, Waterloo, ON N2L 3G1, Canada}

\author{David N. Spergel}
\affiliation{Center for Computational Astrophysics, Flatiron Institute,\\ 162 Fifth Avenue, New York, NY 10010, USA}
\affiliation{Department of Astrophysical Sciences, Princeton University,\\ 8 Ivy Lane, Princeton, NJ 08540, USA}%


\date{\today}

\begin{abstract}
The marked power spectrum is capable of placing far tighter constraints on cosmological parameters (particularly the neutrino mass) than the conventional power spectrum. What new information does it contain beyond conventional statistics? Through the development of a perturbative model, we find that the mark induces a significant coupling between non-Gaussianities, \resub{which are usually found on small scales}, and large scales, leading to the additional information content. The model is derived in the context of one-loop perturbation theory and validated by comparison to $N$-body simulations across a variety of mark parameters. At moderate redshifts, including for massive neutrino cosmologies, the theory is in good agreement with the simulations. The \resub{importance of non-linear gravitational effects on the large-scale spectra} complicates the modeling as there is no well-defined convergence radius of the theory at low $z$. Extension to higher perturbative order and biased tracers is possible via a similar approach, and a simple model of the latter is shown to yield promising results. The theory becomes non-perturbative at redshift zero for small smoothing scales, \resub{with important contributions from higher-order terms}:
these will need to be studied before the full power of this tool can be realized.
\end{abstract}

\maketitle


\section{Introduction}

How can we best extract information from large scale structure? At early times, deviations from Gaussianity in the Universe are slight, thus all relevant statistical information is encoded in the two-point correlators of the matter density field, in configuration or Fourier space. At late times, this is not the case. Non-linear effects become significant, and deviations from Gaussianity shift information into higher-point statistics, most notably the bispectrum \citep{1999ApJ...527....1S,2006PhRvD..74b3522S}. Whilst a number of recent surveys have included these correlators \citep[e.g.,][]{2017MNRAS.465.1757G,2018MNRAS.478.4500P}, their measurement requires sophisticated techniques \citep[e.g.,][]{2001ApJ...546..652S,2004ApJ...605L..89S,2012PhRvD..86f3511F,2013PhRvD..88f3512S,2019arXiv190903248H,2020arXiv200501739P} and complex modeling \citep[e.g.,][]{2006PhRvD..74b3522S,2015JCAP...10..039A,2015PhRvD..92d3514B,2018JCAP...04..055L}. Further, the expansion in $n$-point functions is not closed; there is additional information lying in the higher correlators, computation of which is currently infeasible.

Is there a more powerful and robust statistical tool for studying large-scale structure? A vast swathe of literature exists attempting to answer this question, with candidate statistics including reconstructed density fields \citep{2007ApJ...664..675E}, Gaussianized density fields \citep{1992MNRAS.254..315W,2011ApJ...731..116N,2011ApJ...742...91N}, and log-normal transforms \citep{2009ApJ...698L..90N,2011ApJ...735...32W}. Furthermore, statistics not based on $n$-point correlators have been proposed, for example counts-in-cells \citep[e.g.,][]{1980lssu.book.....P} and void statistics \citep{2019BAAS...51c..40P}. 
Of particular interest is the `marked' density field, formalized in \citet{doi:10.1002/mana.19841160115}. At heart, this is simply a weighted density field \citep{2005MNRAS.364..796S}, where the weights can represent galaxy properties \citep{2005astro.ph.11773S,2006MNRAS.369...68S,2000ApJ...545....6B}, halo merger history \citep{2002A&A...387..778G}, or environmental density \citep{2016JCAP...11..057W}. The latter of significant cosmological relevance, since it can be used to up-weight low-density regions (\textit{i.e.} voids) that contribute little to conventional power spectra. Since these are unvirialized, they are not subject to the degradation of information caused by shell-crossing and are thus expected to be an important probe of cosmology \citep{2019BAAS...51c..40P}. Recent works have found statistics based on the marked density field to be of great use when probing modified gravity \citep{2016JCAP...11..057W,2018PhRvD..97b3535V,2018MNRAS.478.3627A,2018MNRAS.479.4824H,2020JCAP...01..006A} and searching for massive neutrinos \citep{2020arXiv200111024M}.

This work follows the treatment of \citet{2020arXiv200111024M}, which, via a simulation-based approach, demonstrated the efficacy of the \textit{marked power spectrum} in constraining late-time cosmological parameters. Previous works have adopted statistics based on the ratio of the marked field correlators to those of the density field; instead the authors forewent this normalization to ensure that the information contained within the matter correlators is not lost. This was shown to give a Fourier-space statistic capable of improving constraints on all standard cosmological parameters by a factor of at least $2$, using modes up to $k = 0.5\hMpc$ at redshift zero. Perhaps the most striking conclusions are those pertaining to the neutrino mass sum; from a $1h^{-3}\mathrm{Gpc}^3$ box, constraints of $\sigma(M_\nu) = 0.017\,$eV were possible from the marked statistics alone. This is a factor of 80 tighter than that obtained from the matter power spectrum, with further improvements seen from combining statistics.

Here, we consider the marked power spectrum, $M(k)$, from a theoretical standpoint, developing a perturbative framework that (a) provides an accurate model for the spectrum in certain regimes, and (b) gives insight into the sources of cosmological information in $M(k)$, and hence sheds light on its astonishing constraining power. This is done in the context of the Effective Field Theory of Large Scale Structure (hereafter EFT) \citep{2012JCAP...07..051B,2012JHEP...09..082C}; a perturbative theory developed from the underlying cosmological fluid equations that has been shown to provide accurate models of matter and biased tracer statistics on quasi-linear scales. Our EFT model is validated with a suite of $N$-body simulations, and shown to have substantially different properties to the matter EFT, particularly due to scale-mixing, with large non-Gaussian contributions appearing on scales which are linear-theory dominated in $P(k)$. Here, we consider only the marked statistic in real-space; the theory model presented herein may be extended to biased tracers and redshift-space, analogous to the conventional EFT \citep{2014arXiv1409.1225S,2015JCAP...11..007S,2015JCAP...09..029A,2016arXiv161009321P}, paralleling work done in configuration-space studies \citep{2016JCAP...11..057W,2020JCAP...01..006A}.

The structure of this work is as follows. In Sec.\,\ref{sec: theory} we introduce the marked density field, discussing its physical form, perturbative expansion and theoretical power spectra. Theory models are compared to data in Sec.\,\ref{sec: sim-comparison}, and Sec.\,\ref{sec: discussion} is devoted to an exploration into the information content of the mark and a discussion of the theory applicability, including its extension to biased tracers. We conclude in Sec.\,\ref{sec: summary}, with Appendices \ref{appen: simplif-and-limits},\,\ref{appen: prac-eval}\,\&\,\ref{appen: UVIRLimit} containing material pertaining to model simplifications, practical power spectrum evaluation and convergence properties.


\section{Theory Model}\label{sec: theory}
\subsection{Definition of the Marked Overdensity}\label{subsec: defn-of-mark}
The key statistic of this work is the \textit{marked density field}, defined as a weighted sum over particle positions;
\beq\label{eq: discrete-rho}
    \rho_M(\vec x) = \sum_i\delta_D(\vec x-\vec x_i)m(\vec x_i) = \int d\vec x' \left[\sum_i\delta_D(\vec x'-\vec x_i) m(\vec x')\right]\delta_D(\vec x-\vec x'),
\eeq
where $\delta_D$ is a Dirac delta and $i$ runs over all matter particles. In the above expression, $m(\vec x)$ is the mark, defined as a local overdensity as in Refs.\,\citep{2016JCAP...11..057W,2020JCAP...01..006A,2020arXiv200111024M};
\beq\label{eq: mark-def}
    m(\vec x) = \left(\frac{1+\delta_s}{1+\delta_s+\delta_R(\vec x)}\right)^p \equiv \left(1+ \frac{\delta_R(\vec x)}{1+\delta_s}\right)^{-p},
\eeq
where $\delta_R(\vec x)$ is the matter overdensity filtered on scale $R$, with $R$, the bias $\delta_s$ and the exponent $p$ \resub{being} user-defined parameters. In particular, the mark up-weights underdense regions if $p>0$. Defining the sample density field $n(\vec x') = \sum_i\delta_D(\vec x'-\vec x_i)$, Eq.\,\ref{eq: discrete-rho} can be rewritten
\beq\label{eq: rhoM-def}
    \rho_M(\vec x) = m(\vec x)n(\vec x) = m(\vec x)\bar{n}\left[1+\delta(\vec x)\right],
\eeq
where $\bar{n}=\av{n(\vec x)}$ is the average density. 

In order to convert Eq.\,\ref{eq: rhoM-def} into an \textit{overdensity} field, we require the mean density;
\beq
    \av{\rho_M(\vec x)} = \av{n(\vec x)m(\vec x)} = \bar{n}\bar{m},
\eeq
where we have defined $\bar{m}$ as $\av{n(\vec x)m(\vec x)}/\av{n(\vec x)}$, \textit{i.e.} the average of $m(\vec x)$ weighted by the number density field. The marked overdensity field is thus
\beq
    \delta_M(\vec x) \equiv \frac{\rho_M(\vec x) - \av{\rho_M}}{\av{\rho_M}} = \frac{1}{\bar{m}}m(\vec x)\left[1+\delta(\vec x)\right] - 1.
\eeq

\subsection{Perturbative Expansion}\label{subsec: pt-expansion}
We proceed to expand the marked overdensity in powers of the linear density field $\dpt{1}(\vec x)$, which will allow power spectra to be computed perturbatively. To obtain a consistent theory at one-loop accuracy, we must expand to third-order in $\dpt{1}$. First, we approximate the mark $m(\vec x)$ (Eq.\,\ref{eq: mark-def}) by its Taylor series in $\delta_R(\vec x)$, noting that $\delta_R(\vec x)$ is simply a convolution of $\delta(\vec x)$ with a window function $W_R(\vec x)$ on scale $R$;
\beq
    m(\vec x) &=& 1 - \frac{p}{1+\delta_s} \delta_R(\vec x) + \frac{p(p+1)}{2(1+\delta_s)^2}\delta_R^2(\vec x) - \frac{p(p+1)(p+2)}{6(1+\delta_s)^3}\delta_R^3(\vec x) + \mathcal{O}\left(\delta_R^4\right)\\\nonumber
    &\equiv& 1 - C_1 \delta_R(\vec x) + C_2\delta_R^2(\vec x) - C_3\delta_R^3(\vec x) + \mathcal{O}\left(\delta_R^4\right),
\eeq
defining the coefficients $C_j$, which have the general form
\beq\label{eq: mark-coeff}
    C_j = \frac{p(p+1)...(p+j-1)}{j!(1+\delta_s)^j}.
\eeq
Changing the form of the mark will simply lead to a different set of expansion coefficients.\footnote{\new{This expansion is similar to the perturbative treatment of Ref.\,\citep{2020JCAP...01..006A}, though we adopt EFT rather than Lagrangian Perturbation Theory (since we base our analysis in Fourier space) and do not normalize by the unmarked two-point correlator}.}

Before proceeding, it is important to ask ourselves the question; is the above expansion actually valid? The condition for a convergent Taylor series is simply
\beq
    \left|\frac{\delta_R(\vec x)}{1+\delta_s}\right| < 1.
\eeq
Noting that the fluctuation scale of $\delta_R$ is just the usual variance $\sigma^2_{RR}(z) = \av{\delta_R^2(\vec x)}$. We thus expect convergence if
\beq
    \sigma_{RR}(z) \lesssim (1+\delta_s).
\eeq
At high redshifts, $\sigma_{RR}(z)$ is small, thus convergence is expected for $\delta_s\geq0$, though this is not guaranteed at late times, and will depend on the choice of $R$ and $\delta_s$.

Assuming the expansion to be valid, we can write
\beq\label{eq: delta-M-expansion}
    \delta_M(\vec x) = \frac{1}{\bar{m}}\left[1+\delta(\vec x)\right]\left[1 - C_1 \delta_R(\vec x) + C_2\delta_R^2(\vec x) - C_3\delta_R^3(\vec x)  \right] - 1 + \mathcal{O}\left(\delta^4\right).
\eeq
It remains to express this in terms of the \textit{linear} density field $\dpt{1}(\vec x)$. This is achieved by expanding $\delta(\vec x)$ and $\delta_R(\vec x)$ perturbatively, and separating out each order;
\beq\label{eq: delta-expan}
    \delta(\vec x) &=& \dpt{1}(\vec x) + \dpt{2}(\vec x) + \dpt{3}(\vec x) + \dpt{ct}(\vec x) + \mathcal{O}\left(\left[\dpt{1}(\vec x)\right]^4\right)\\\nonumber
    \delta_R(\vec x) &=& \dptR{1}(\vec x) + \dptR{2}(\vec x) + \dptR{3}(\vec x) + \dptR{ct}(\vec x) + \mathcal{O}\left(\left[\dptR{1}(\vec x)\right]^4\right).
\eeq
Here $\dpt{n}$ and $\dptR{n}$ are $n$-th order contributions that include $n$ copies of $\dpt{1}$ or $\dptR{1}$.\footnote{Note that we do not include the non-perturbative effects of long wavelength modes in this work, \textit{i.e.} we do not perform IR resummation.} Furthermore, by distributivity, each $\dptR{n}$ is just the convolution of $\dpt{n}$ with $W_R(\vec x)$. We have additionally introduced the third-order \textit{counterterm} of effective field theory (EFT) in Eq.\,\ref{eq: delta-expan}; this accounts for the effect of short-scale physics on large-scale modes \citep{2012JHEP...09..082C,2012JCAP...07..051B} and will be discussed below.

Inserting these expansions into Eq.\,\ref{eq: delta-M-expansion} and collecting terms of equal perturbative order, we obtain;
\beq\label{eq: delta-expan2}
    \delta_M(\vec x) &\equiv& \left(\frac{1}{\bar{m}}-1\right) + \frac{1}{\bar{m}}\left(\dptM{1}(\vec x) + \dptM{2}(\vec x) + \dptM{3}(\vec x) + \dptM{ct}(\vec x)\right)\\\nonumber
    \dptM{1}(\vec x) &=& \left[\dpt{1} - C_1\dptR{1}\right](\vec x)\\\nonumber
    \dptM{2}(\vec x) &=& \left[\dpt{2} - C_1\dptR{2} - C_1\dpt{1}\dptR{1} + C_2\dptR{1}\dptR{1}\right](\vec x)\\\nonumber 
    \dptM{3}(\vec x) &=& \left[\dpt{3}-C_1\dptR{3} - C_1\dpt{1}\dptR{2} - C_1\dpt{2}\dptR{1} + C_2\dpt{1}\dptR{1}\dptR{1} + 2C_2\dptR{1}\dptR{2} - C_3\dptR{1}\dptR{1}\dptR{1}\right](\vec x)\\\nonumber
    \dptM{ct}(\vec x) &=& \left[\dpt{ct} - C_1 \dptR{ct}\right](\vec x).
\eeq

In Fourier space,\footnote{In this paper, we define the Fourier and inverse Fourier transforms as
\beq
    X(\vec k) &=& \int d\vec x\,e^{-i\vec k\cdot\vec x}X(\vec x),  \qquad
    X(\vec x) = \int \frac{d\vec k}{(2\pi)^3}e^{i\vec k\cdot\vec x}X(\vec k)\nonumber
\eeq
and the Dirac function $\delta_D$ via 
\beq
    \int d\vec x\,e^{i(\vec k_1-\vec k_2)\cdot\vec x} = (2\pi)^3\delta_D(\vec k_1-\vec k_2).\nonumber
\eeq
The correlation function and power spectrum of the density field are defined as 
\beq
    \xi(\vec r) = \av{\delta(\vec x)\delta(\vec x+\vec r)},\qquad (2\pi)^3\delta_D(\vec k+\vec k')P(\vec k) = 
    \av{\delta(\vec k)\delta(\vec k')}\nonumber
\eeq
with the power spectrum as the Fourier transform of the correlation function and higher order correlators being defined similarly.
} these products may be written as convolutions using the convolution operator $\ast$, here defined by
\beq
    \left[X\ast Y\right](\vec k) &=& \int_{\vec p} X(\vec p)Y(\vec k-\vec p)\\\nonumber
    \left[X\ast Y\ast Z\right](\vec k) &=& \int_{\vec p_1\vec p_2} X(\vec p_1)Y(\vec p_2)Z(\vec k-\vec p_1-\vec p_2)
\eeq
where we denote $\int_{\vec p}\equiv (2\pi)^{-3} \int d\vec p $. Noting that $\delta_R(\vec k) = W(kR)\delta(\vec k)$ for Fourier-space window $W(kR)$, this yields
\beq
    \delta_M(\vec k) &\equiv& \delta_D(\vec k)\left(\frac{1}{\bar{m}}-1\right)(2\pi)^3 + \frac{1}{\bar{m}}\left(\dptM{1}(\vec k) + \dptM{2}(\vec k) + \dptM{3}(\vec k) + \dptM{ct}(\vec k)\right)\\\nonumber
    \dptM{1}(\vec k) &=& \left[1-C_1W(kR)\right]\dpt{1}(\vec k)\\\nonumber
    \dptM{2}(\vec k) &=& \left[1-C_1W(kR)\right]\dpt{2}(\vec k) - C_1\left[\dpt{1}\ast\dptR{1}\right](\vec k) + C_2 \left[\dptR{1}\ast\dptR{1}\right](\vec k) \\\nonumber
    \dptM{3}(\vec k) &=& \left[1-C_1W(kR)\right]\dpt{3}(\vec k) + 2C_2\left[\dptR{1}\ast\dptR{2}\right](\vec k) - C_3\left[\dptR{1}\ast\dptR{1}\ast\dptR{1}\right](\vec k)\\\nonumber
    &&\,- C_1\left[\dpt{1}\ast\dptR{2}\right](\vec k) - C_1\left[\dpt{2}\ast\dptR{1}\right](\vec k)+ C_2\left[\dpt{1}\ast\dptR{1}\ast\dptR{1}\right](\vec k) \\\nonumber
    \dptM{ct}(\vec k) &=& \left[1-C_1W(kR)\right]\dpt{ct}(\vec k).
\eeq
In standard perturbation theory (hereafter SPT), the $n$-th order density field $\dpt{n}$ is written in terms of $\dpt{1}$ and the Fourier space kernels $F_{n}$ (tabulated in \citet{2002PhR...367....1B});
\beq\label{eq: SPT-expansion}
    \dpt{n}(\vec k) = \int_{\vec p_1...\vec p_n}\,F_{n}(\vec p_1,...,\vec p_n)\dpt{1}(\vec p_1)...\dpt{1}(\vec p_n)\delta_D(\vec p_1+...+\vec p_n-\vec k),
\eeq
with the counterterm being given by $\dpt{ct}(\vec k) = -c_s^2k^2\dpt{1}(\vec k)$ for effective sound-speed $c_s^2$ \citep{2012JHEP...09..082C,2012JCAP...07..051B}. By analogy with Eq.\,\ref{eq: SPT-expansion}, it is convenient to defined the marked kernels $H_n$ such that
\beq
    \dptM{n}(\vec k) = \int_{\vec p_1...\vec p_n}\,H_{n}(\vec p_1,...,\vec p_n)\dpt{1}(\vec p_1)...\dpt{1}(\vec p_n)\delta_D(\vec p_1+...+\vec p_n-\vec k);
\eeq
the resulting forms are given by
\beq\label{eq: H-kernels}
    H_1(\vec p_1) &=& 1-C_1W(p_1R)\\\nonumber
    H_2(\vec p_1,\vec p_2) &=& \left[1-C_1W(p_{12}R)\right]F_2(\vec p_1,\vec p_2) + C_2W(p_1R)W(p_2R) - \frac{C_1}{2}\left[W(p_1R)+W(p_2R)\right]\\\nonumber
    H_3(\vec p_1,\vec p_2,\vec p_3) &=& \left[1-C_1W(p_{123}R)\right]F_3(\vec p_1,\vec p_2,\vec p_3) - C_3W(p_1R)W(p_2R)W(p_3R) \\\nonumber
    &&\,+\frac{1}{3}\left\{_{}^{}C_2W(p_2R)W(p_3R)-C_1\left[W(p_1R)+W(p_{23}R)\right]F_2(\vec p_2,\vec p_3)\right.\\\nonumber
    &&\,\quad\quad\left._{}^{}+2C_2W(p_1R)W(p_{23}R)F_2(\vec p_2,\vec p_3)+\text{2 cyc.}\right\}
\eeq
where $\vec p_{i..k} \equiv \vec p_i+...+\vec p_k$, $p_i\equiv |\vec p_i|$ and `cyc.' denotes cyclic permutations over $\{\vec p_1,\vec p_2,\vec p_3\}$. Note also that the counterterm becomes $\dptM{ct}(\vec k) = -c_s^2k^2H_1(\vec k)\dpt{1}(\vec k)$.

Before continuing, it is important to discuss whether the above results remain valid for cosmologies incorporating massive neutrinos, given that searching for neutrinos is one of the prime applications of the marked spectrum. Massive neutrinos alter the perturbation theory in a number of ways, most notably by giving scale dependence to the linear growth factor $D(z)$, as a consequence of the scale of neutrino damping being set by the redshift at which neutrinos become non-relativistic. This complicates the formalism as the temporal- and spatial-parts are no longer separable, properly requiring computation of scale-dependent Green's functions 
\citep{2009PhRvD..80h3528S,2014JCAP...11..039B,2015JCAP...03..046F,2017arXiv170704698S}. However, a simpler approach has been adopted by a number of works \citep{2008PhRvL.100s1301S,2014JCAP...11..039B,2019JCAP...11..034C}, evaluating the perturbation theory with the usual (massless neutrino) perturbative kernels but using the full linear power spectrum computed in the presence of massive neutrinos present. The leading-order differences between this and the full approach can absorbed into the $c_s^2$ counterterm, as shown in the full EFT calculation of Ref.\,\citep{2017arXiv170704698S}. This approach will be adopted herein.

\subsection{Power Spectrum of $\delta_M(\vec x)$}\label{subsec: pk-def}
Given the above Fourier-space decomposition, the marked power spectrum can be computed as
\beq
     M(\vec k) = \left|\delta_M(\vec k)\right|^2 &=& \frac{1}{\bar{m}^2}\left|\dptM{1}(\vec k)+\dptM{2}(\vec k)+\dptM{3}(\vec k)+\dpt{ct}(\vec k)\right|^2 + M_\mathrm{shot},
\eeq
ignoring the zero-lag terms contributing only to $\vec k = \vec 0$, and including a shot-noise term that will be discussed below. As for the matter power spectrum, this may be written as a sum of linear contributions, one-loop contributions and counterterms. To see this, first note that $\dptM{n}$ contains $n$ copies of the Gaussian field $\dpt{1}$ and the expectation of any odd number of fields is zero, via Wick's theorem. The general expansion leads to the following form;
\beq
    M(\vec k) = \frac{1}{\bar{m}^2}\left[M_{11}(\vec k)+M_{22}(\vec k)+2M_{13}(\vec k) + 2M_{ct}(\vec k)\right] + M_\mathrm{shot},
\eeq
subject to the definitions
\beq\label{eq: M-ij-def}
    M_{11}(\vec k) &=& H_1^2(\vec k)P_L(\vec k)\\\nonumber
    M_{22}(\vec k) &=& 2\int_{\vec p}\left|H_2(\vec p,\vec k-\vec p)\right|^2P_L(\vec p)P_L(\vec k-\vec p)\\\nonumber
    M_{13}(\vec k) &=& 3H_1(\vec k)P_L(\vec k)\int_{\vec p}H_3(\vec p,-\vec p,\vec k)P_L(\vec p)\\\nonumber
    M_{ct}(\vec k) &=& -c_s^2k^2H_1^2(\vec k)P_L(\vec k).
\eeq
This makes further use of Wick's theorem, and uses the definition of the linear power spectrum; $P_L(\vec k) = \left|\dpt{1}(\vec k)\right|^2$.

It is instructive to consider the linear piece, which, in full is given by\footnote{Here, the `linear' spectrum refers to that computed from the first-order pieces of $\delta_M$, i.e. $\delta_M(\vec x) = \dptM{1}(\vec x)$. Due to the presence of higher-order terms in the expansion of the mark, this is not equal to the spectrum obtained by assuming the underlying density fields $\delta(\vec x)$ and $\delta_R(\vec x)$ to be linear (except at $\mathcal{O}\left(\left[\dpt{1}(\vec x)\right]^2\right)$).}
\beq\label{eq: M_lin-def}
    M_L(\vec k) = \frac{1}{\bar{m}^2}\left[1-C_1W(kR)\right]^2P_L(\vec k).
\eeq
Given that this is a a simple prefactor multiplying the linear matter power spectrum, this may be simply estimated using a Boltzmann solver code (e.g., \texttt{CAMB} or \texttt{CLASS} \citep{2011ascl.soft02026L,2011JCAP...07..034B}). A curious feature of this piece is that it vanishes exactly at a specific value of $k$, satisfying $C_1W(kR)=1$. Since $0\leq W(kR)\leq 1$ for all $k$, this occurs if $C_1 \geq 1$. Providing $\delta_s > -1$ (as would be physically reasonable) this will not occur for negative $p$, but is a notable feature for marked spectra with positive indices and small $\delta_s$, such that $p\geq (1+\delta_s)$ (from Eq.\,\ref{eq: mark-coeff}). In practice, $M(\vec k)$ will not be exactly zero, since some higher-loop terms (and shot-noise) will be non-zero. 

It remains to consider the shot-noise term, here idealized as Poissonian (but see the discussion in Sec.\,\ref{sec: sim-comparison}). This is most easily derived in configuration space, noting that $M(\vec k)$ is the Fourier-transform of the correlation function $V^{-1}\int d\vec x\,\av{\delta_M(\vec x)\delta_M(\vec x+\vec r)}$. Focusing on the piece containing $\rho_M(\vec x)\rho_M(\vec x+\vec r)$ this may be written
\beq
    \rho_M(\vec x)\rho_M(\vec x+\vec r) = \left(\sum_i\delta_D(\vec x-\vec x_i)m(\vec x_i)\right)\left(\sum_j\delta_D(\vec x+\vec r-\vec x_j)m(\vec x_j)\right)
\eeq
in the discrete form of Eq.\,\ref{eq: discrete-rho}. We now isolate the $i=j$ term of the summation, yielding
\beq
    \rho_M(\vec x)\rho_M(\vec x+\vec r) = \sum_{i\neq j}\delta_D(\vec x-\vec x_i)\delta_D(\vec x+\vec r-\vec x_j)m(\vec x_i)m(\vec x_j) + \sum_i \delta_D(\vec r)m^2(\vec x_i).
\eeq
Returning to continuous form and averaging over $\vec x$, we obtain
\beq
    \int \frac{d\vec x}{V}\av{\rho_M(\vec x)\rho_M(\vec x+\vec r)} = \int \frac{d\vec x}{V}\,\av{n(\vec x)n(\vec x+\vec r)m(\vec x)m(\vec x+\vec r)} + \int \frac{d\vec x}{V}\,\av{n(\vec x)m^2(\vec x)}\delta_D(\vec r).
\eeq
The second term, equal to $\av{m^2n}\delta_D(\vec r) = \overline{m^2}\bar{n}\delta_D(\vec r)$, is the previously ignored shot-noise term. Returning to Fourier-space, and inserting the correct normalizing factors, it has the simple $k$-independent form $M_\mathrm{shot} = \overline{m^2}/\left[\bar{m}^2\bar{n}\right]$, which can be simply computed from the data. For the matter power spectrum, $k$-independent Poisson shot-noise is often found to be a poor model; we discuss this in Sec.\,\ref{sec: sim-comparison}. 

\subsection{Cross-Spectrum with the Matter Field}\label{subsec: cross-spec-def}
An additional statistic of interest is the cross-spectrum of $\delta_M(\vec k)$ and $\delta(\vec k)$. This is given by
\beq
    \mathcal{C}(\vec k) &=& \delta_M(\vec k)\delta^*(\vec k)\\\nonumber
    &=& \frac{1}{\bar m}\left[\mathcal{C}_{11}(\vec k)+\mathcal{C}_{22}(\vec k)+\mathcal{C}_{13}(\vec k)+\mathcal{C}_{31}+2\mathcal{C}_{ct}(\vec k)\right] + \mathcal{C}_\mathrm{shot},
\eeq
defining
\beq\label{eq: Cross-spec-terms}
    \mathcal{C}_{11}(\vec k) &=& H_1(\vec k)P_L(\vec k)\\\nonumber
    \mathcal{C}_{22}(\vec k) &=& 2\int_{\vec p} H_2(\vec p,\vec k-\vec p)F_2(\vec p,\vec k-\vec p)P_L(\vec p)P_L(\vec k-\vec p)\\\nonumber
    \mathcal{C}_{13}(\vec k) &=& 3H_1(\vec k)P_L(\vec k)\int_{\vec p}F_3(\vec k,\vec p,-\vec p)P_L(\vec p)\\\nonumber
    \mathcal{C}_{31}(\vec k) &=& 3P_L(\vec k)\int_{\vec p}H_3(\vec k,\vec p,-\vec p)P_L(\vec p)\\\nonumber
    \mathcal{C}_{ct}(\vec k) &=& -k^2c_s^2H_1(\vec k)P_L(\vec k),
\eeq
in an analogous manner to before, though with a broken $\mathcal{C}_{13}$ and $\mathcal{C}_{31}$ symmetry. The Poissonian shot-noise term can be estimated in a similar fashion to the above, here arising from the term $\av{\rho_M(\vec x)n(\vec x+\vec r)}$, which gives
\beq
    \rho_M(\vec x)n(\vec x+\vec r) &=& \sum_{i\neq j}\delta_D(\vec x-\vec x_i)\delta_D(\vec x+\vec r-\vec x_j)m(\vec x_i) + \sum_i\delta_D(\vec r)m(\vec x_i)\\\nonumber
    \int \frac{d\vec x}{V}\av{\rho_M(\vec x)n(\vec x+\vec r)} &=& \int \frac{d\vec x}{V}\av{n(\vec x)n(\vec x+\vec r)m(\vec x+\vec r)} + \int \frac{d\vec x}{V}\av{n(\vec x)m(\vec x)}\delta_D(\vec r).
\eeq
The second term, equal to $\bar{n}\bar m\delta_{D}(\vec r)$, leads to the Poisson shot-noise contribution $\mathcal{C}_\mathrm{shot} = 1/\bar{n}$.

\subsection{Practical Evaluation}
\new{Some simplification of the above results is needed before they can be robustly compared to data. As shown in Appendix \ref{appen: simplif-and-limits}, the marked spectra may be written in terms of a set of convolution integrals that can be efficiently evaluated using FFTLog methods \citep{2000MNRAS.312..257H,2016PhRvD..93j3528S,2016JCAP...09..015M}, discussion of which is found in Appendix \ref{appen: prac-eval}. Furthermore, as with any calculation in EFT, it is important to ensure that the loop integrals appearing in the one-loop terms are well defined; \textit{i.e.} that they are independent of the momentum cut-off applied. If this is not the case, counterterms are required. As shown in Appendix \ref{appen: UVIRLimit}, the limiting behavior of the marked power spectrum with `hard' internal momenta, $p\gg k$, is given by}
\beq\label{eq: UV-limit}
    \left(\frac{M_{22}(\vec k)}{\left(1-C_1W(kR)\right)^2}\right)_\mathrm{UV} &=& \frac{9k^4}{98}\int \frac{dp}{2\pi^2}\frac{P_L^2(p)}{p^2} + (\text{Gaussian suppressed})\\\nonumber
    \left(\frac{M_{13}(\vec k)}{\left(1-C_1W(kR)\right)^2P_L(\vec k)}\right)_\mathrm{UV} &=& -\frac{61}{210}k^2\sigma_v^2 + (\text{Gaussian suppressed}),
\eeq
assuming a Gaussian window function \resub{and defining $\sigma_v^2\equiv (6\pi^2)^{-1}\int dp\,P_L(p)$}. For a power-law cosmology with $P_L(p)\propto p^{n}$ these expressions are convergent for $n<-1$, which holds in our universe, with $n \approx -2.1$ beyond the non-linear scale. Notably, due to the smoothing window, the marked power spectrum does \textit{not} contain any additional UV divergences relative to $P(k)$, and thus requires no additional counterterms or renormalization.

\section{Results}\label{sec: sim-comparison}
We may now compare the predictions of the above model to data, and test our assumptions. For this, we make use of simulations from the \texttt{Quijote} project \citep{2019arXiv190905273V}, a suite of over $40,000$ $N$-body simulations spanning a wide variety of cosmologies. Here, we use 50 $1h^{-3}\mathrm{Gpc}^3$ boxes of the fiducial cosmology $\{\Omega_m = 0.3175, \Omega_b = 0.049, h = 0.6711, n_s = 0.9624, \sigma_8 = 0.834, M_\nu = 0\,\mathrm{eV}, w = -1\}$, each of which contains $512^3$ cold dark matter (CDM) particles evolved from $z = 127$ with initial conditions obtained using second-order Lagrangian perturbation theory (2LPT). We principally consider redshifts $0.5, 1,$ and $2$, as these are of greatest relevance for upcoming large scale structure surveys. To examine our treatment of massive neutrinos, we additionally consider 50 \texttt{Quijote} simulations with the same cosmology except with three degenerate neutrinos of total mass $M_\nu = 0.1$\,eV. These boxes contain $512^3$ each of CDM and neutrino particles, with initial conditions generated at $z = 127$ using Zel`dovich perturbation theory. Only the power spectrum of CDM is used here, again considering the eventual application to large scale structure data-sets.

For each realization, marked power spectra are computed as in \citet{2020arXiv200111024M}, additionally recording the value of the density-weighted $\bar{m}$ parameter (needed for normalization) which is simply the mean mark across all simulation particles.
On large scales (and at moderate-to-large redshifts) shot-noise in $N$ body simulations is known to exhibit strong scale dependence, and is far from Poissonian on all but the smallest scales, due to memory of the original particle grid \citep{2005ApJ...634..728S}. For this reason, we set the shot-noise terms of Sec.\,\ref{sec: theory} to zero henceforth; a valid assumption for the $k$-range probed in this work. Note that shot noise could affect the marked spectrum on \textit{large} scales, unlike for the full power spectrum, since the low-$k$ amplitude is suppressed.

\subsection{Choice of Mark Parameters}\label{subsec: mark-param-choice}
The mark adopted in this work (Eq.\,\ref{eq: mark-def}) has three free parameters; the smoothing scale $R$, the exponent $p$ and the shift $\delta_s$. As noted in \citet{2020arXiv200111024M}, using $p>0$ up-weights low density regions, leading to greater cosmological constraining power, thus this will be assumed herein. For simplicity, we consider only three sets of parameters in this work; $\{p = 1, R = 15\Mpch, \delta_s = 0.25\}$, $\{p = 2, R = 15\Mpch,\delta_s = 0.25\}$ and $\{p = 1, R = 30\Mpch, \delta_s = 0.25\}$. The second of these is similar to the optimal configuration found in Ref.\,\citep{2020arXiv200111024M}, with others used to highlight important parameter dependencies.\footnote{The effective smoothing scale used here is greater than that of Ref.\,\citep{2020arXiv200111024M} due to the use of a different window function. Expanding the Fourier-space Gaussian and top-hat window functions to quadratic order shows that the characteristic width of a Gaussian filter is $\sqrt{5}$ times that of a top-hat filter with the same value of $R$. The density field variances $\sigma_{R}^2$ and $\sigma_{RR}^2$ are thus larger with the top-hat window function. Using $R_\mathrm{TH} = \sqrt{5}R_\mathrm{Gaussian}$ largely accounts for this difference.} As noted in Sec.\,\ref{subsec: pt-expansion}, our theoretical model relies on the Taylor expansion of the mark being convergent, requiring $\sigma_{RR}(z)\leq (1+\delta_s)$. For the above parameters (using a Gaussian window function and assuming linear physics), we obtain $\sigma_{RR}(z) = 0.261\times D(z)$ ($0.015\times D(z)$) for $R = 15\Mpch$ ($30\Mpch$) where $D(z)$ is the usual scale-independent growth function. In practice, we also need the function $\sigma_R(z) = \av{\delta(\vec x)\delta_R(\vec x)}$ to be small; this is equal to $\sigma_{R}(z) = 0.365\times D(z)$ ($0.034\times D(z)$) for $R = 15\Mpch$ ($30\Mpch$). Whilst all parameter sets satisfy the convergence inequality, we expect greater contribution from higher loop terms at low-$z$, and percent-level agreement between data and observations is not expected for the smaller value of $R$. 

A second issue becomes clear when considering the models for $M(k)$ and $\mathcal{C}(k)$ at linear order. As previously noted, these vanish exactly at the wavenumber satisfying $(1 - C_1W(kR)) = 0$, which has non-trivial solutions for $p\geq (1+\delta_s)$. This also gives zero contributions from the all the $13$-type components of $M(k)$ and $\mathcal{C}(k)$, as well as the $P_{22}$ component. Physically, this occurs due to a cancellation between the non-linear density field $\delta$ and one power of the non-linear smoothed field $\delta_R$, which occurs at all orders in perturbation theory. At these locations, the power spectrum is set by contributions from terms including higher powers of $\delta_R$, thus its accuracy is less clear. For the parameter sets given above, only the second satisfies this inequality, thus it will be expected to show a zero point at finite $k$. 

\subsection{Marked Spectrum Components}
\begin{figure}
    \centering
    \subfloat[$p = 1, R = 15\Mpch, \delta_s = 0.25$]{\includegraphics[width=0.33\linewidth]{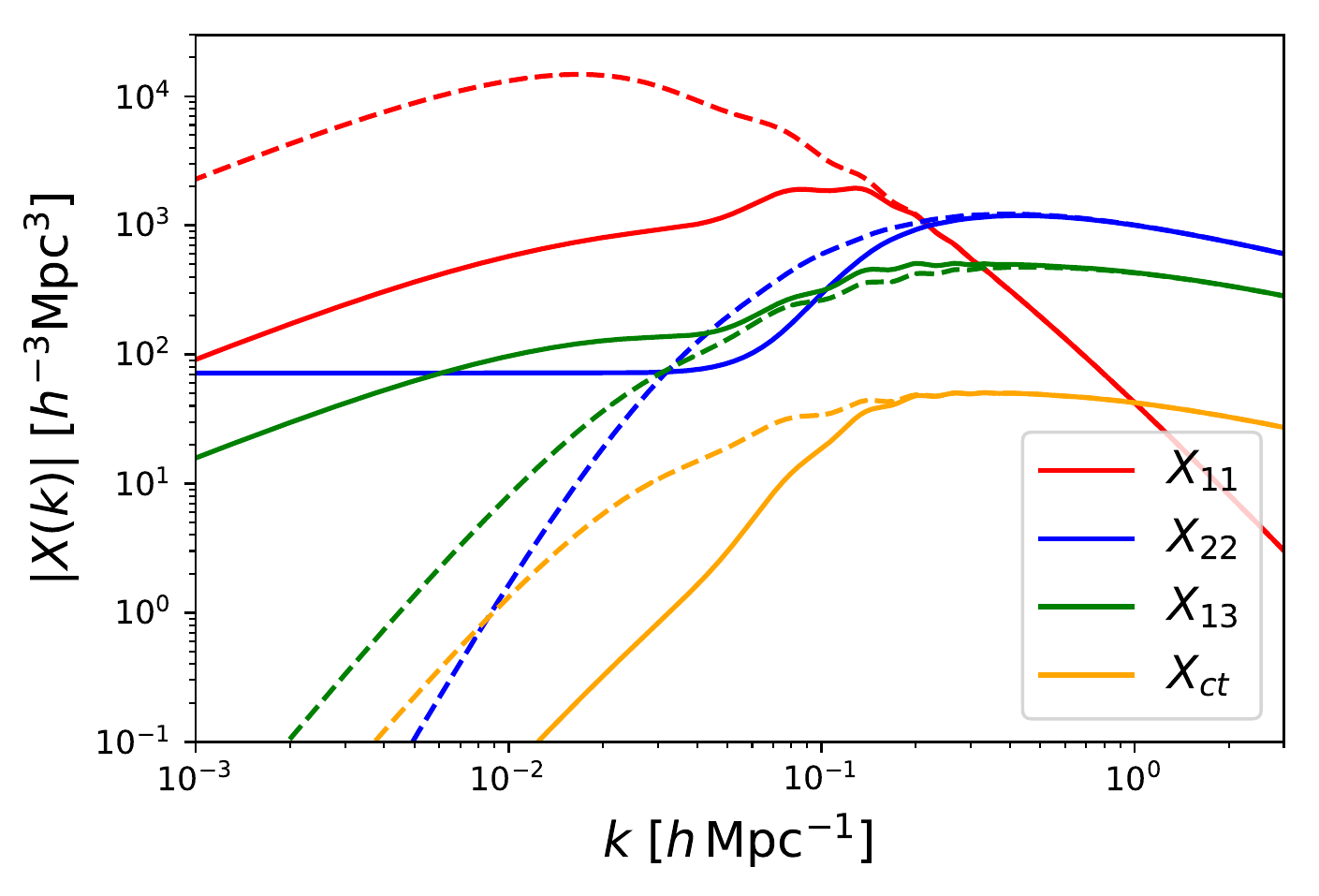}}
    \subfloat[$p = 2, R = 15\Mpch, \delta_s = 0.25$]{\includegraphics[width=0.33\linewidth]{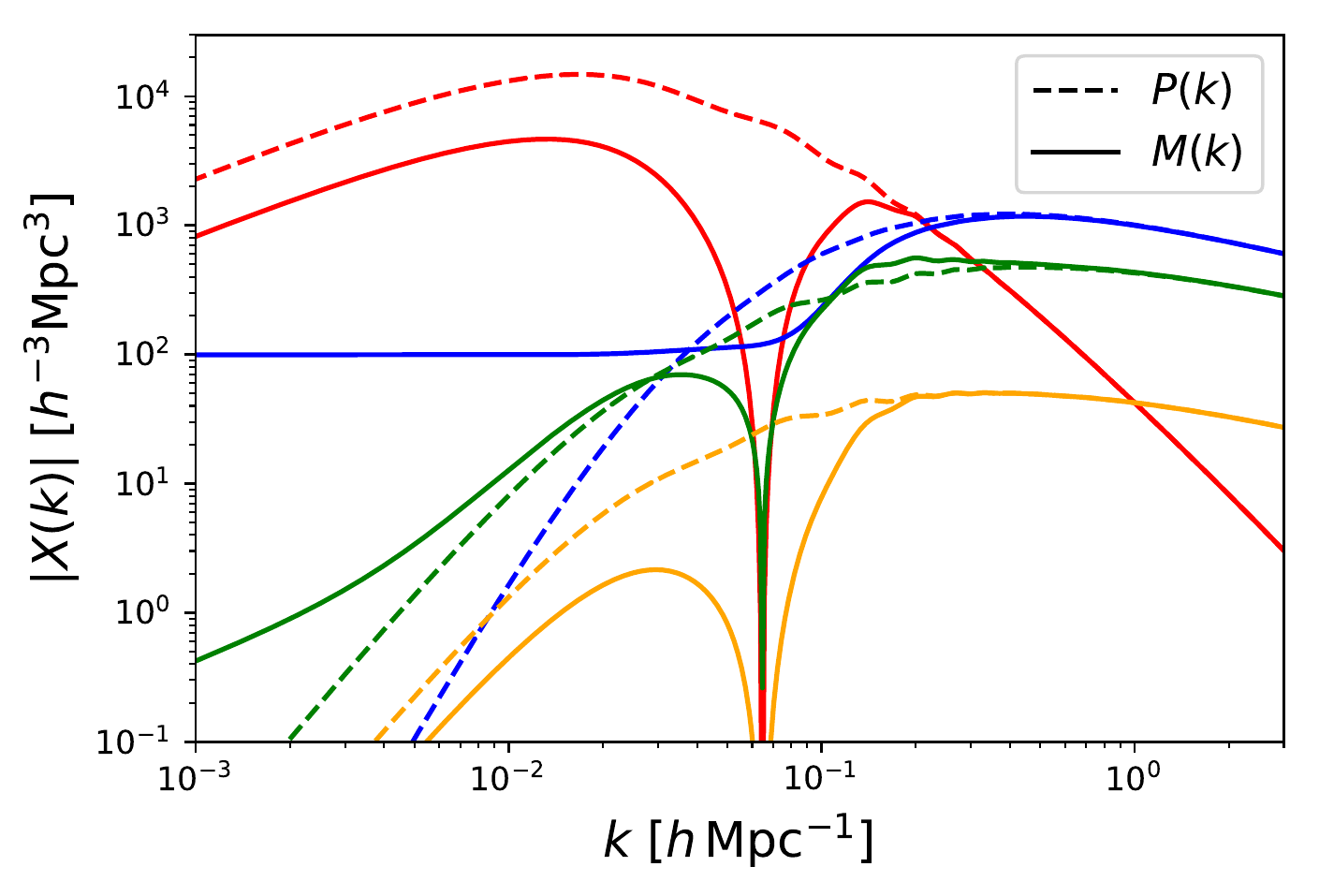}}
    \subfloat[$p = 1, R = 30\Mpch, \delta_s = 0.25$]{\includegraphics[width=0.33\linewidth]{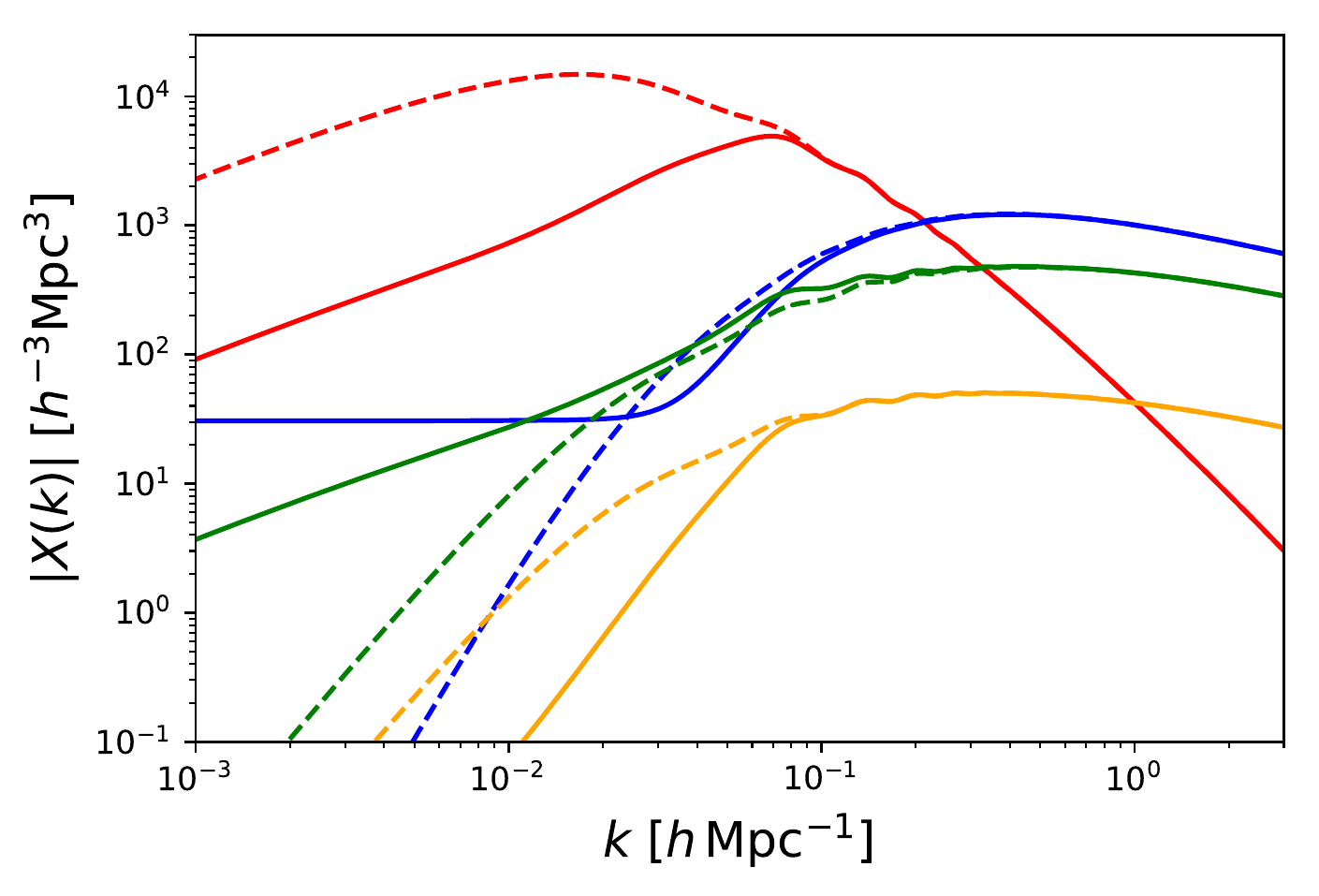}}\\
    \caption{Components of the marked (complete lines) and full (dashed lines) matter power spectrum at $z = 0.5$ for three sets of mark parameters, indicated by the captions. The linear $11$-type contributions are marked in red, whilst the one-loop $22$-type and $13$-type pieces are shown in blue and green respectively. The counterterms are shown in orange, and do not include the $-c_s^2$ prefactor. Spectra are computed for the fiducial cosmology given in the text. Note that we plot the modulus of the components, \resub{since the $13$-type contributions are negative}. For the second set of parameters, the $11$- and $13$-type marked power spectrum components vanish at $k\approx 0.07\hMpc$ due to cancellation between the smoothed and unsmoothed density fields.}\label{fig: spectral_components}
\end{figure}

Given the above parameters, the theory model of Sec.\,\ref{sec: theory} can be computed. Before comparing the results with simulations, it is instructive to consider the magnitudes of the various $M(k)$ and $P(k)$ terms, which are shown in Fig.\,\ref{fig: spectral_components} for the three parameter sets at $z = 0.5$. We note good agreement between the linear ($11$-type) pieces of the marked and full power spectrum at large-$k$; this is expected as the $(1-C_1W(kR))^2$ prefactor in Eq.\,\ref{eq: M_lin-def} asymptotes to unity, or, more physically, because the marked density field is dominated by the unsmoothed component, which sources $P(k)$. At lower $k$, there is a significant reduction in power, with a fractional reduction by $(1-C_1)^2$ as $k\rightarrow0$ due to the density field smoothing. Comparison between Figs.\,\ref{fig: spectral_components}a\,\&\,\ref{fig: spectral_components}c show the impact of increasing the smoothing window; the linear theory is less suppressed on large scales, though has the same asymptotic limit. Furthermore, as noted in Sec.\,\ref{subsec: mark-param-choice}, the linear component of $M(k)$ in the second parameter set shows a clear zero at $k\sim 0.07\hMpc$, due to $(1-C_1W(kR))$ crossing zero.

Similar considerations apply for the one-loop components $M_{13}(k)$ and $M_{22}(k)$. In all cases, the small-scale behavior matches that of $P(k)$, whilst there are significant differences in the lower $k$ regime. In particular, the presence of additional non-linear contributions gives \resub{excess power for $k\lesssim 0.1\hMpc$}, and we see that $M_{13}$ reaches zero at $k\sim 0.07\hMpc$ with the second set of mark parameters. Of note is the low-$k$ behavior; the full spectrum is \textit{not} dominated by linear theory at low-$k$, as the $M_{22}$ term becomes large and constant. As discussed in Sec.\,\ref{subsec: where-is-information?}, \resub{this is due to the 
\textit{contact terms} (involving two fields evaluated at the same location) in $M_{22}$ (Eq.\,\ref{eq: M22-simpl}), which asymptote to constants in the limit of $k\rightarrow 0$, as discussed in Appendix \ref{appen: simplif-and-limits}. Furthermore, $M_{13}$ scales as $P(k)$ at low $k$, thus the one-loop terms always remain important (unlike for the unmarked spectrum).} On the largest scales, accurate predictions can only be made if we have an understanding of higher-order terms, \resub{both Gaussian and non-Gaussian}. This is very different to standard perturbative analyses, and complicates the analysis, though the effect is reduced by using larger \resub{smoothing scale} $R$.\footnote{\resub{An additional approach to ameliorate this would be to add a free constant encapsulating the low-$k$ limit, analogous to the $\delta^2$ counterterm in the (power spectrum) one-loop EFT of biased tracers. Whilst this is not strictly necessary, as the contact terms are not formally divergent due to the smoothing, it may significantly help with the modeling.}}

\subsection{Comparison to Simulations}
\begin{figure}
    \centering
    \subfloat[$z = 0.5$]{\includegraphics[width=0.33\linewidth]{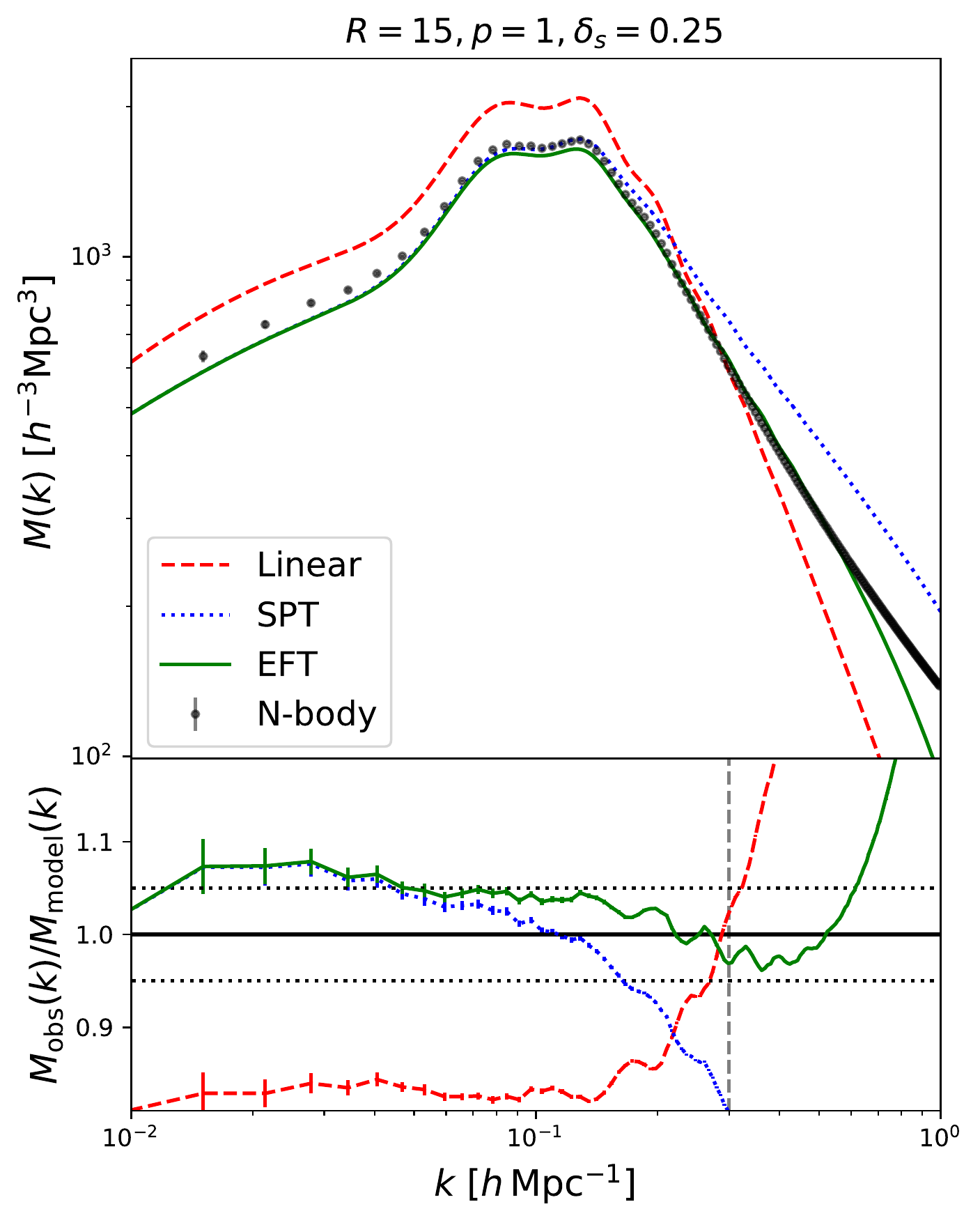}}
    \subfloat[$z = 1$]{\includegraphics[width=0.33\linewidth]{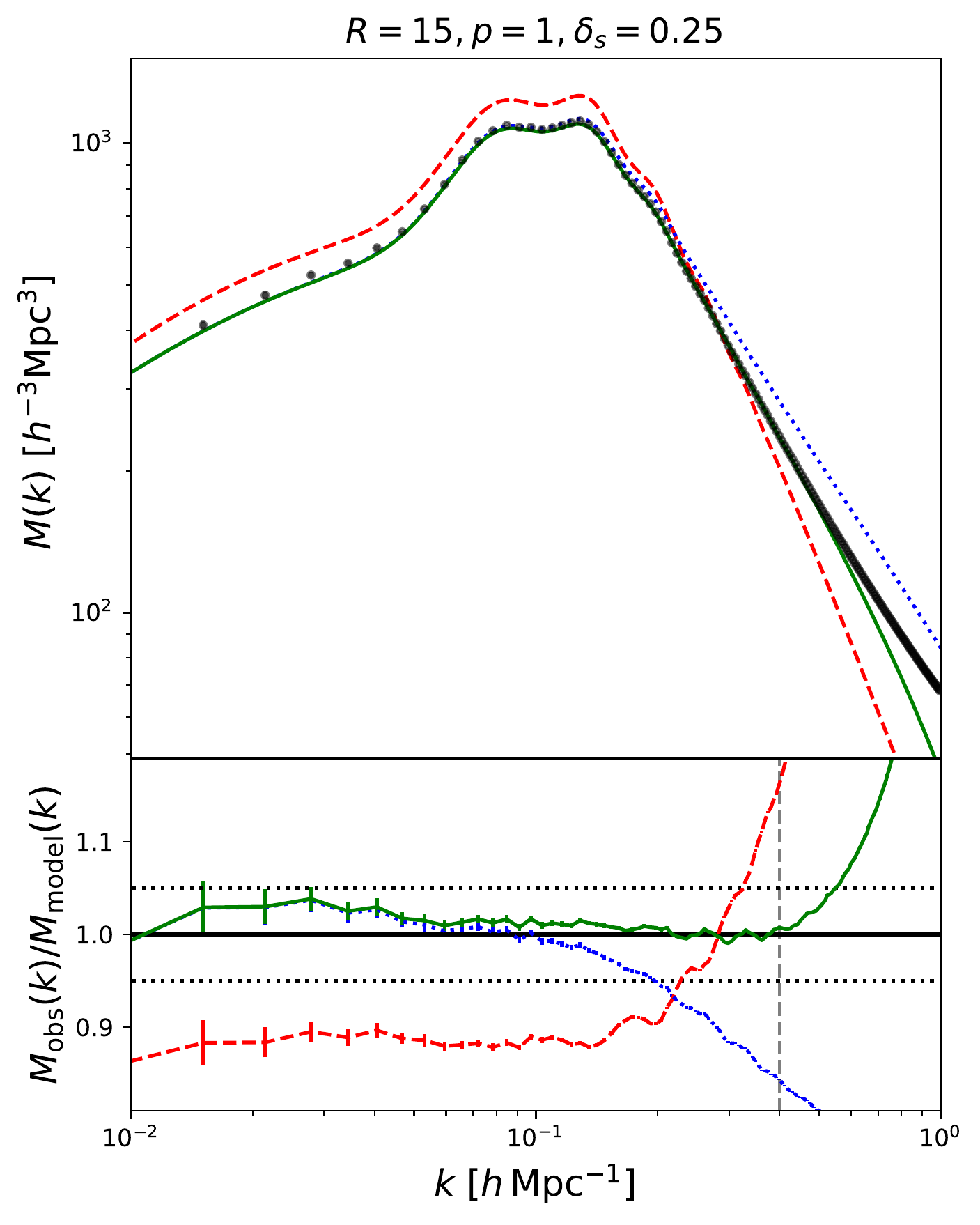}}
     \subfloat[$z = 2$]{\includegraphics[width=0.33\linewidth]{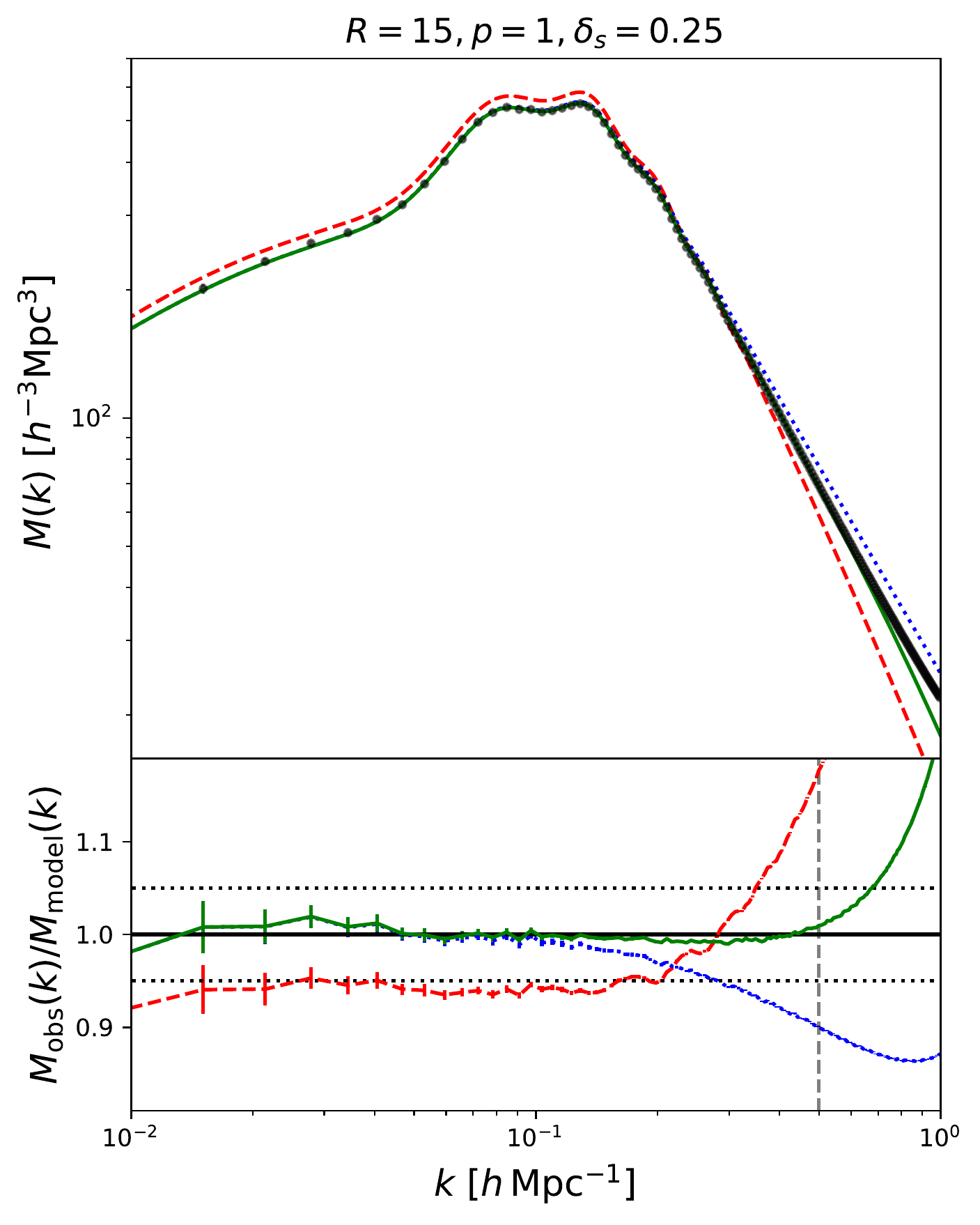}}\\
    \caption{Comparison of simulated (points) and model (lines) marked power spectra across a variety of redshifts, assuming the parameter set $\{p = 1, R = 15\Mpch, \delta_s = 0.25\}$. Predictions using linear theory, one-loop SPT and one-loop EFT are shown in dashed red, dotted blue and solid green respectively, and the simulation results are taken from 50 \texttt{Quijote} simulations. The top plots show the complete spectra, with the ratio of simulation to model shown in the bottom panel, with dashed lines indicating $5\%$ agreement. For the EFT models, we fit the counterterm $c_s^2$ using all \resub{(unmarked) power spectrum modes with wavenumbers} to the left of the vertical dashed line in the lower plots. We find $c_s^2 = 1.00$, $0.51$ and $0.21$ for the three panels respectively, in $h^{-2}\mathrm{Mpc}^2$ units.}\label{fig: param_1_M}
\end{figure}

\begin{figure}
    \centering
    \subfloat[$z = 0.5$]{\includegraphics[width=0.33\linewidth]{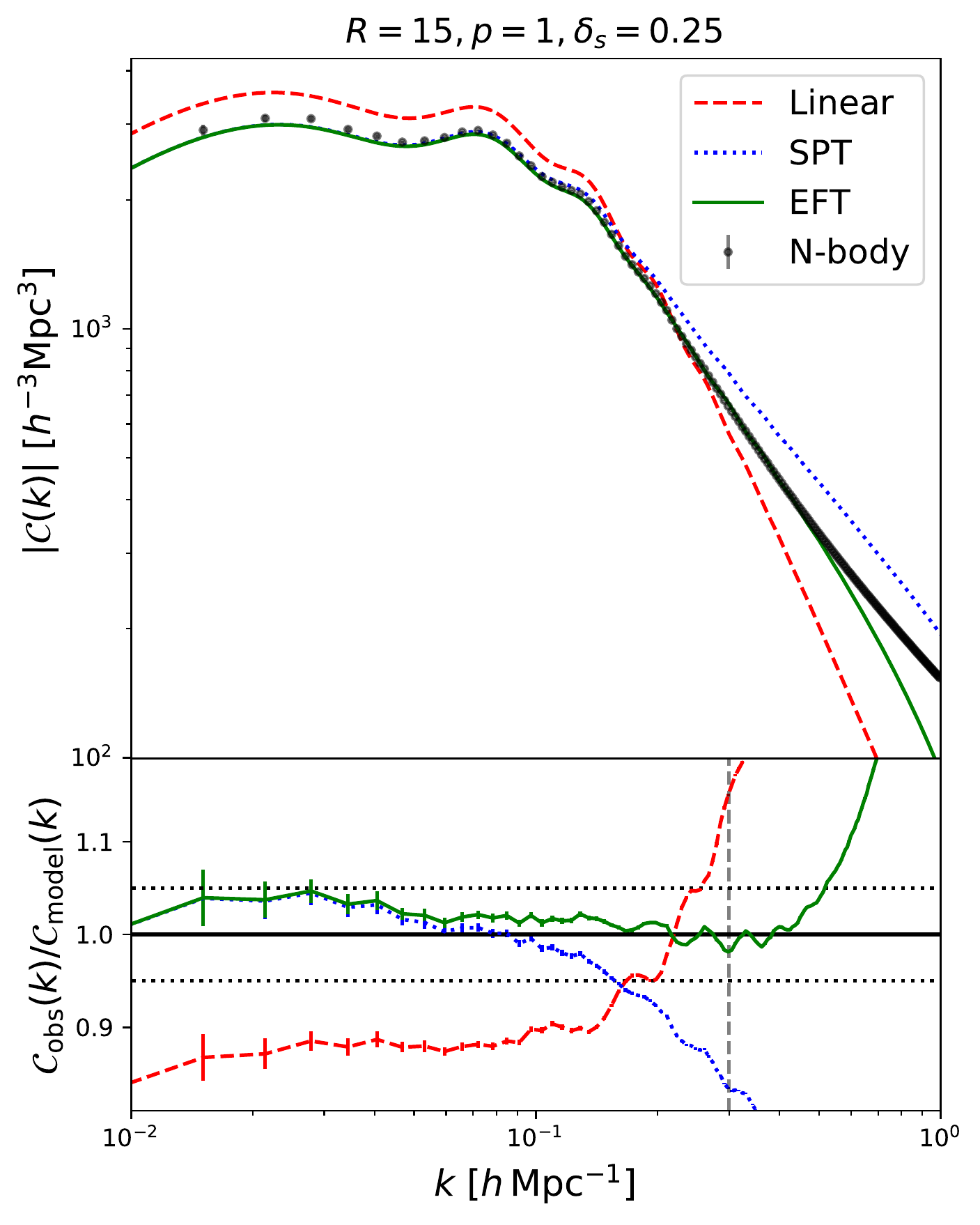}}
    \subfloat[$z = 1$]{\includegraphics[width=0.33\linewidth]{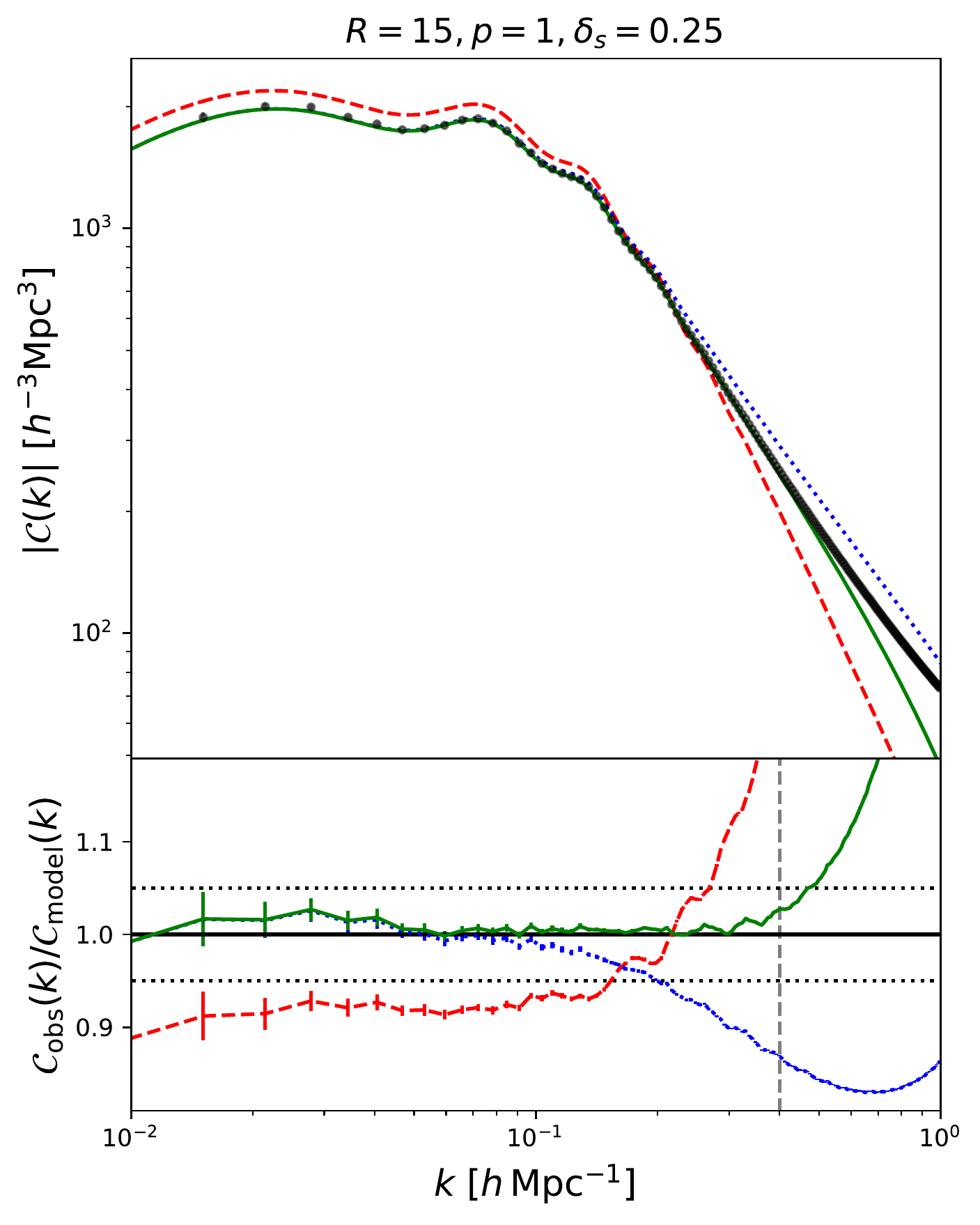}}
     \subfloat[$z = 2$]{\includegraphics[width=0.33\linewidth]{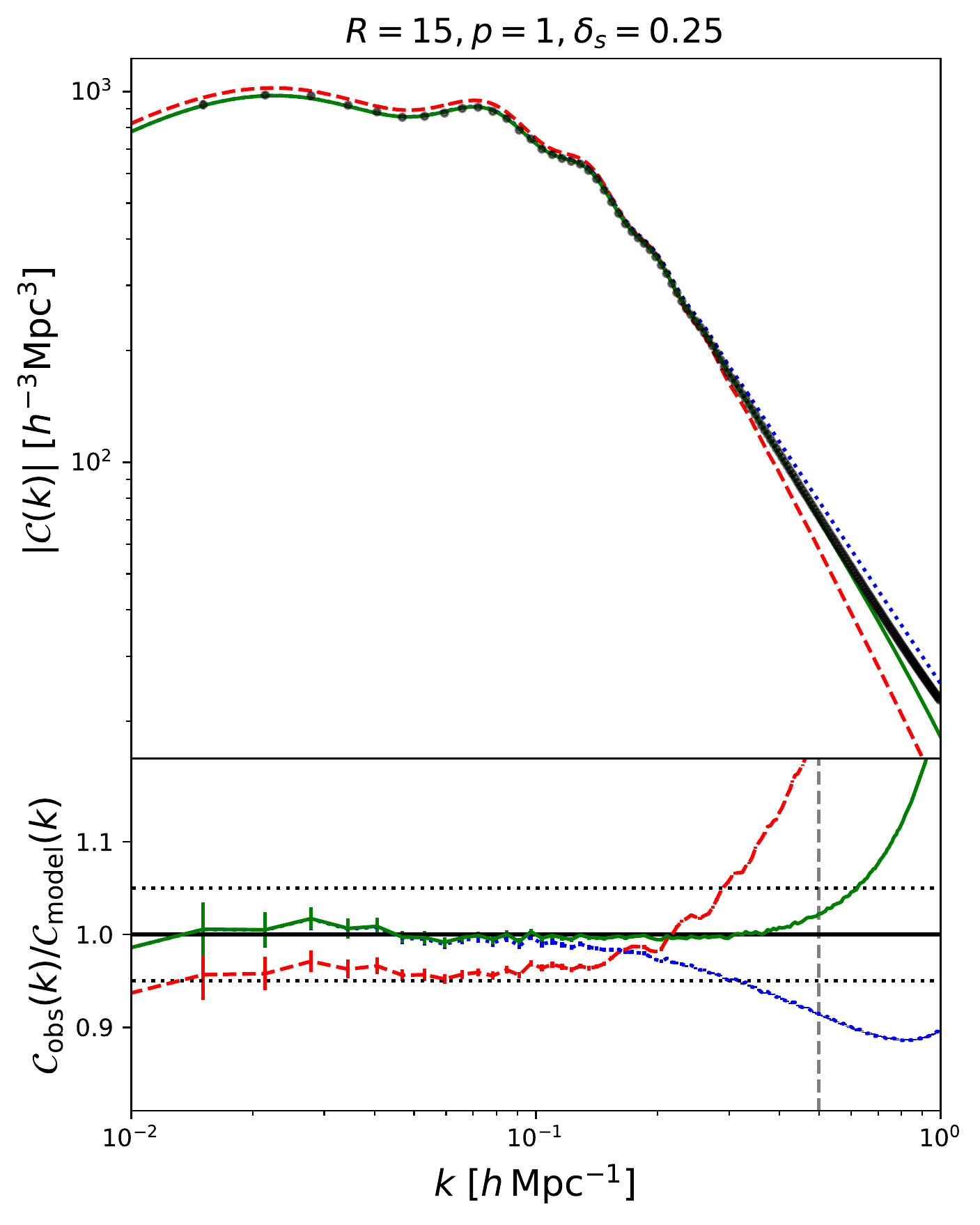}}\\
    \caption{As Fig.\,\ref{fig: param_1_M}, but for the cross spectrum between the marked and standard density fields, $\mathcal{C}(k)$. With this choice of parameters, the cross spectrum is everywhere positive.}\label{fig: param_1_C}
\end{figure}

Figs.\,\ref{fig: param_1_M}\,\&\,\ref{fig: param_1_C} show the marked spectrum measurements from \texttt{Quijote} at a range of redshifts using the first set of parameters: $\{p = 1, R = 15\Mpch, \delta_s = 0.25\}$ (Sec.\,\ref{subsec: mark-param-choice}). This is plotted alongside perturbative models from linear theory, SPT and EFT. For the latter, we fit the counterterm parameter, $c_s^2$, \resub{to the (unmarked) matter power spectrum} using all modes up to $k_\mathrm{fit}$, where $k_\mathrm{fit}$ is fixed to $0.3$, $0.4$, and $0.5\hMpc$ for $z = 0.5,1,2$ respectively, and note that the SPT prediction is simply given by setting $c_s^2 = 0$. \resub{By using $P(k)$ to fit $c_s^2$, we are performing a rigorous test of the analysis, since no additional free parameters are generated to fit the marked or cross spectra.} From the figure, it is clear that linear theory does \textit{not} provide an accurate model for $M(k)$ for any of the scales or redshifts tested. Whilst it becomes more accurate at high $z$, there still a $\sim 5\%$ error for $z = 2$, even at $k = 10^{-2}\hMpc$, showing the importance of the higher order terms at low-$k$.

For the one-loop models, we report much better consistency between simulation and theory, with visual inspection showing that the model captures the overall shape and amplitude of the auto- and cross-spectra well across down to small $k$. EFT modeling shows sub-percent agreement at $z = 2$ up to $k\approx 0.5\hMpc$. As for linear theory, the accuracy of the $M(k)$ (and to a lesser extent $\mathcal{C}(k)$) model reduces as we move to lower $z$. Whilst on quasi-linear scales, the free counterterm $c_s^2$ substantially improves the fit, this is not true for low-$k$, where the counterterm (which scales as $k^2P(k)$) is necessarily small. As seen in Fig.\,\ref{fig: spectral_components}a, the $k\rightarrow 0$ behavior is strongly affected by loop contributions, and this error could be reduced either by the addition of two-loop perturbative terms or by increasing the smoothing scale $R$ (which damps the higher order contributions). 

\begin{figure}
    \centering
    \subfloat[$z = 0.5$]{\includegraphics[width=0.33\linewidth]{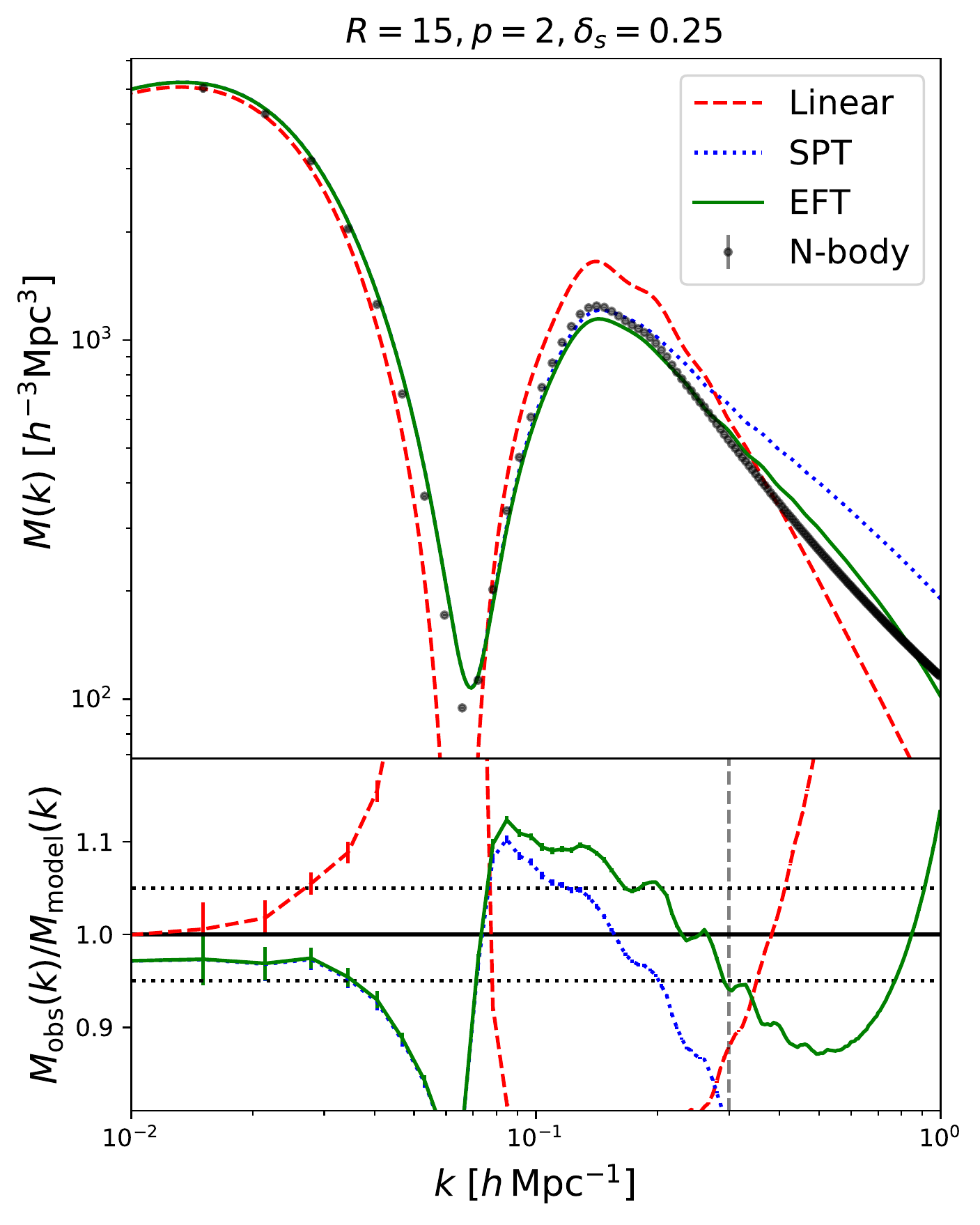}}
    \subfloat[$z = 1$]{\includegraphics[width=0.33\linewidth]{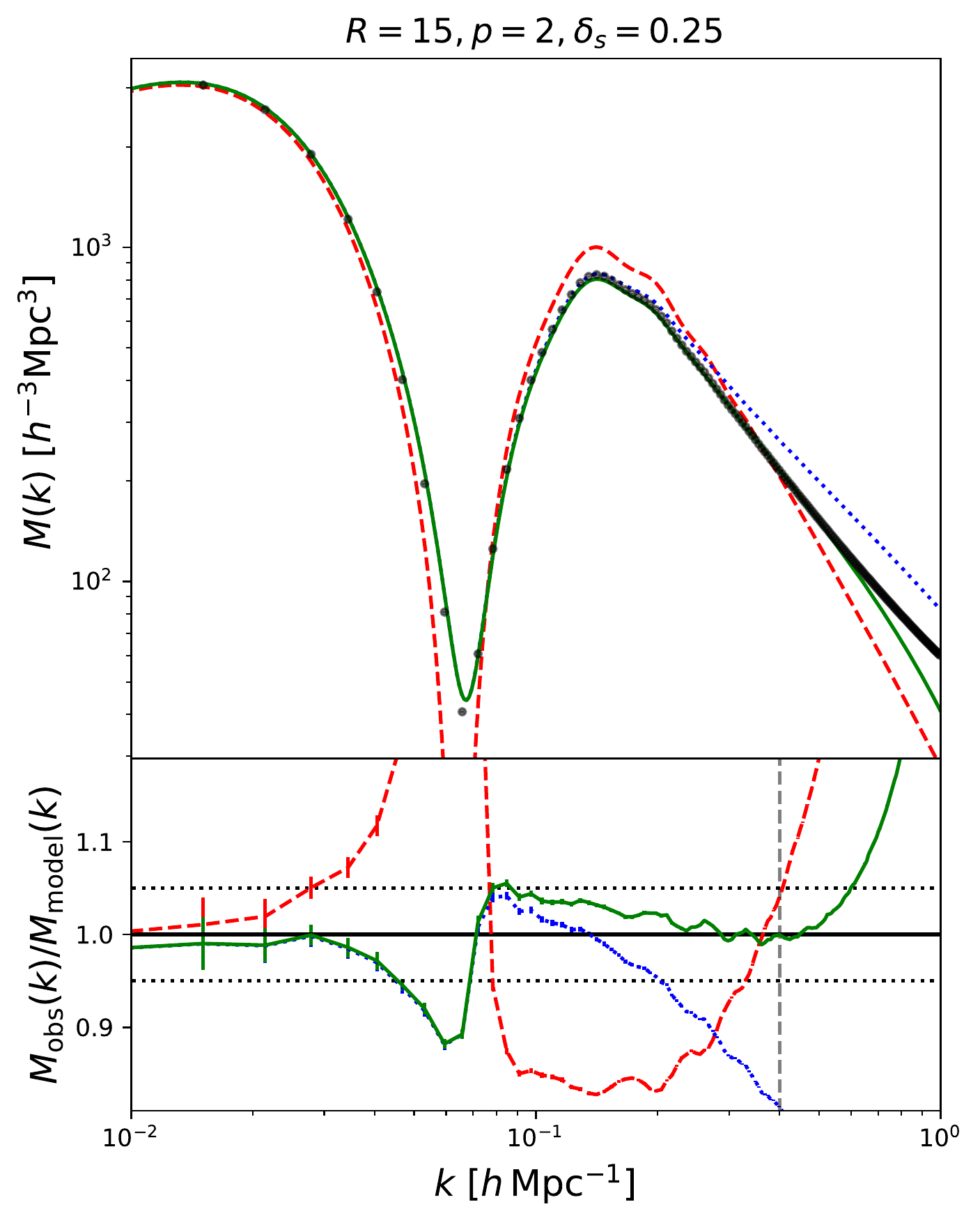}}
     \subfloat[$z = 2$]{\includegraphics[width=0.33\linewidth]{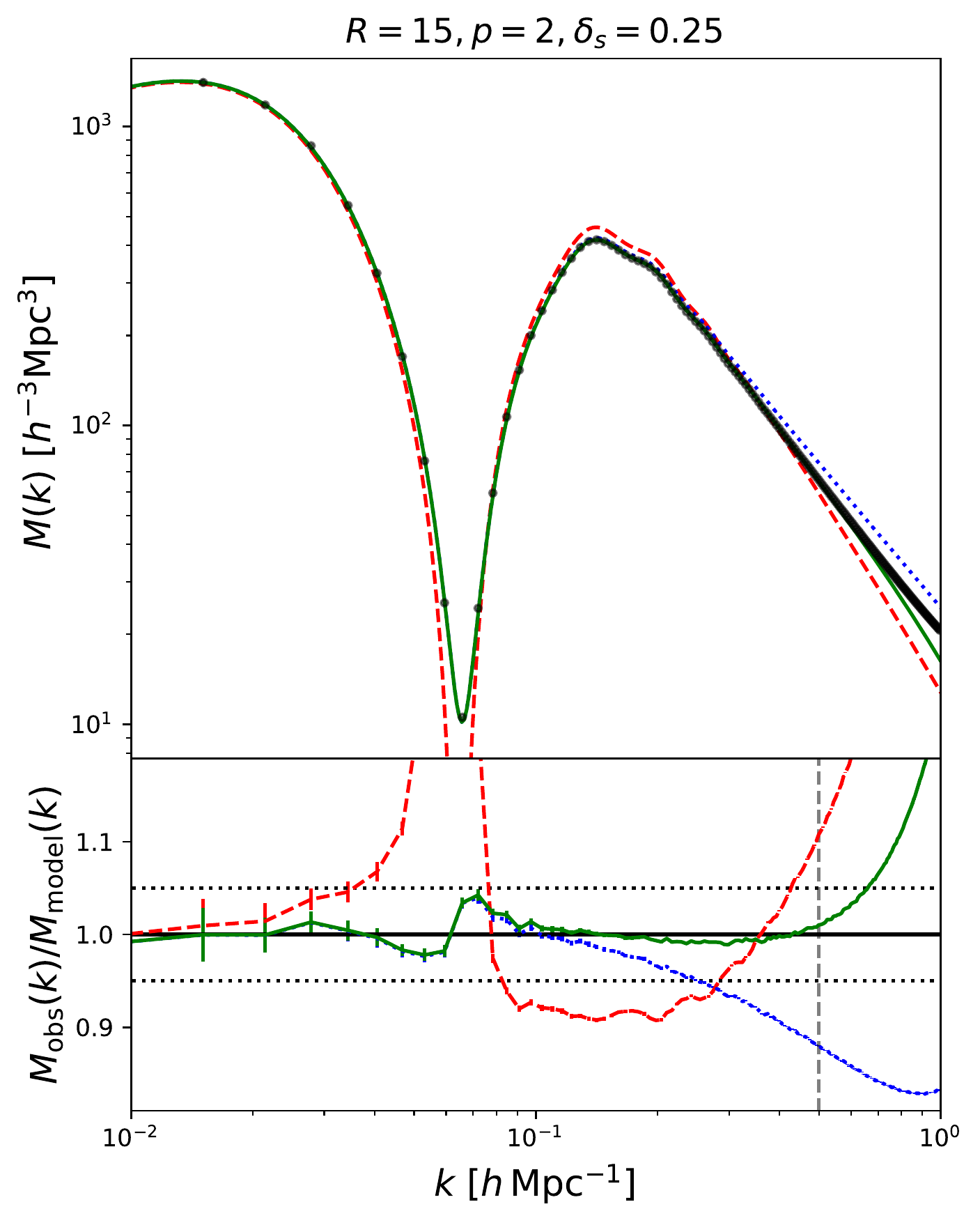}}\\
    \caption{As Fig.\,\ref{fig: param_1_M}, but for the second parameter set, $p = 2, R = 15\Mpch, \delta_s = 0.25$. Note the strong suppression of power at $k\sim 0.07\hMpc$ due to the density field cancellation, as discussed in the text.}\label{fig: param_2_M}
\end{figure}


Fig.\,\ref{fig: param_2_M} 
shows the corresponding results for the marked spectrum using the second parameter set, \textit{i.e.} $p = 2$. Our conclusions are qualitatively similar to those for the first set; the EFT model is accurate at high-$z$ with an important contribution from the $c_s^2$ counterterm at moderate $k$, whilst the accuracy declines at lower redshifts. The first clear distinction between the parameter sets is in the behavior around $k = 0.07\hMpc$; as expected, there is a large suppression of power due to the $(1-C_1W(kR))$ prefactor crossing zero. It can also be shown that this causes $\mathcal{C}(k)$ to change sign, indicating that the marked and full density fields are anticorrelated on the largest scales. The model error significantly increases across this transition, in part due to our lack of inclusion of $k$-space binning in the theory model, but predominantly due to influence of higher order terms that would be expected to dominate in this regime. We caution that the plotted $M_\mathrm{obs}(k)/M_\mathrm{model}(k)$ function is poorly defined near this point, since it is a ratio of small quantities.

A second observation of note is that, unlike for $p = 1$, linear theory provides an accurate model for the spectra as $k\rightarrow0$. From Fig.\,\ref{fig: spectral_components}b, we see that the large scale ratio of one-loop to linear terms (set by the factor $(1-C_1)^2$) is far smaller for this parameter set than the previous, causing this effect. 
Whilst this work focuses on a Gaussian kernel for density field smoothing, a top-hat kernel with the same truncation scale $R$ leads to a significantly narrower dip in $M(k)$, which would make the spectra easier to model; this is however driven by the narrower width of a Fourier-space top-hat window, and thus will give larger one-loop contributions. As previously mentioned, setting $R_\mathrm{TH} = \sqrt{5}R_\mathrm{Gaussian}$ largely accounts for this, giving similar theoretical predictions for the two choices of smoothing.

\begin{figure}
    \centering
    \subfloat[$z = 0.5$]{\includegraphics[width=0.33\linewidth]{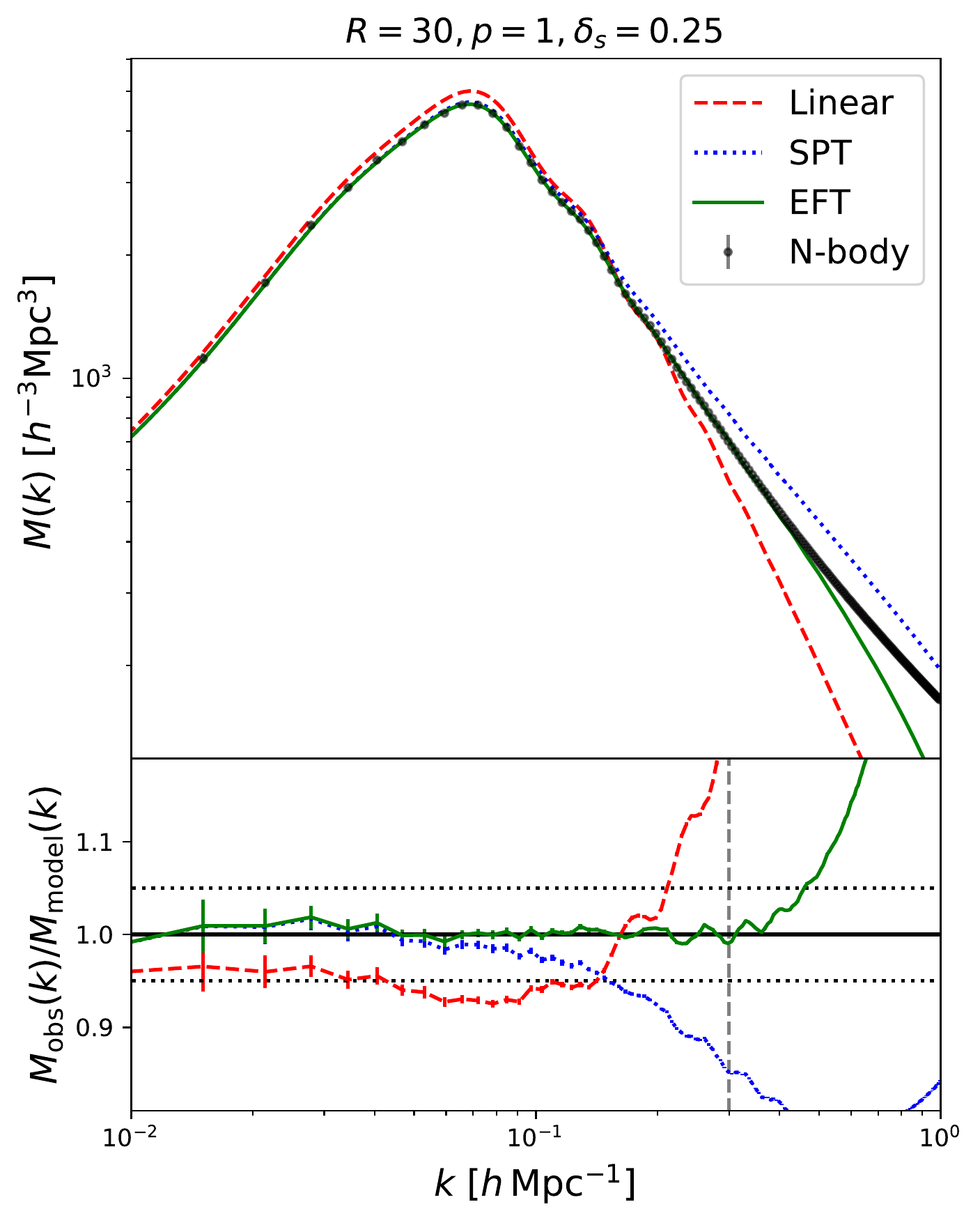}}
    \subfloat[$z = 1$]{\includegraphics[width=0.33\linewidth]{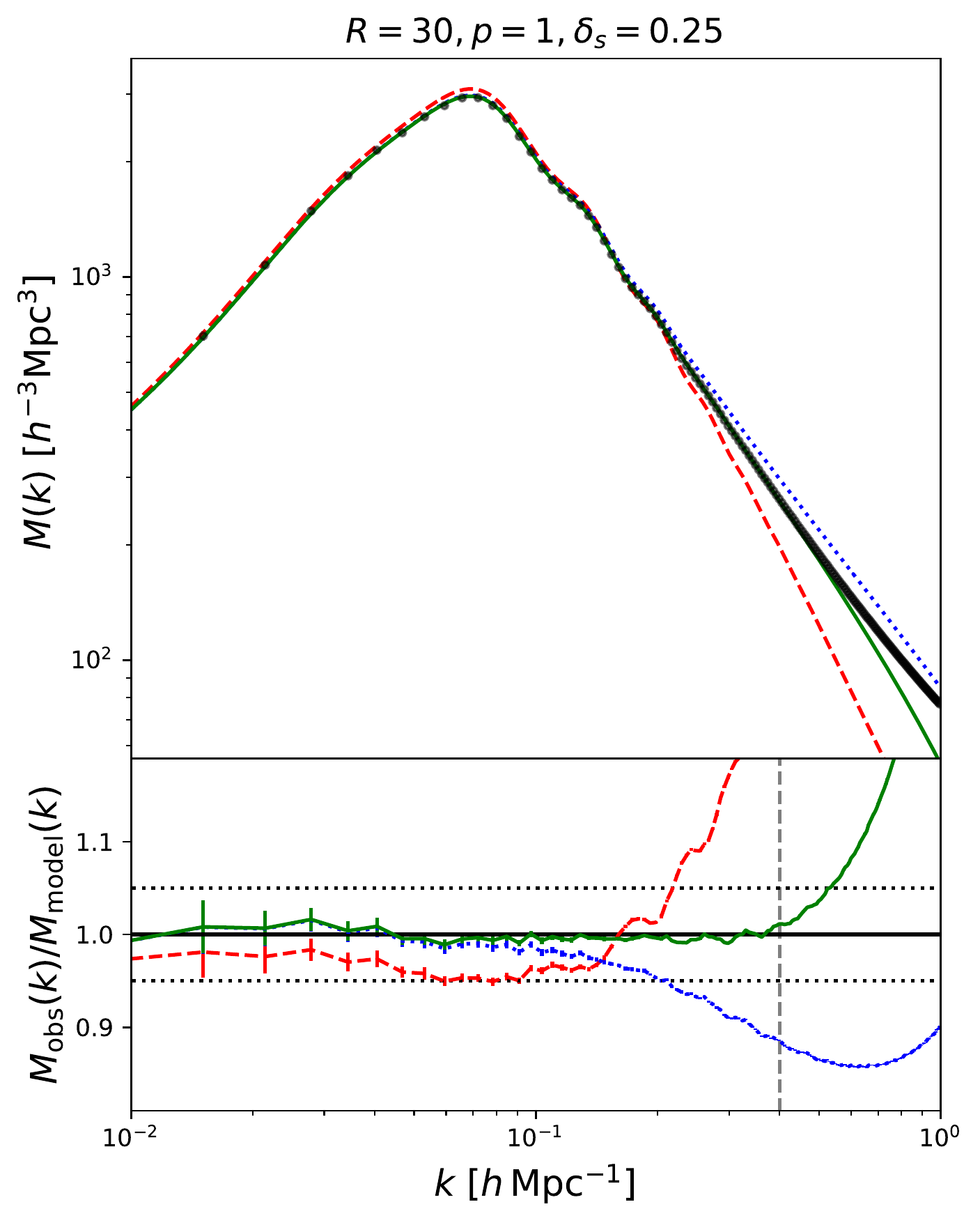}}
     \subfloat[$z = 2$]{\includegraphics[width=0.33\linewidth]{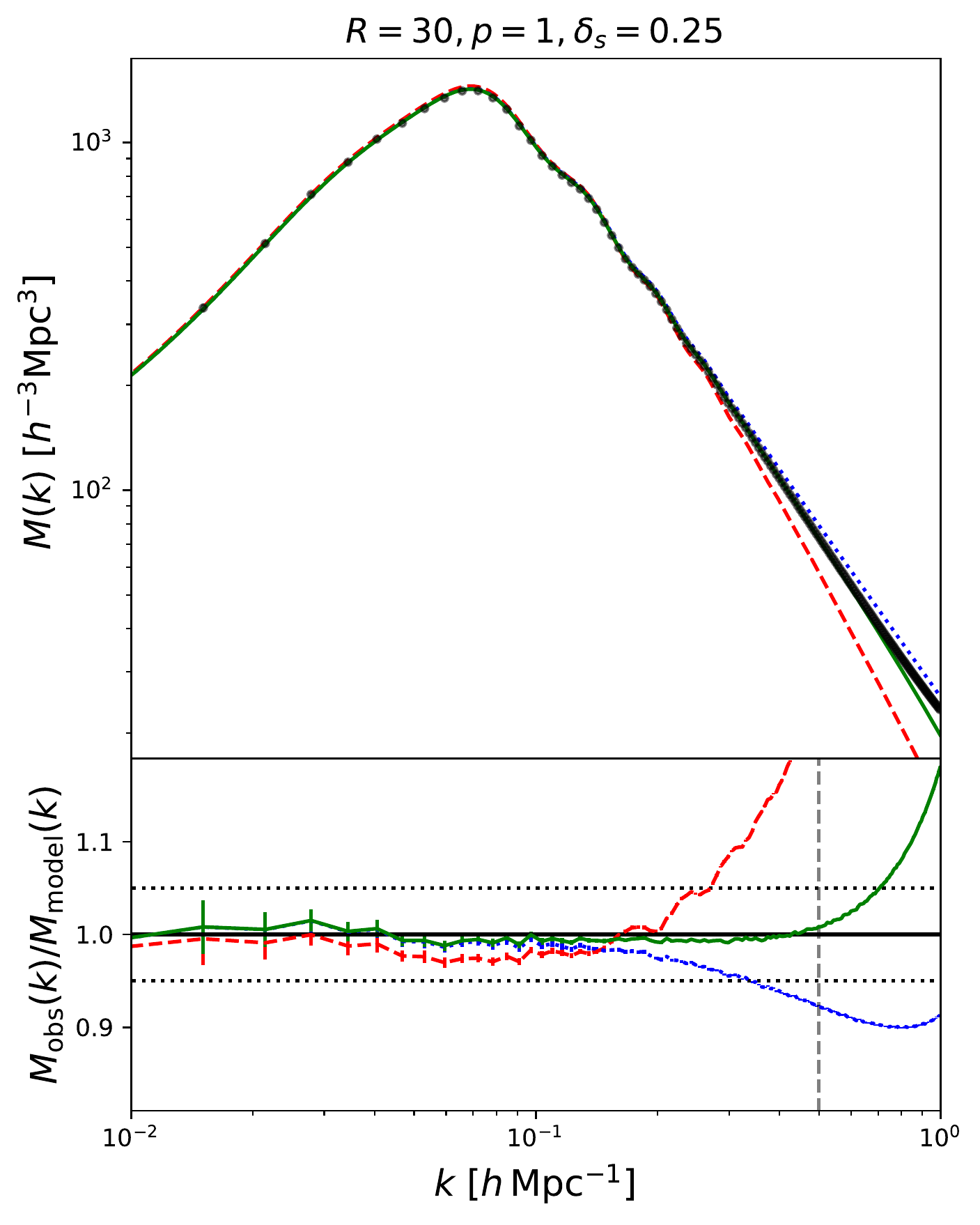}}\\
    \caption{As Fig.\,\ref{fig: param_1_M}, but for the third parameter set, $p = 1, R = 30\Mpch, \delta_s = 0.25$. Due to the increased smoothing scale, the agreement between model and simulations is significantly greater at all $z$.}\label{fig: param_3_M}
\end{figure}


Results for $M(k)$ using the third set of parameters are plotted in Fig.\,\ref{fig: param_3_M}
; these show the effects of doubling the smoothing radius $R$. Notably, we observe percent-level agreement up to $k_\mathrm{fit}$ for all redshifts tested for the EFT theory. As before, the utility of the counterterm is clear, with SPT seen to overpredict the spectrum on quasi-linear scales, as is conventionally found for the unmarked power spectrum. At linear order, the marked power spectrum is closer to the simulated results at low-$k$ than for $R = 15\hMpc$, though it is still inaccurate for low redshifts. This can be rationalized by noting that the amplitude of the higher loop terms is significantly reduced by the smoothing, as in Fig.\,\ref{fig: spectral_components}c. In general, it is clear that using a larger smoothing radius suppresses the higher-order terms, giving better agreement between theory and observations, and extending the model applicability to lower redshifts. One additional feature is of note; the appearance of wiggles in the ratio of simulated to EFT mark and cross spectra on quasi-linear scales at low $z$. This is attributed to the overprediction of BAO wiggles in Eulerian perturbation theories at low redshift, due to large long-wavelength modes being treated perturbatively. Including IR resummation in our mode would alleviate this \citep[e.g.,][]{2015JCAP...02..013S,2015PhRvD..92d3514B}, though this is beyond the scope of this work. 

\begin{figure}
    \centering
    \subfloat[$z = 0.5$]{\includegraphics[width=0.33\linewidth]{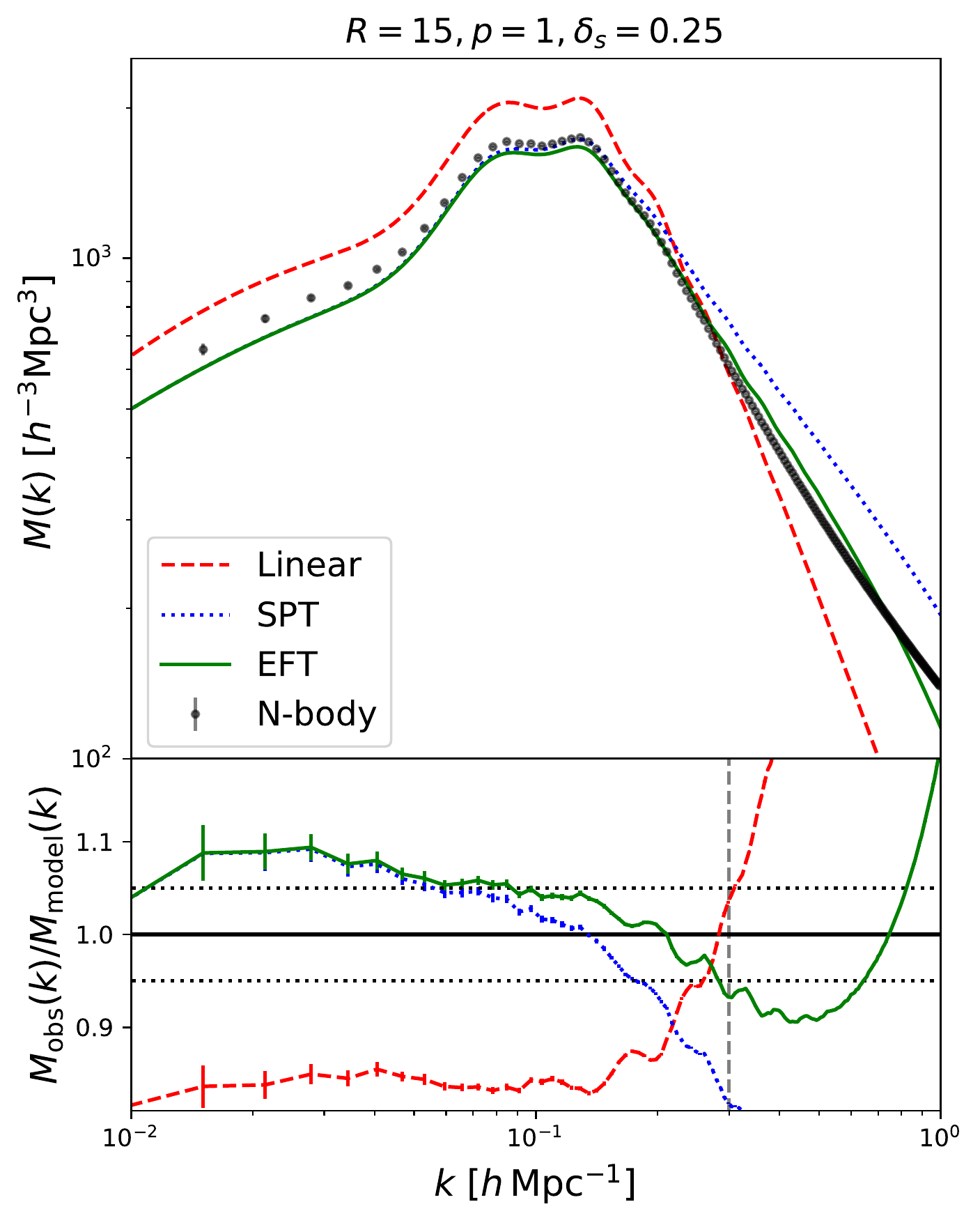}}
    \subfloat[$z = 1$]{\includegraphics[width=0.33\linewidth]{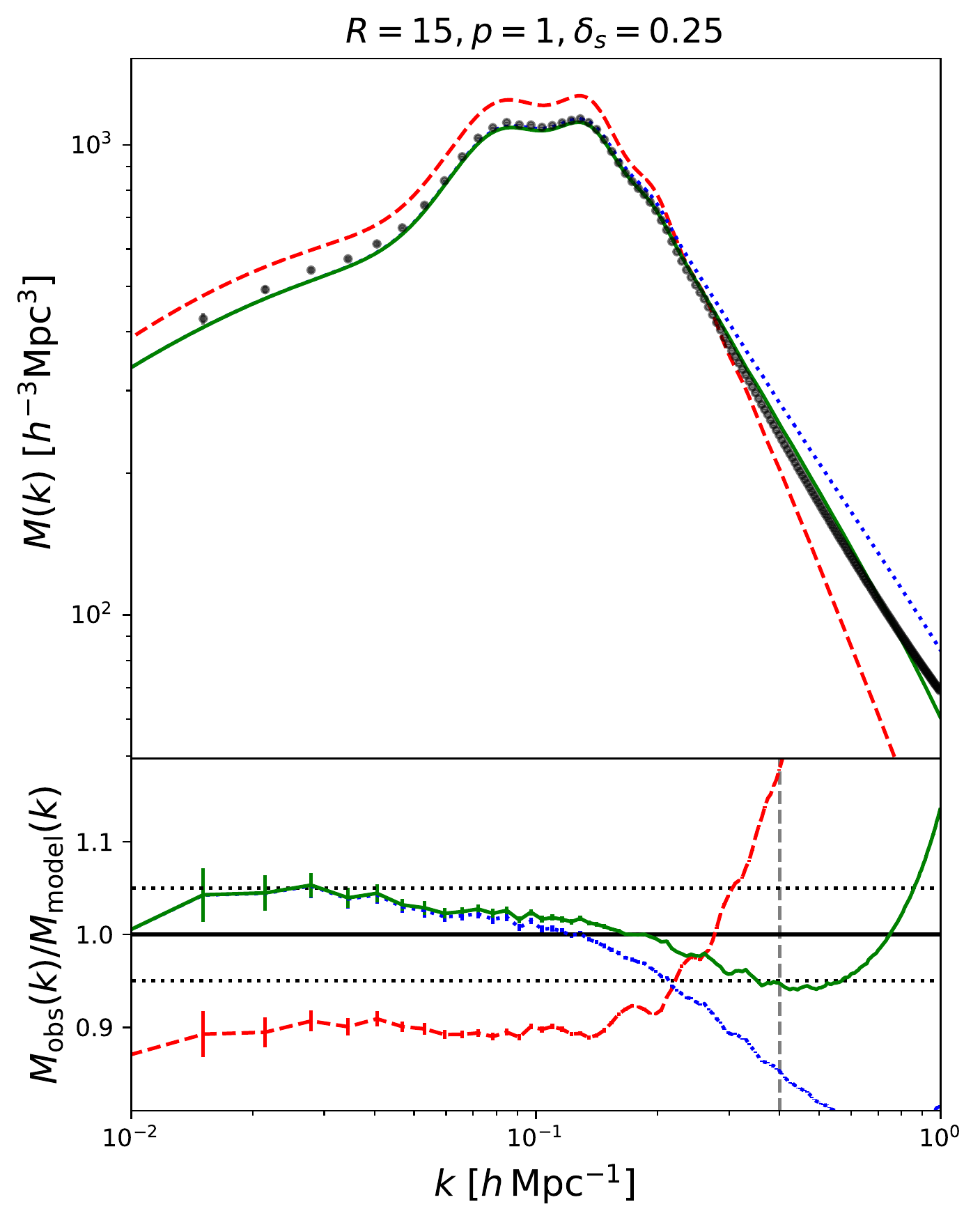}}
     \subfloat[$z = 2$]{\includegraphics[width=0.33\linewidth]{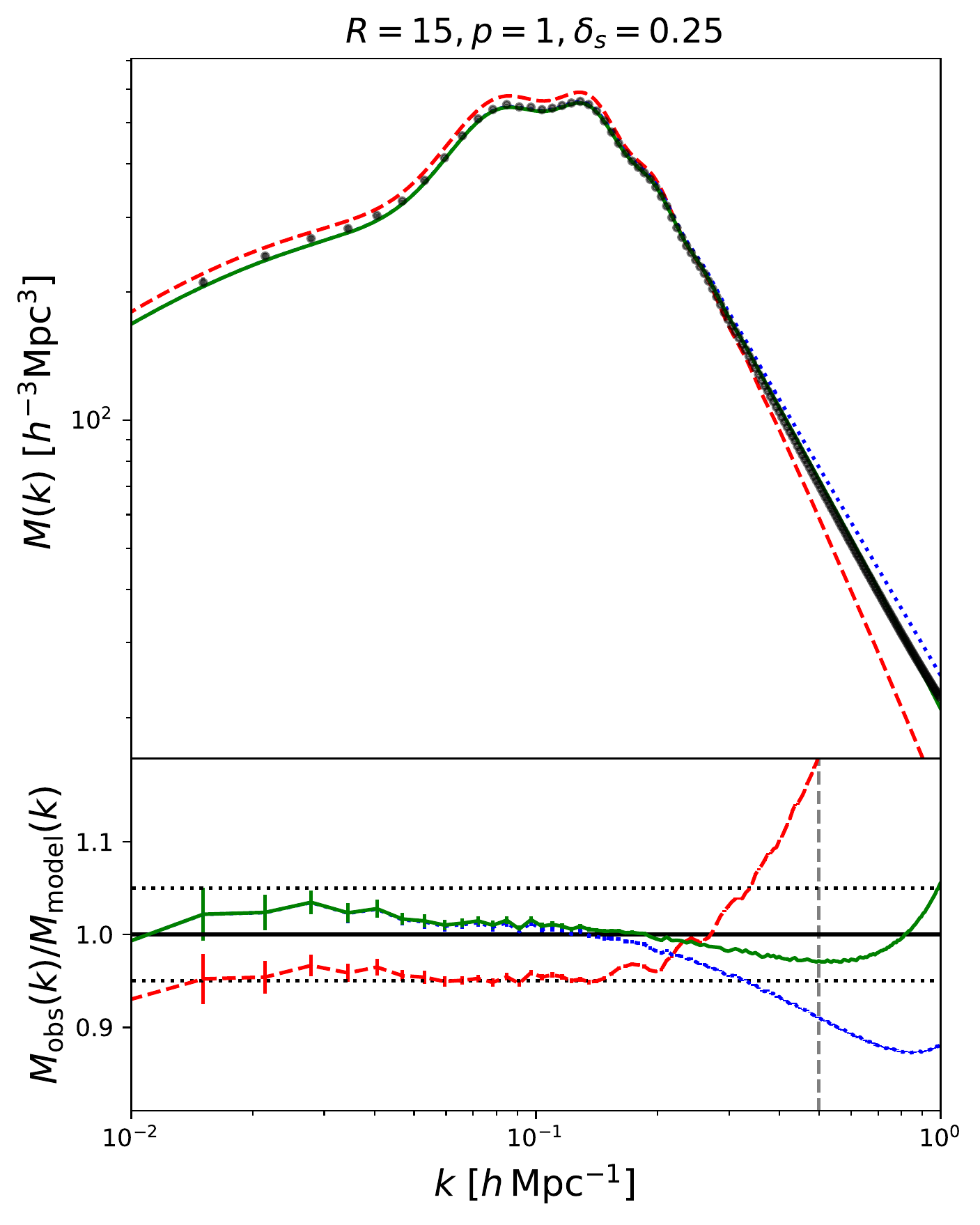}}\\
    \caption{As Fig.\,\ref{fig: param_1_M}, but for massive neutrino cosmologies with a total neutrino mass $M_\nu = 0.1$ eV. This uses the first parameter set, $p = 1, R = 15\Mpch, \delta_s = 0.25$, with best-fit counterterms $0.87$, $0.43$ and $0.16h^{-2}\mathrm{Mpc}^2$ for the three redshifts respectively. We note slightly reduced agreement at low-$k$, likely arising from non-Gaussian effects not captured by our simple neutrino treatment.}\label{fig: param_1_M_mnu}
\end{figure}

We finally consider application to massive neutrino cosmologies, using the $M_\nu = 0.1\,$eV simulations discussed above. These are plotted in Fig.\,\ref{fig: param_1_M_mnu} for the first set of mark parameters, and can be compared to Fig.\,\ref{fig: param_1_M} for the massless case. Similar conclusions can be again drawn, with linear theory giving inaccurate predictions on all scales tested and one-loop EFT performing significantly better up to the fitting wavenumber. The introduction of massive neutrinos is seen to slightly degrade the low-$k$ agreement of simulations and model, which, whilst small, may indicate failings of our approximate neutrino treatment (discussed in Sec.\,\ref{subsec: pt-expansion}). This is unsurprising since the low-$k$ regions of the spectra receive non-negligible contributions from higher-order perturbative terms (as in Fig.\,\ref{fig: spectral_components}) but are not affected by the $c_s^2$ counterterm, due to its $k^2$ dependence. For low redshifts and small smoothing scales, more work is needed to understand the neutrino contributions.

\section{Discussion}\label{sec: discussion}
\subsection{How is Cosmological Information Encoded in $M(k)$?}\label{subsec: where-is-information?}
Having developed a theoretical model for the marked spectrum, we may now ask the question; `why is this statistic a powerful probe of cosmology?' Via simulation-based Fisher analyses, Ref.\,\citep{2020arXiv200111024M} clearly demonstrated the utility of $M(k)$ in constraining fundamental parameters, showing it to be far superior to the usual power spectrum at $z = 0$. At linear order, however, the information content of $M(k)$ and $P(k)$ is identical. This occurs since $M_L(k)$ is simply $P_L(k)$ multiplied by a cosmology-independent prefactor ($H_1(k) = \left[1-C_1W(kR)\right]$), thus the two spectra share a Fisher matrix.\footnote{Neglecting shot-noise, the linear order covariance of $M_L(k)$ is diagonal and proportional to $M_L^2(k)$, and thus contains two factors of $H_1(k)^2$. Since $H_1(k)$ is cosmology independent, the factors cancel between the parameter derivatives and the precision matrix, giving the same Fisher matrix for $M_L(k)$ and $P_L(k)$.} To understand the utility of the marked statistic we must go beyond second order in the linear density field.

The breakdown of Fig.\,\ref{fig: spectral_components} shows that the contributions of the `22' and `13' terms in $M(k)$ are important on both large and small scales. In the latter case, however, they asymptote to the one-loop contributions of $P(k)$, which leads us to posit that the additional information in $M(k)$ \resub{starts to be sourced at} \textit{large} scales. This is aided by two factors: (a) the higher-order contributions in $M(k)$ at low-$k$ are larger than those for $P(k)$ and does not decay as $k\rightarrow0$; (b) the linear term of $M(k)$ is suppressed at low-$k$, giving a larger signal-to-noise for the other components. \new{This is assisted by the positive mark exponent $p$, which up-weights low-density regions and gives the linear-theory suppression. Using a negative mark would instead up-weight the high-density regions, which naturally have greater contributions to two-point correlators, since the density contrast is higher in peaks than in voids. In this case, the linear theory is enhanced by a factor $\left(1 - C_1W(kR)\right)$ (for $C_1<0$) and the effect of higher-order terms is instead a small suppression of power.}

Whilst the above argument is appealing, it does not tell us the source of the low-$k$ plateau, \resub{and indeed, whether it includes small-scale information}. Since this is a one-loop effect, it must occur at $\mathcal{O}(\left[\dpt{1}\right]^4)$, giving three possible contributions (all of which are included in the model of Sec.\,\ref{sec: theory}):
\begin{enumerate}
    \item \textbf{Higher order terms in the mark.} For a Universe which is Gaussian and linear, both the density field $\delta(\vec x)$ and its smoothed counterpart $\delta_R(\vec x)$ are equal to their first-order parts, \textit{i.e.} $\delta(\vec x) = \dpt{1}(\vec x)$, $\delta_R(\vec x) = \dptR{1}(\vec x)$, in the language of Sec.\,\ref{sec: theory}. Since we Taylor expand the mark in powers of $\delta_R(\vec x)$, the quadratic and cubic terms source one-loop contributions to $M(k)$, \resub{though these do not depend on small-scale physics (see Appendix \ref{appen: simplif-and-limits})};
    \item \textbf{Non-linearities in $\delta(\vec x)$.} These terms ($\dpt{2}(\vec x)$, $\dpt{3}(\vec x)$, $\dpt{ct}(\vec x)$) arise from non-linear structure formation and generate non-Gaussian terms in the one-loop spectrum, both from self-couplings and couplings to powers of the linear field $\delta_R(\vec x)$;
    \item \textbf{Non-linearities in $\delta_R(\vec x)$.} For a finite smoothing radius $R$, the smoothed field can also be non-linear, giving an additional set of contributions to the marked spectrum.
\end{enumerate}

\begin{figure}
    \centering
    \includegraphics[width=0.8\textwidth]{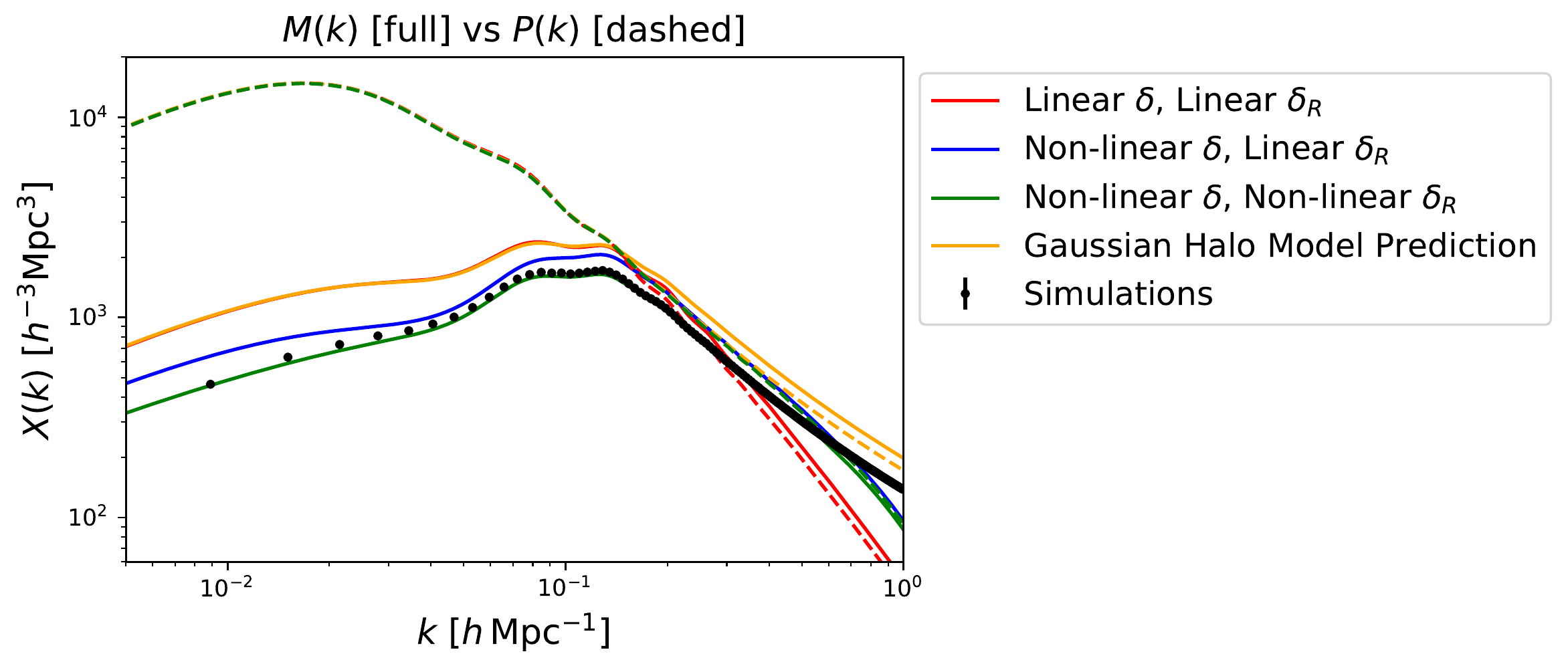}
    \caption{The marked power spectrum at $z = 0.5$ using a variety of physical models, all evaluated to fourth-order in the density fields. This uses the parameter set $\{p = 1, R = 15\Mpch, \delta_s = 0.25\}$ with data from 50 $N$-body simulations as in Fig.\,\ref{fig: param_1_M}. Predictions are given for one-loop theories where the overdensity field $\delta$ and the smoothed field $\delta_R$ are allowed to be linear or non-linear. Additionally, we show predictions for a Gaussian model in which the correlators of $\delta$ and $\delta_R$ are evaluated with the fully non-linear halo model power spectrum of Ref.\,\citep{2020PhRvD.101l3520P} rather than linear theory. (Note that the red and yellow curves are coincident at low-$k$). Corresponding results for the matter spectrum are shown as dashed lines.}
    \label{fig: low-k-NGs}
\end{figure}

Fig.\,\ref{fig: low-k-NGs} shows the marked power spectra for a variety of theoretical predictions at $z = 0.5$ with each evaluated at $\mathcal{O}(\left[\dpt{1}\right]^4)$. We use three models: (a) linear $\delta$ and $\delta_R$, \textit{i.e.}
\beq\label{eq: deltaMa}
    \delta_M(\vec x) = \frac{1}{\bar{m}}\left[1+\dpt{1}(\vec x)\right]\left[1 - C_1 \dpt{1}_R(\vec x) + C_2\left(\dpt{1}_R(\vec x)\right)^2 - C_3\left(\dpt{1}_R(\vec x) \right)^3 \right] - 1;
\eeq
(b) non-linear $\delta$ and linear $\delta_R$, \textit{i.e.}
\beq\label{eq: deltaMb}
    \delta_M(\vec x) = \frac{1}{\bar{m}}\left[1+\dpt{1}(\vec x)+\dpt{2}(\vec x)+\dpt{3}(\vec x)\right]\left[1 - C_1 \dpt{1}_R(\vec x) + C_2\left(\dpt{1}_R(\vec x)\right)^2 - C_3\left(\dpt{1}_R(\vec x) \right)^3 \right] - 1;
\eeq
(c) non-linear $\delta$ and $\delta_R$, as in Eq.\,\ref{eq: delta-expan2}. These correspond to adding each of the three above contributions in turn.\footnote{Each of model uses some subset of the perturbative terms in Eqs.\,\ref{eq: M22-simpl}\,\&\,\ref{eq: M13-exp} and may be derived by repeating the derivation starting from Eqs.\,\ref{eq: deltaMa}\,\&\,\ref{eq: deltaMb}.} To begin, note that all models for $P(k)$ are co-incident until $k\sim 0.1\hMpc$, since non-linear effects are restricted to small scales. Clearly, this is not true for $M(k)$, shown by the significant overprediction by the model which uses linear theory to evaluate $\delta$ and $\delta_R$. Note that this is not the same model as the `linear' predictions in Sec.\,\ref{sec: sim-comparison}, since the former worked only to second-order in the density fields. This contributes a difference of only $\sim 10\%$ on these scales. The model including non-linearities in the unsmoothed field is certainly an improvement, but still an overprediction, indicating that $M(k)$ is receiving significant contributions from non-linearities \textit{both} in $\delta$ and $\delta_R$. Na\"ively one might expect the level of non-linearity in $\delta_R$ to be small (given that it is smoothed), however this is found to be an important contributor to large-scale modes. This provides an explanation for the fall-off of cosmological information contained within the mark for large $R$ \citep{2020arXiv200111024M}.

Given that non-linear effects are important for the large-scale marked spectrum, it is interesting to ask whether these are sourced solely from non-linear two-point correlators (\textit{i.e.} the matter power spectrum) or have contributions from higher statistics (such as bispectra) created by these non-linearities. To test this, we can compute a model for $M(k)$ where any two-point statistic, for example $\av{\delta_R(\vec k_1)\delta_R(\vec k_2)}$ is evaluated using the (suitably smoothed) non-linear power spectrum $P_{NL}$, but with other connected correlators, e.g., $\av{\delta(\vec k_1)\delta(\vec k_2)\delta_R(\vec k_3)}$ set to zero.\footnote{\new{Note this contains an implicit contradiction; non-linearities necessarily generate non-Gaussianities, but we find it useful to separate the two here in order to assess the importance of higher order correlators.}} In practice, this corresponds to taking the above `linear $\delta$, linear $\delta_R$` model but using $P_{NL}(k)$ in place of of the linear power spectrum. To evaluate this, we require an accurate model for $P_{NL}$ up to large $k$ (in order to compute convolutions), thus we here use the fitted `Effective Halo Model' of Ref.\,\citep{2020PhRvD.101l3520P}. The resulting predictions are shown in Fig.\,\ref{fig: low-k-NGs}, and provide inaccurate predictions on all scales, matching the `linear $\delta$, linear $\delta_R$' model at low-$k$. Even though this model encapsulates \textit{all} non-linearity in the two-point correlator (including halo formation effects), it is unable to fit large-scale data, implying that the assumption of Gaussianity is invalid. \new{Note that this is not guaranteed by the limit of $P_{NL}(k)\rightarrow P_L(k)$ at low $k$; the marked spectrum contains convolutions between power spectra that also contribute at low-$k$ and will have contributions from non-linearities.} \resub{However, as discussed in Appendix \ref{appen: simplif-and-limits}, these convolutions always include a smoothing kernel, which removes any small-scale modes from the analysis, ensuring the agreement between the two approaches.}

Collecting results, it is clear that any model without consideration of non-linear and non-Gaussian effects cannot provide accurate predictions of the marked power spectrum on large scales. It is these low-$k$ contributions \resub{(in particular, those from the non-Gaussian terms)} that lead to the additional constraining power in $M(k)$, relative to that of $P(k)$. Whilst the signal-to-noise of the large-scale modes is slight, the low-$k$ visibility of non-Gaussian information allows subtle effects (such as those provided by neutrinos) to be probed within $M(k)$ on relatively large scales. \resub{Furthermore, we surmise that the addition of non-Gaussian information in the two-point function is complemented by its removal from the higher point statistics, 
which would lead to a suppressed trispectrum, and thus a more diagonal covariance matrix, as observed in Ref.\,\citep{2020arXiv200111024M}. This would ensure that $M(k)$ measurements in different bins are less correlated, further increasing the information content.} 

\subsection{Validity of the Perturbation Theory}
As noted in Sec.\,\ref{sec: theory}, a necessary condition for any perturbative theory of $M(k)$ to be valid is that the variance $\sigma_{RR}(z)$ must be less than $(1+\delta_s)$. This ensures convergence of the Taylor expansion of the mark, though does not control the validity of the EFT expansion of the unsmoothed density field $\delta$. For the matter power spectrum, EFT is valid down to $z = 0$ on quasi-linear scales, with a larger radius of convergence obtained by introducing higher loop terms.\footnote{There is some evidence that the radius of convergence of the matter EFT saturates at three-loop order, e.g., Ref.\,\citep{2019JCAP...11..027K}.} As shown above, the EFT of the marked density field does not have a well-defined radius of convergence due to the \resub{importance of non-linear and non-Gaussian} terms at low $k$. However, the results of Sec.\,\ref{sec: sim-comparison} indicate that the one-loop theory is convergent and accurate in certain regimes, and, given that the one-loop predictions are closer to the truth than those of linear theory (\textit{i.e.} zero-loop), it is expected that the inclusion of two-loop terms will further increase the accuracy. This should be verified with simulations, \resub{which can rigorously test the effect impact on $M(k)$ from scales not under perturbative control.}

The aforementioned low-$k$ behavior makes the perturbation theory more difficult to apply in practice; we cannot simply use our model for $M(k)$ up to some maximum wavenumber $k_\mathrm{fit}$, since when one-loop theory becomes inaccurate, it does so on all scales. If we restrict ourselves to relatively large smoothing radii $R$ and/or high redshifts, one-loop EFT is a good predictor of $M(k)$, though it remains to see whether this has cosmological use. As demonstrated in Sec.\,\ref{subsec: where-is-information?}, the low-$k$ behavior of $M(k)$ is strongly affected by non-linearities and non-Gaussianities in $\delta_R$ that are expected to source the additional information content of the marked field. At large $R$ and high-$z$, these effects are suppressed, thus it stands to reason that the constraining power of $M(k)$ relative to $P(k)$ is reduced. To utilize the mark to its fullest extent, higher-order, or non-perturbative approaches may be required. 

In Ref.\,\citep{2020arXiv200111024M} it was shown that the information content of $M(k)$ was maximized using a top-hat smoothing of $R_\mathrm{TH} = 10\Mpch$ at $z = 0$, corresponding to a Gaussian smoothing of $R_\mathrm{TH}/\sqrt{5}\approx 4.5\Mpch$. Whilst this also uses a different exponent ($p = 2$) to that considered principally in this work and works at a lower redshift than those used by upcoming cosmological surveys, we expect that for a survey-specific optimal mark, the non-linear contributions will be large at low-$k$, \resub{requiring that the perturbation theory be evaluated at high order.} \resub{Some form of coefficient resummation may be of use here; this could ameliorate the unusual feature of the theory that higher-loops renormalize the amplitude of lower loop contributions, as well as providing new spectral shapes.}

\subsection{Biased Tracers and the Mark}\label{subsec: bias-tracers}
Given that the perturbation theory for $M(k)$ is essentially just a Taylor series in the (smoothed) matter density field, certain parallels can be drawn with the the EFT of biased tracers. In the latter case, the tracer field, $\delta_h$ contains a similar expansion in powers of the density field, with the important distinction being that the bias parameters are now free. Further, the biased tracer EFT can contain a number of UV divergences from integrating powers of $\delta$ across non-linear regimes, which disappear for the marked EFT (aside from the $c_s^2$ counterterm divergence), due to the density field smoothing. One useful property of biased tracer EFT lies in its renormalizability; the expansion of $\delta_h$ can be cast in terms of \textit{renormalized} operators \citep{2009JCAP...08..020M,2014JCAP...08..056A} that simplify the expansions. This is not possible with our approach as it requires modifying the bias coefficients, which are here set by the Taylor expansion. The expressions arising in the EFT of marked matter can be thought of as a subset of those in the biased tracer EFT, except without the UV counterterms or free bias parameters, nor the simplification afforded by operator renormalization. Furthermore, the suppression of linear power afforded by the density field smoothing terms ensures that the non-linear terms dominate at low-$k$, unlike for the biased tracer EFT.

\begin{figure}
    \centering
    \includegraphics[width=0.7\textwidth]{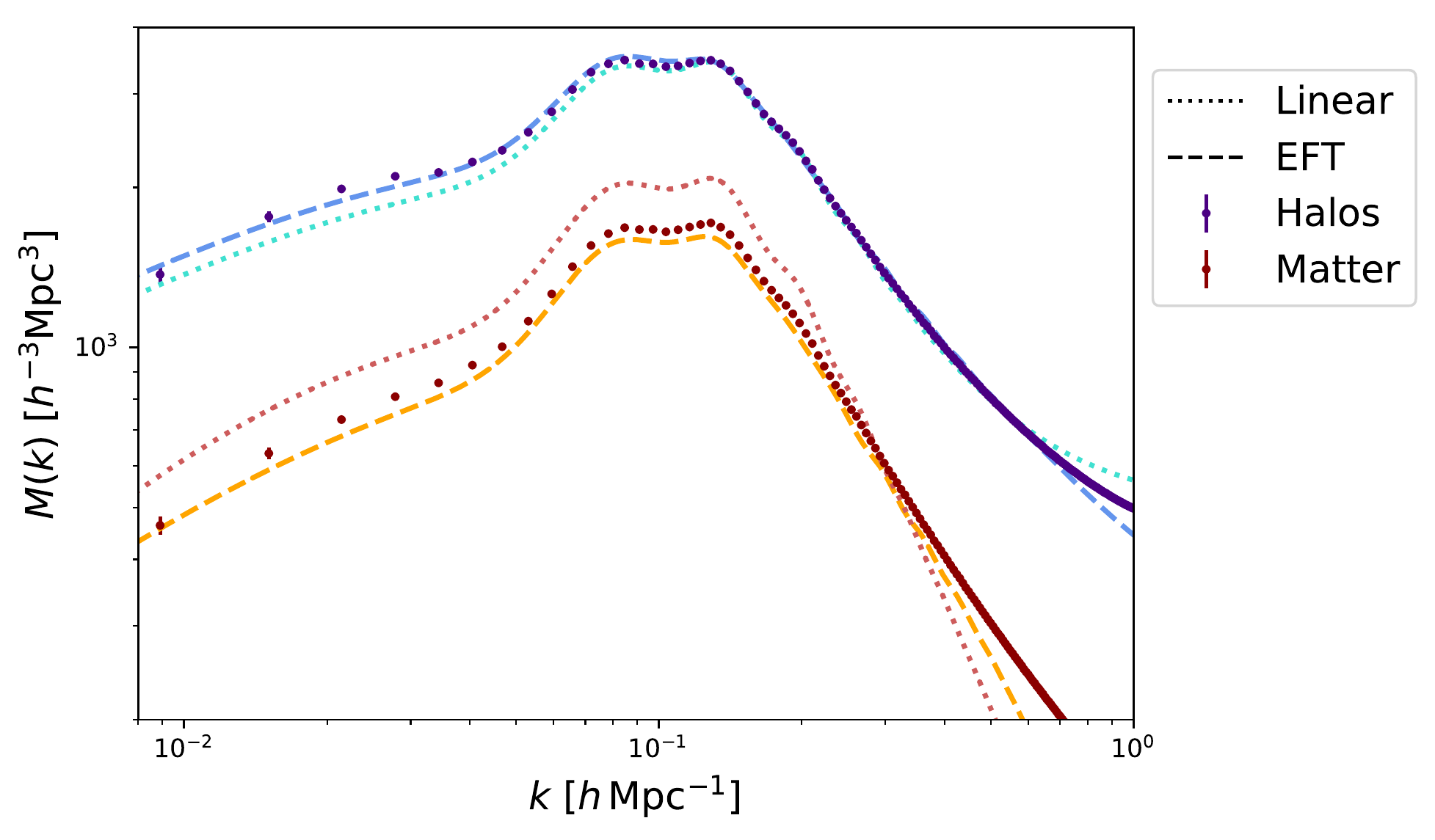}
    \caption{Comparison of the marked power spectrum of halos and matter at $z = 0.5$. This uses the mark parameters $\{p = 1, R = 15\Mpch, \delta_s = 0.25\}$. The halo spectrum uses a total of $\sim 3\times 10^{6}$ halos with mass above $1.6\times 10^{12}h^{-1}M_\odot$ drawn from 50 high-resolution \texttt{Quijote} simulations, and, to generate the marked power spectrum models, we fit for linear bias, shot-noise, and $c_s^2$ (for the EFT model only), using all modes up to $k = 0.3\hMpc$. The matter spectrum is identical to that of Fig.\,\ref{fig: param_1_M}.}
    \label{fig: bias_tracer_mk}
\end{figure}

A related question is whether one can formulate a marked spectrum of biased tracers, for example galaxies. In principle, this is straightforward; the density fields $\delta$ and $\delta_R$ would simply be replaced by the galaxy density fields, $\delta_g$ and $\delta_{g,R}$ \new{(as in Ref.\,\citep{2020JCAP...01..006A} for the configuration-space correlator)}. Given that $\delta_g$ can be written as an expansion in terms of the matter field $\delta$ (and, at higher order, the gradient and tidal fields), we obtain a similar perturbative expansion;
\beq
    \delta_{M,g}(\vec x) &=& \frac{1}{\bar{m}}\left[1+\delta_g(\vec x)\right]\left[1-C_1\delta_{g,R}(\vec x)+...\right]\\\nonumber
    \delta_g(\vec x) &=& b_1\delta(\vec x) + ...\\\nonumber
    \delta_{g,R}(\vec x) &=& \ast\left[\delta_g,W_R\right](\vec x),
\eeq
truncating at first order for brevity. This may be treated similarly to the marked overdensity of matter, with the caveat that we must include additional second- and third-order operators in $\delta_g$ such as the stochastic and tidal fields. However, the end-point will be an expansion in powers of $\dpt{1}(\vec x)$, (cf.\,Eq.\,\ref{eq: delta-expan2}) which can be modeled using EFT. The full spectrum of biased tracers is substantially more complex than that of matter, due to the higher-order operators and increased number of terms.

\resub{The general case proceeds by replacing the $F_n$ gravitational kernels in the perturbative expansion of Eq.,\,\ref{eq: H-kernels} with those for biased tracers, and evaluating the theory as before.} For this exercise, we consider a simple model, obtained by assuming linear bias for the galaxy field, such that $\delta_g(\vec x) = b\,\delta(\vec x)$, $\delta_{g,R}(\vec x) = b\,\delta_R(\vec x)$, where the redshift-dependent bias $b$ depends on the halo or galaxy sample in question. In this limit, the theory model is straightforward;
\beq
    \delta_{M,g}(\vec x) \rightarrow b\,\delta_M(\vec x), \quad C_k \rightarrow b^{k-1}C_k,
\eeq
where $C_k$ is the $k$-th order Taylor expansion coefficient of the mark (Eq.\,\ref{eq: mark-coeff}). This leads to a simple power spectrum; $M_g(\vec k)\rightarrow b^2M(\vec k)$ additionally using the bias-rescaled mark coefficients (and noting that the mean mark, $\bar{m}$, is also expected to change). Since the number density of galaxies is low compared to that of simulation matter particles, we expect a significantly increased contribution from shot-noise. Whilst Sec.\,\ref{sec: theory} presented simple Poissonian predictions for the shot-noise, the largest (smallest) halos are known to be somewhat sub- (super-)Poissonian due to exclusion effects, making it necessary to fit the shot noise as a free parameter, alongside the linear bias $b$, and the effective sound-speed $c_s^2$ (for models including the EFT counterterm).

Fig.\,\ref{fig: bias_tracer_mk} compares the marked spectra of matter and halos at $z = 0.5$, using the first set of mark parameters. For each simulation, we use a total of $\sim 3\times 10^6$ halos, selected as described in Ref.\,\citep{2019arXiv190905273V}, with a linear bias $b\sim 1.3$. It is immediately clear that the amplitude of the halo spectrum exceeds that of matter on all scales; this is a consequence of the significant shot-noise contributions to the latter. We also display comparison to theory models computed using linear and one-loop EFT, computed using the simple linear bias prescription above. In this instance, both models perform remarkably well (and outperform those applied to the matter spectrum), though this is likely principally due to the free shot-noise parameter. Whilst a full model of the biased tracer spectrum should necessarily include higher order biases, it is gratifying to see that a good approximation can be wrought with only the simplest term. Given that the difference between linear and non-linear models is reduced for biased tracers, it is unclear to what extent the mark still provides additional cosmological information, and this should be probed in future work.

\section{Summary}\label{sec: summary}
In this paper, we have developed a perturbation theory for the power spectrum of the \textit{marked density field}; a local-overdensity-weighted field recently shown to produce strong constraints on cosmological parameters, especially the neutrino mass sum. The theory has been rigorously developed for the matter density field in the context of the Effective Field Theory of Large Scale Structure (EFT) and compared to simulations across a range of redshifts and mark parameters. At higher redshifts, or on large smoothing scales, the one-loop marked spectrum (and the cross-spectrum with the usual density field) is seen to be in excellent agreement with the data on all scales where the matter EFT is valid, whilst linear theory is systematically biased. For small smoothing scales and at low redshift, the theory becomes less accurate, but, in contrast to the matter power spectrum, the theory is inaccurate for all $k$, rather than just those above some limiting scale.

Using our model for the marked spectrum, $M(k)$, we are able to understand the reason for the increased cosmological information in this statistic compared to that of the usual power spectrum, $P(k)$. On small scales, the perturbative contributions of the two are similar, thus any additional information must be coming from \resub{intermediate-to-large} scale behavior. This is the case in practice, since the linear component of $M(k)$ is suppressed on large scales, and the higher-loop terms become large. Furthermore, at low $k$, we have shown there to be important contributions from \resub{non-Gaussianities}; a feature not found in $P(k)$. \new{In essence, the up-weighting of low-density regions shifts information from higher-point statistics into the two-point correlator, sourcing} the additional constraining power of $M(k)$. This mixing between scales complicates the perturbative modeling, as there is no well-defined convergence radius of the theory. Further, it is likely that the easiest to model parameter regimes are not those with the greatest information content, since the information content will be maximized with significant signal from non-Gaussianities on large-scales, \resub{which requires high-order perturbative expansions}. 

\resub{It is important to place these results in the context of other alternative density statistics. For the marked spectrum, in common with most other cases, the cosmological utility is sourced by the addition of higher-order correlators, which we show to give non-negligible contributions to $M(k)$ at low-$k$. The manner in which these contributions enter differs between statistics; reconstructed spectra \citep{2007ApJ...664..675E} partially recover the \textit{early-time} power spectrum via a shift of gravitational information from the bispectrum \citep{2017PhRvD..96b3505S}, whilst the position-dependent power spectrum \citep{2014JCAP...05..048C} is simply a squeezed bispectrum. In contrast, bispectrum and trispectrum contributions arise in the marked spectrum due to the mode coupling induced by the local-density weighting. As such we are free to adjust the amplitude of these effects by changing the mark parameters \citep{2020arXiv200111024M}, a freedom not usually found in other statistics, though we note that $M(k)$ does \textit{not} preserve the two-point function, instead damping it by a significant factor at low $k$. It is further interesting to compare this to the log-normal density field \citep{2009ApJ...698L..90N}, which also receives non-Gaussian contributions from the statistic definition, rather than a specific physical effect. As with the marked density field, it can be written as a Taylor expansion in matter overdensity and its perturbation theory treated in a similar manner \citep{2011ApJ...735...32W}, though there is an important difference; the expansion is in $\delta(\vec x)$ rather than the \textit{smoothed} field $\delta_R(\vec x)$. For this reason, it is not guaranteed to be convergent (unlike for the marked field, where this is ensured by using a suitably large smoothing scale $R$), thus requiring much more careful treatment of the UV limits and counterterms.}

A number of extensions to this work are possible. Firstly, this work has focused on the statistics of matter in real space; for application to survey data, one must rigorously consider biased tracers and redshift space, both in terms of a perturbative model and with simulations. The simple biased tracer results of Sec.\,\ref{subsec: bias-tracers} indicate that such models are within reach. Furthermore, whilst our model performed well across a range of redshifts, a more sophisticated treatment of massive neutrinos would be useful, as well as inclusion of infra-red resummation, to fully account for the suppression of BAO wiggles by long-wavelength modes. 

In summary, the development of a model for the marked density field is of significant use, both for understanding the statistic itself, and for performing cosmological inference.  Expansion of the theory to higher orders as well as application to more realistic cosmological scenarios may allow us to increase the information yield from large scale structure surveys, though much remains to be learnt. 

\begin{acknowledgments}
We thank Marko Simonovi\'c for assistance with FFTLog. We additionally thank \resub{Alejandro Avil\'es}, Simon Foreman, \resub{Henrique Rubira}, Francisco Villaescusa-Navarro,  and Martin White for useful feedback. \resub{Furthermore, we acknowledge the referee for a prompt and insightful report.} OHEP acknowledges funding from the WFIRST program through NNG26PJ30C and NNN12AA01C. The Flatiron Institute is supported by the Simons Foundation.
\end{acknowledgments}

\appendix

\section{Simplifications of $M(k)$ and $\mathcal{C}(k)$}\label{appen: simplif-and-limits}
Below, we simplify the expressions derived in Sec.\,\ref{sec: theory} for the marked power spectrum $M(\vec k)$ and cross-spectrum $\mathcal{C}(\vec k)$ (Eqs.\,\ref{eq: M-ij-def}\,\&\,\ref{eq: Cross-spec-terms}). These may be written in terms of convolution integrals, using the definitions of the $F$ and $H$ coupling kernels (Eq.\,\ref{eq: H-kernels}). Firstly, inserting $H_2$ into $M_{22}$ gives
\beq
    M_{22}(\vec k) = 2\int_{\vec p}&&\left[\left(1-C_1W(kR)\right)F_2(\vec p,\vec k-\vec p) - \frac{C_1}{2}\left(W(pR) + W(|\vec k-\vec p|R)\right)\right.\\\nonumber
    &&+\left.\frac{}{} C_2W(pR)W(|\vec k-\vec p|R)\right]^2P_L(\vec p)P_L(\vec k-\vec p).
\eeq

After significant rearrangement, this may be written as a set of convolutions;
\beq\label{eq: M22-simpl}
    \frac{1}{2}M_{22}(\vec k) &=& \left(1-C_1W(kR)\right)^2 \ast_{F_2^2}\left[P_L,P_L\right](\vec k) - 2C_1\left(1-C_1W(kR)\right)\ast_{F_2}\left[W_RP_L,P_L\right](\vec k)\\\nonumber
    &&\,+\, 2C_2\left(1-C_1W(kR)\right)\ast_{F_2}\left[W_RP_L,W_RP_L\right](\vec k) +\frac{C_1^2}{2}\ast\left[W_RP_L,W_RP_L\right](\vec k)\\\nonumber
    &&\,+\,C_2^2\ast\left[W_R^2P_L,W_R^2P_L\right](\vec k)\\\nonumber
    &&\,+\,\frac{C_1^2}{2}\ast\left[W_R^2P_L,P_L\right](\vec k)-2C_1C_2\ast\left[W_R^2P_L,W_RP_L\right](\vec k),
\eeq
where $\left(W_RP_L\right)(\vec k)\equiv W(kR)P_L(\vec k)$ and we have defined the general operator
\beq\label{eq: conv-integral}
    \ast_X\left[A,B\right](\vec k) = \int_{\vec p}X(\vec p,\vec k-\vec p)A(\vec p)B(\vec k-\vec p),
\eeq
which is the convolution of $A$ and $B$ with kernel $X$.\footnote{Note that the unkerneled convolution $\ast\left[A,B\right](\vec k)$ is equal to $\left[A\ast B\right](\vec k)$ in the notation of Sec.\,\ref{sec: theory}.} This is symmetric under $A\leftrightarrow B$ if $X$ is symmetric in its arguments. Furthermore, $\ast_{F_2^2}\left[P_L,P_L\right](\vec k)$ is equal to half the $P_{22}(\vec k)$ term of the perturbative matter power spectrum. The remaining terms are simply convolutions of the windowed and unwindowed power spectra, as well as convolutions with a kernel $F_2$ that are familiar from the one-loop perturbation theory of biased tracers in real space.

\resub{From \ref{eq: M22-simpl}, we may also obtain the low-$k$ behavior of $M_{22}$ (and thus the full one-loop spectrum, since $M_{13}(\vec k)\propto P_L(\vec k)$ which decays at large $|\vec k|$). In particular, we note that the convolution terms involving $F_2$ kernels decay as $k \rightarrow 0$, whilst those without a kernel asymptote to a constant. This can be seen from the limit of Eq.\,\ref{eq: conv-integral};}
\beq\label{eq: conv-integral-limit}
    \lim_{|\vec k|\rightarrow 0}\ast_X\left[A,B\right](\vec k) = \int_{\vec p}X(\vec p,-\vec p)A(\vec p)B(-\vec p)
\eeq
\resub{which is constant for $X = 1$ and vanishes for $X = F_2$ and $X = F_2^2$ since $F_2(\vec p,-\vec p) = 0$. In general, the non-zero asymptotes arise from contact terms in the marked density field, \textit{i.e.} terms involving products of operators at the same location, such as $\delta(\vec x)\delta_R(\vec x)$. Since the matter field does not contain these contact terms (except from biased tracers), the power spectrum model does not contain these asymptotes, giving the different low-$k$ limits of Fig.\,\ref{fig: spectral_components}. From Eq.\,\ref{eq: M22-simpl}, we see that the non-trivial limits arise only from terms involving one or more window functions $W_R$; this removes any contributions from modes with $p>1/R$, and hence small-scale physics. }

In a similar vein, the $M_{13}(\vec k)$ term may be written
\beq
    M_{13}(\vec k) &=& 3\left(1-C_1W(kR)\right)P_L(\vec k)\int_{\vec p}P_L(\vec p)\\\nonumber
    &&\,\times\left\{\left(1-C_1W(kR)\right)F_3(\vec p,-\vec p,\vec k)+ \frac{1}{3}C_2\left(W^2(pR)+2W(kR)W(pR)\right)\right.\\\nonumber
    &&\quad\,- \frac{C_1}{3}\left[
    \left(W(pR)+W(|\vec k-\vec p|R)\right)F_2(\vec k,-\vec p) + \left(W(pR)+W(|\vec k+\vec p|R)\right)F_2(\vec k,\vec p)\right]\\\nonumber
    &&\quad\,\left.+\frac{2C_2}{3}\left[W(pR)W(|\vec k-\vec p|R)F_2(\vec k,-\vec p)+W(pR)W(|\vec k+\vec p|R)F_2(\vec k,\vec p)\right]-C_3W^2(pR)W(kR)\frac{}{}\right\},
\eeq
using $F_2(\vec p,-\vec p)= 0$. By comparison to the perturbation theory of $P(\vec k)$, we note that the first term is simply $\left(1-C_1W(kR)\right)P_{13}(\vec k)$. After some rearrangement, this can be written
\beq\label{eq: M13-exp}
    M_{13}(\vec k) &=& \left(1-C_1W(kR)\right)^2P_{13}(\vec k)+\left(1-C_1W(kR)\right)P_L(\vec k)\left[C_2\sigma^2_{RR}+2C_2W(kR)\sigma^2_{R}-3C_3\sigma^2_{RR}W(kR)\right]\\\nonumber
    &&\,+\,2\left(1-C_1W(kR)\right)P_L(\vec k)\int_{\vec p}\left[2C_2W(pR)W(|\vec k-\vec p|R)-C_1W(pR)-C_1W(|\vec k-\vec p|R)\right]F_2(\vec k,-\vec p)P_L(\vec p),
\eeq
where we have combined the $F_2(\vec k,-\vec p)$ and $F_2(\vec k,\vec p)$ terms by symmetry. This uses the density field variances
\beq\label{eq: s2def}
    \sigma_R^2 &=& \int_{\vec p}P_L(\vec p)W(pR)\quad \text{and} \quad \sigma_{RR}^2 = \int_{\vec p}P_L(\vec p)W^2(pR).
\eeq
Whilst the first two lines of Eq.\,\ref{eq: M13-exp} can be simply computed from usual routines for the matter power spectrum, the terms in the final line are more complex since each depends on a two dimensional integral (comprising the modulus of $\vec p$ and its separation from $\vec k$) of Fourier-space window function. As in Appendix \ref{appen: prac-eval}, these can be reduced to one-dimensional integrals when we insert the definitions of $F_2(\vec k,-\vec p)$ and the (Gaussian or top-hat) window function.

For $\mathcal{C}_{22}(\vec k)$, we instead obtain;
\beq
    \frac{1}{2}\mathcal{C}_{22}(\vec k) &=& \left(1-C_1W(kR)\right)\ast_{F_2^2}\left[P_L,P_L\right](\vec k) - C_1\ast_{F_2}\left[W_RP_L,P_L\right](\vec k)\\\nonumber
    &&\,+C_2\ast_{F_2}\left[W_RP_L,W_RP_L\right](\vec k),
\eeq
whilst for $\mathcal{C}_{13}(\vec k)$ and $\mathcal{C}_{31}(\vec k)$;
\beq
    \mathcal{C}_{13}(\vec k) &=& \left(1-C_1W(kR)\right)P_{13}(\vec k)\\\nonumber
    \mathcal{C}_{31}(\vec k) &=& \frac{M_{13}(\vec k)}{1-C_1W(kR)}.
\eeq

\resub{One important feature of these expansions requires note. As seen from Eq.\,\ref{eq: M13-exp}, spectral shapes arising from the linear expansion of $M(\vec k)$ and $\mathcal{C}(\vec k)$ receive contributions from higher loops, e.g., for the term proportional to $P_L(\vec k)$;}
\beq
    M(\vec k) \supset \begin{cases} P_L(\vec k) & \text{ linear theory}\\ P_L(\vec k)\left[1+2C_2\sigma_{RR}^2\right] & \text{ one-loop theory.}\end{cases} 
\eeq
\resub{This is a generic feature of all shapes ($k$-dependencies) and arises from the presence of contact terms such as $\delta_R^3(\vec x)$ in the mark expansion. The higher loops thus have two effects; (a) contribution of new $k$-dependencies, (b) a \textit{renormalization} of the amplitudes of the lower loop shapes.\footnote{Technically, this arises since the perturbative expansion includes diagrams that are not one-particle irreducible.} Whilst the higher-order contributions are small for sufficiently large $R$, this raises the importance of the loop terms, enhancing the difficulty of the calculation. In the power spectrum EFT of biased tracers, a similar effect is seen due to the contact terms, however this can be absorbed into the free bias parameters \citep{2006PhRvD..74j3512M,2014JCAP...08..056A}, simplifying the calculation. Whilst a similar approach may be possible here, we must be careful since the Taylor expansion coefficients are fixed. We defer proper consideration of this to future work.}

\section{Evaluating 22- and 13-type Terms}\label{appen: prac-eval}
We briefly comment on the practicalities of evaluating the convolution integrals appearing in the one-loop formulae of Appendix \ref{appen: simplif-and-limits}. Note that the linear power spectrum can be efficiently computed using \texttt{CLASS} \citep{2011JCAP...07..034B}. Furthermore, whilst the normalization and shot-noise terms $\av{m(\vec x)n(\vec x)}$ and $\av{m^2(\vec x)n(\vec x)}$ can be computed perturbatively (e.g., $\bar{m} \equiv \av{m(\vec x)n(\vec x)}/\bar{n} = 1 - C_1\sigma_R^2+C_2\sigma_{RR}^2$ at linear order), it is far easier to measure these directly from the data, and this does not affect our numerical results.

\subsection{22-type Terms}
The expression for $M_{22}$ (Eq.\,\ref{eq: M22-simpl}) may be computed remarkably easily via Fourier methods, \citep{2016PhRvD..93j3528S,2016JCAP...09..015M} in particular the FFTLog procedure \citep{2000MNRAS.312..257H}. Firstly, note that unkerneled convolutions can be easily computed via Fourier transforms, \textit{i.e.}
\beq
    \ast\left[A,B\right](\vec k) &=& \mathcal{F}\left[\mathcal{F}^{-1}\left[A\right](\vec r)\mathcal{F}^{-1}\left[B\right](\vec r)\right](\vec k)\\\nonumber
\eeq
where $\mathcal{F}$ and $\mathcal{F}^{-1}$ represent forward and inverse Fourier transforms. Since $A$ and $B$ are here isotropic functions (as $P_L(\vec k) = P_L(k)$ in real-space), the transforms have a simple form $Y(r) = (2\pi^2)^{-1}\int p^2dp\,Y(p)j_0(pr)$ (for $Y$ equal to $A$ or $B$), where $j_0$ is the zeroth order spherical Bessel function. This gives
\beq
    \ast\left[A,B\right](k) = 4\pi\int r^2dr\,j_0(kr)A(r)B(r),
\eeq
which is simply evaluated with FFTLog-based packages such as \texttt{mcfit}.\footnote{\href{https://github.com/eelregit/mcfit}{github.com/eelregit/mcfit}}

For evaluating the $\ast_{F_2}$ and $\ast_{F_2^2}$ terms, we make use of the method of \citet{2018JCAP...04..030S}, developed for computing loop integrals for the EFT of matter and biased tracers. Considering a general convolution $\ast_{X}\left[A,B\right](k)$, we first expand $A$ and $B$ as discrete Fourier transforms in $\log k$;
\beq\label{eq: FFT-decomp}
    \bar{A}(k) = \sum_{m=-N/2}^{m=N/2}c_m^Ak^{\nu+i\eta_m},
\eeq
using $N$ points spanning $\log k_\mathrm{min}$ to $\log k_\mathrm{max}$. $\nu$ is known as the `bias' and the coefficients $c_m^A$ and frequencies $\eta_m$ are given by 
\beq
    c_m^A\equiv \frac{1}{N}\sum_{l=0}^{N-1}A(k_l)k_l^{-\nu}k_\mathrm{min}^{-i\eta_m}e^{-2\pi iml/N},\quad \eta_m = \frac{2\pi m}{\log\left(k_\mathrm{max}/k_\mathrm{min}\right)}.
\eeq
With this approximation, the $A$ and $B$ functions are simply (complex) power laws, thus the kerneled convolution integral may be written
\beq
    \ast_X\left[A,B\right](k) \approx \sum_{m_1m_2}c_{m_1}^Ac_{m_2}^B\int_{\vec p}X(\vec p,\vec k-\vec p)p^{-2\nu_1}|\vec k-\vec p|^{-2\nu_2}
\eeq
where $\nu_j\equiv -\tfrac{1}{2}\left(\nu+i\eta_{m_j}\right)$. In Ref.\,\citep{2018JCAP...04..030S} the following result is proved;
\beq\label{eq: I-def}
    \int_{\vec p}\frac{1}{p^{2\nu_1}|\vec k-\vec p|^{2\nu_2}} &\equiv& k^{3-2\nu_{12}}\mathsf{I}(\nu_1,\nu_2) \qquad 
    \mathsf{I}(\nu_1,\nu_2) = \frac{1}{8\pi^{3/2}}\frac{\Gamma\left(\frac{3}{2}-\nu_1\right)\Gamma\left(\frac{3}{2}-\nu_2\right)\Gamma\left(\nu_{12}-\frac{3}{2}\right)}{\Gamma(\nu_1)\Gamma(\nu_2)\Gamma(3-\nu_{12})},
\eeq
where $\nu_{12}=\nu_1+\nu_2$ and $\Gamma$ is the Gamma function. 

In this work, we consider two choices of the kernel; $X = F_2$ and $X = F_2^2$. Using the definition
\beq\label{eq: f2-kernel}
    F_2(\vec p,\vec k-\vec p) = \frac{5}{14} + \frac{3k^2}{28p^2} + \frac{3k^2}{28|\vec k-\vec p|^2} - \frac{5p^2}{28|\vec k-\vec p|^2} - \frac{5|\vec k-\vec p|^2}{28p^2} + \frac{k^4}{14|\vec k-\vec p|^2p^2}
\eeq
\citep{2002PhR...367....1B}, both can be written as a sum over powers of $p^2$ and $|\vec k-\vec p|^2$, \textit{i.e.}
\beq
    X(\vec p,\vec k-\vec p) = \sum_{n_1n_2}f_X(n_1,n_2)p^{2n_1}|\vec k-\vec p|^{2n_2}k^{-2(n_1+n_2)},
\eeq
where $f_X(n_1,n_2)$ is the (constant) coefficient of the term containing $n_1$ powers of $p^2$ and $n_2$ powers of $|\vec k-\vec p|^2$. This allows the convolution integral to be written 
\beq\label{eq: kernel-conv-form}
    \ast_{X}\left[A,B\right](k) &\approx& \sum_{m_1m_2}c_{m_1}^Ac_{m_2}^B\sum_{n_1n_2}f_X(n_1,n_2)k^{-2(n_1+n_2)}\int_{\vec p}\frac{1}{p^{2\nu_1-2n_1}|\vec k-\vec p|^{2\nu_2-2n_2}}\\\nonumber
    &=& k^3\sum_{m_1m_2}\left(c_{m_1}^Ak^{-2\nu_1}\right)\mathcal{M}_{X}(\nu_1,\nu_2)\left(c_{m_2}^Bk^{-2\nu_2}\right),
\eeq
which is a matrix product with coupling
\beq
    \mathcal{M}_X(\nu_1,\nu_2) = \sum_{n_1,n_2}f_X(n_1,n_2)\mathsf{I}(\nu_1-n_1,\nu_2,n_2).
\eeq
in terms of the $\mathsf{I}$ matrix of Eq.\,\ref{eq: I-def}. For the $F_2$ and $F_2^2$ kernels, the couplings can be shown to equal
\beq
    \mathcal{M}_{F_2}(\nu_1,\nu_2) &=& \frac{(3-2\nu_{12})(4-7\nu_{12})}{34\nu_1\nu_2}\mathsf{I}(\nu_1,\nu_2)\\\nonumber
    \mathcal{M}_{F_2^2}(\nu_1,\nu_2) &=& \frac{(\tfrac{3}{2}-\nu_{12})(\tfrac{1}{2}-\nu_{12})\left[\nu_1\nu_2\left(98\nu_{12}^2-14\nu_{12}+36\right)-91\nu_{12}^2+3\nu_{12}+58\right]}{392\nu_1(1+\nu_1)(\tfrac{1}{2}-\nu_1)\nu_2(1+\nu_2)(\tfrac{1}{2}-\nu_2)}\mathsf{I}(\nu_1,\nu_2)
\eeq
\citep[Eqs.\,2.38\,\&\,2.24]{2018JCAP...04..030S},\footnote{Note that our normalization differs from that of Ref.\,\citep{2018JCAP...04..030S} by a factor of two.} which depends only on a single $\mathsf{I}$ matrix and a $\nu$-dependent prefactor (from application of the recursion relations of $\mathsf{I}(\nu_1,\nu_2)$). Computation of the $M_{22}$ convolution terms thus reduces to a weighted sum over the discrete FFT components of the functions $A$ and $B$. This is a slight generalization of Ref.\,\citep{2018JCAP...04..030S}, in which we allow $A$ and $B$ to be distinct.

Some care is needed to ensure that these integrals are convergent. From the discussions in Appendix \ref{appen: UVIRLimit}, we find that the $P_{22}$ part of $M_{22}$ is convergent for power law cosmologies $P_L(k)\sim k^n$ for $-1<n<\tfrac{1}{2}$ with all other pieces fully convergent (assuming a Gaussian window function). As discussed in Ref.\,\citep{2018JCAP...04..030S}, if the bias $\nu$ is chosen to be in this range, the integral is finite for every term in the FFT decomposition of Eq.\,\ref{eq: FFT-decomp}, and hence the FFTLog expansion is well-defined. For practical evaluation, it is often useful to use a bias outside this range. This can be simply achieved by adding on the true UV or IR limit from the divergent piece manually. As an example, if we wished to compute the $P_{22}$ piece using $\nu=-2$ (giving an IR-divergent integral), we would use Eq.\,\ref{eq: kernel-conv-form} as normal, then add the IR limit to this result. \resub{In this paper, we use $\nu = -1.6$ as in Ref.\,\citep{2018JCAP...04..030S} and add on the IR limits manually, with $M_{22,\mathrm{IR}}(k) = -2M_{13,\mathrm{IR}}(k) = \left(1-C_1W(kR)\right)^2P_{22,\mathrm{IR}}(k) = -2\left(1-C_1W(kR)\right)^2P_{13,\mathrm{IR}}(k) = -k^2\sigma_v^2P_L(k)$.}

\subsection{Evaluation of $M_{13}$}
Similar treatment is possible for the $13$-type integrals. First, the $P_{13}(k)$ term can be written in the FFTLog formalism;
\beq
    P_{13}(k) &\approx& k^3P_L(k)\sum_{m_1}c^{P_L}_{m_1}k^{-2\nu_1}\mathcal{M}_{13}(\nu_1)\\\nonumber
    \mathcal{M}_{13}(\nu_1) &=& \frac{1+9\nu_1}{8}\frac{\tan({\nu_1\pi})}{28\pi(\nu_1+1)\nu_1(\nu_1-1)(\nu_1-2)(\nu_1-3)},
\eeq
where $\{c^{P_L}\}$ are the discrete FFT coefficients of $P_L(k)$. The remaining non-trivial integrals are of the form
\beq\label{eq: m13-integs}
    \int_{\vec p}P_L(p)W(pR)F_2(\vec k,-\vec p)\quad \text{and}\quad\int_{\vec p}A(p)W(|\vec k-\vec p|R)F_2(\vec k,-\vec p)
\eeq
for isotropic $A(p)$. Whilst these can be computed via FFTLog methods, it is simpler to perform the angular integration analytically, making use of the simple form of the window function. Parametrizing by the angle $\mu = \hat{\vec k}\cdot\hat{\vec p}$, such that $|\vec k-\vec p|^2 = k^2 + p^2 - 2kp\mu$, the kernel is given by
\beq\label{eq: f2-kernel2}
    F_2(\vec k,-\vec p) &=& \frac{5}{14} + \frac{3|\vec k-\vec p|^2}{28p^2}+\frac{3|\vec k-\vec p|^2}{28k^2}-\frac{5k^2}{28p^2}-\frac{5p^2}{28k^2}+\frac{|\vec k-\vec p|^4}{14k^2p^2}\\\nonumber
    &=& \frac{1}{14}\left[10+4\mu^2-7\mu\left(\frac{k}{p}+\frac{p}{k}\right)\right].
\eeq
For a Gaussian window function, this yields
\beq
    \int_{\vec p}P_L(p)W(pR)F_2(\vec k,-\vec p) &=& \frac{1}{14}\int\frac{p^2dp}{2\pi^2}P_L(p)W(pR)\int_{-1}^1 \frac{d\mu}{2}\left[10+4\mu^2-7\mu\left(\frac{k}{p}+\frac{p}{k}\right)\right]\\\nonumber
    &=& \frac{17}{21}\int \frac{p^2dp}{2\pi^2}P_L(p)W(pR) \equiv \frac{17}{21}\sigma^2_R\\\nonumber
    \int_{\vec p}A(p)W(|\vec k-\vec p|R)F_2(\vec k,-\vec p) &=& \frac{1}{14}\int\frac{p^2dp}{2\pi^2}A(p)e^{-R^2(k^2+p^2)/2}
    \int_{-1}^1 \frac{d\mu}{2}e^{R^2kp\mu}\left[10+4\mu^2-7\mu\left(\frac{k}{p}+\frac{p}{k}\right)\right]\\\nonumber
    &=& \int\frac{p^2dp}{2\pi^2}A(p)g(pR,kR),
\eeq
where we define 
\beq
    g(P,K) = \frac{1}{28K^3P^3}&&\left\{\frac{}{}e^{-\frac{1}{2} (K-P)^2} \left(-7 K^3 P+7 K^2\left(2 P^2+1\right)-K P \left(7 P^2+8\right)+7P^2+8\right)\right.\\\nonumber
    &&\,\left.\frac{}{}-e^{-\frac{1}{2} (K+P)^2} \left(7 K^3
   P+7 K^2 \left(2 P^2+1\right)+K P \left(7
   P^2+8\right)+7 P^2+8\right)\right\},
\eeq
\resub{which goes as $g(P,K) = (17/21-7P^2/42)W(pR)$ in the low-$k$ limit.} The convolution integrals thus become one-dimensional, allowing for simple computation.\footnote{As shown in \citet{2011ApJ...735...32W}, a similar computation is possible when using a top-hat window function with $W(K) = 3j_1(K)/K$; in this case we obtain \beq
    g(P,K) = \frac{8}{21}W(K)W(P)-\frac{1}{21}\left[PW'(P)W(K)+KW'(K)W(P)\right].
\eeq}  

Once again, we must consider convergence. As shown in Appendix \ref{appen: UVIRLimit}, most $M_{13}(\vec k)$ terms are convergent for any power law cosmology $P(k)\sim k^n$, though this is not true for the FFTLog term involving $P_{13}(\vec k)$. This is UV divergent for $n>-1$ and IR divergent for $n<-1$, thus the correct (IR or UV) limit must be added to the FFTLog expansion depending on the value of the bias $\nu$. Note however that this divergence will necessarily cancel when combined with $M_{22}(\vec k)$. 
Using a bias in the range $-3<\nu<\tfrac{1}{2}$, the FFTLog procedure is thus guaranteed to give the correct spectrum when the two one-loop terms are combined, though for $\nu<-1$, the individual $M_{22}$ and $M_{13}$ terms will require the addition of their IR limits.

\section{Limiting Behavior of the One-Loop Integrands}\label{appen: UVIRLimit}
Here, we give the limiting behavior of the power spectrum integrands for large and small internal momenta which is important for ensuring that (a) the perturbation theory is well-posed and does not require additional counterterms, and (b) the FFTLog expansion is convergent. 

\subsection{UV Limit: $p\gg k$}
Ensuring that the integrand is UV-safe (\textit{i.e.} convergent for `hard' internal momenta $p\gg k$) is important for evaluation of the loop integrals; since we only integrate up to some large cut-off $\Lambda$ in $|\vec p|$, we require the result to be independent of $\Lambda$. In general, any divergences present will be captured by counterterms in the theory. 
For this purpose, consider a power-law cosmology with $P_L(p)\propto p^{n}$, noting that $n\approx -2.1$ at $k\sim 0.6h\,\mathrm{Mpc}^{-1}$ (beyond the non-linear scale). Our expressions are significantly simplified by the use of a Gaussian window function in the mark, since any expression containing $W(pR)$ or $W(|\vec k-\vec p|R)$ will asymptote to zero for all choices of the power-law index $n$. Importantly, this ensures that all loop integrals containing an internal $\delta_R$ field are convergent. This is not true for a generic window function; in the case of a top-hat window function, $W(kR)$ contains spherical Bessel functions, and the UV limits should be carefully considered.

For the Gaussian window, the only integrand terms which can contain UV divergences are thus $P_{13}(\vec k)$ and $P_{22}(\vec k)$, as in the matter EFT. These can be shown to have the following contributions;
\beq\label{eq: usual-UV-lim}
    \left[P_{13}\right]_\mathrm{UV}(\vec k) &=& -\frac{61}{210}k^2P_L(\vec k) \sigma_v^2\,,\qquad (n\geq -1)\\\nonumber
    \ast_{F_2^2}\left[P_L,P_L\right]_\mathrm{UV}(\vec k) &=& \frac{9}{196}k^4\int \frac{p^2dp}{2\pi^2}\frac{P_L^2}{p^4}\,,\qquad (n\geq \frac{1}{2})
\eeq
where the parentheses show the values of the power law index $n$ for which the integrals are UV divergent and we define $\sigma_v^2\equiv (6\pi^2)^{-1}\int dp\,P_L(p)$. This uses the kernel relations
\beq
    \lim_{p\rightarrow\infty} F_3(\vec k,-\vec p,\vec p) &\propto& \lim_{p\rightarrow\infty} F_2(\vec p,\vec k-\vec p) \propto \frac{k^2}{p^2}.
\eeq
Assuming Gaussian smoothing, the spectrum is thus UV convergent for all $n<-1$.\footnote{In the corresponding calculation for the unmarked power spectrum of biased tracers in real space, we obtain a UV divergence in $P_{22}$ since there is no regularizing $W_R$ function. This can be dealt with by renormalizing the bias parameters to remove the cut-off dependence.} By design, all divergences in the $P_{22}$ and $P_{13}$ parts (for any power-law index $n$) are exactly cancelled by the EFT counterterms arising from smoothing the theory on scales smaller than $\Lambda$. For a linear power spectrum which is at steeper than $P_L(p)\sim p^{-1}$ at large $p$ (as in our Universe), the integrals are convergent and thus the theory is well defined.

In full, the limiting behavior of the marked power spectrum with UV internal momentum is given by
\beq\label{eq: UV-limit2}
    \frac{1}{2}\left(M_{22}\right)_\mathrm{UV}(k) &=& \frac{9k^4}{196}\left(1-C_1W(kR)\right)^2\Pi^{00}_4 + \frac{k^2}{21}C_1\left(1-C_1W(kR)\right)\Pi^{10}_2\\\nonumber
    &&\,- \frac{k}{2R^2}C_2\left(1-C_1W(kR)\right)\Pi_3^{11}(k)+\frac{1}{8R^2k}C_1^2\Pi_1^{11}(k)+\frac{1}{4R^2k}C_2^2\Pi_1^{22}(k)\\\nonumber
    &&\,+\frac{1}{2}C_1^2\Pi_0^{20} - \frac{1}{R^2k}C_1C_2\Pi_{1}^{21}(k)\\\nonumber
    \left(\frac{M_{13}}{\left(1-C_1W(kR)\right)}\right)_\mathrm{UV}(k) &=& -\frac{61k^2}{210}\left(1-C_1W(kR)\right)\sigma_v^2P_L(k)\\\nonumber
    &&\,+P_L(k)\left[C_2\sigma^2_{RR}+2C_2W(kR)\sigma^2_{R}-3C_3\sigma^2_{RR}W(kR)\right]\\\nonumber
    &&\,+P_L(k)\left[-\frac{1}{k^2R^2}C_2\Sigma^{11}(k)-\frac{34}{21}C_1\sigma_R^2+\frac{1}{2k^2R^2}C_1\Sigma^{01}(k)\right]\\\nonumber
\eeq
with all terms except the first receiving Gaussian suppression in the UV. These use the definitions
\beq
    \Pi^{\alpha\beta}_\gamma(k) &\equiv& \int \frac{p^2dp}{2\pi^2}\frac{P_L^2(p)}{p^\gamma}W^\alpha(pR)\left[W^\beta((p-k)R)-W^\beta((p+k)R)\right]\\\nonumber
    \Sigma^{\alpha\beta}(k) &\equiv& \int \frac{p^2dp}{2\pi^2}P_L(p)W^\alpha(pR)\left[W^\beta((p-k)R)+W^\beta((p+k)R)\right]
\eeq
for integer $\alpha,\beta$ (with UV sensitivity only for $\alpha=\beta=0$). Note that $\Sigma^{10}\equiv \sigma^2_R, \Sigma^{20}\equiv \sigma^2_{RR}$ and $\Pi^{\alpha\beta}_{\gamma}$ is a function of $k$ only for $\beta>0$. The cross-spectra have similar forms;
\beq
    \frac{1}{2}\left(\mathcal{C}_{22}\right)_\mathrm{UV}(k) &=& \frac{9k^4}{196}\left(1-C_1W(kR)\right)\Pi^{00}_4 + \frac{k^2}{42}C_1\Pi^{10}_2 - \frac{k}{4R^2}C_2\Pi_3^{11}(k)\\\nonumber
    \left(\frac{\mathcal{C}_{13}}{\left(1-C_1W(kR)\right)}\right)_\mathrm{UV}(k) &=& -\frac{61k^2}{210}\sigma_v^2P_L(k)\\\nonumber
    \left(\mathcal{C}_{31}\right)_\mathrm{UV}(k) &=& \left(\frac{M_{13}}{\left(1-C_1W(kR)\right)}\right)_\mathrm{UV}(k),
\eeq
again with only the terms proportional to $P_{22}(\vec k)$ and $P_{13}(\vec k)$ containing UV divergences, which are regularized by the $c_s^2$ counterterm.

\subsection{IR Limit: $p \ll k, |\vec p-\vec k|\ll k$}
The behavior of one-loop terms with internal momenta in the infrared (IR) regime (\textit{i.e.} `soft') is important for practical evaluation of the integrals. For $M_{22}$ there are two cases to consider; $|\vec k-\vec p|\ll k $ and $p \ll k$, though, if the convolution has symmetric arguments, these are identical due to the relabelling symmetry of the integrand under $\vec p\rightarrow \vec k-\vec p$. It can be shown that all $p$ dependence in the IR limit is encapsulated by the terms
\beq
   \sigma^2, \sigma_R^2, \sigma_{RR}^2, \sigma_v^2
\eeq
where $\sigma^2 \equiv (2\pi^2)^{-1}\int p^2dp\,P_L(p)$. For a power-law cosmology with $P_L(p)\sim p^n$, the first three terms all scale as $\sigma^2$ and are IR-divergent only for $n\leq -3$. For $\sigma_v^2 \sim \int dp\,P_L(p)$, divergences occur instead for $n\leq -1$. However, these terms appear both in $M_{22}$ and $M_{13}$ and, due to the Galilean invariance and the equivalence principle \citep{1996ApJS..105...37S,1996ApJ...456...43J,2013JCAP...09..024B,2014JCAP...07..056C,2014JCAP...01..010B}, cancel identically when $M_{22}$ and $M_{13}$ are combined (and similarly for the cross-spectrum). In our Universe, $n\approx 0.96$, thus the individual terms are also convergent, though this has consequences for the choice of FFTLog bias (Appendix \ref{appen: prac-eval}).

In full, the IR limits are given by
\beq
    \frac{1}{2}\left[M_{22}\right]_\mathrm{IR}(k) &=& \frac{k^2}{2}\left(1-C_1W(kR)\right)^2P_L(\vec k)\sigma_v^2\\\nonumber
    &&\,- \frac{1}{21}C_1\left(1-C_1W(kR)\right)\left[\left(13P_L(\vec k)-7kP_L'(\vec k)\right)\sigma_R^2+7k^2R^2\sigma^2\right]\\\nonumber
    &&\,+ \left[\frac{14k^2R^2}{21}C_2\left(1-C_1W(kR)\right)+C_1^2W(kR)-2C_1C_2W^2(kR)\right]P_L(\vec k)\sigma_{R}^2\\\nonumber
    &&\,+\frac{1}{2}C_1^2W^2(kR)P_L(\vec k)\sigma^2+\left[2C_2^2W^2(kR)+\frac{1}{2}C_1^2 - 2C_1C_2W(kR)\right]P_L(\vec k)\sigma_{RR}^2\\\nonumber
    \left(\frac{M_{13}}{\left(1-C_1W(kR)\right)}\right)_\mathrm{IR}(k) &=& -\frac{k^2}{2}\left(1-C_1W(kR)\right)\sigma_v^2P_L(k)\\\nonumber
    &&\,+\left[C_2\sigma^2_{RR}+2C_2W(kR)\sigma^2_{R}-3C_3\sigma^2_{RR}W(kR)\right]P_L(k)\\\nonumber
    &&\,+\left[\frac{k^2R^2}{3}W(kR)\left(C_1\sigma^2-2C_2\sigma_{R}^2\right)-\frac{34}{21}\sigma_R^2\right]P_L(k)
\eeq
for $M(k)$ and 
\beq
    \frac{1}{2}\left[\mathcal{C}_{22}\right]_\mathrm{IR}(k) &=& \frac{k^2}{2}\left(1-C_1W(kR)\right)P_L(\vec k)\sigma_v^2- \frac{1}{42}C_1\left[\left(13P_L(\vec k)-7kP_L'(\vec k)\right)\sigma_R^2+7k^2R^2\sigma^2\right]\\\nonumber
    &&\,+ \frac{14k^2R^2}{42}C_2P_L(\vec k)\sigma_{R}^2\\\nonumber
    \left(\frac{\mathcal{C}_{13}}{\left(1-C_1W(kR)\right)}\right)_\mathrm{IR}(k) &=& -\frac{k^2}{2}\sigma_v^2P_L(k)\\\nonumber
    \left(\mathcal{C}_{31}\right)_\mathrm{IR}(k) &=& \left(\frac{M_{13}}{\left(1-C_1W(kR)\right)}\right)_\mathrm{IR}(k)
\eeq
for $\mathcal{C}(k)$.


\bibliography{adslib,otherlib} 

\begin{thebibliography}{67}%
\makeatletter
\providecommand \@ifxundefined [1]{%
 \@ifx{#1\undefined}
}%
\providecommand \@ifnum [1]{%
 \ifnum #1\expandafter \@firstoftwo
 \else \expandafter \@secondoftwo
 \fi
}%
\providecommand \@ifx [1]{%
 \ifx #1\expandafter \@firstoftwo
 \else \expandafter \@secondoftwo
 \fi
}%
\providecommand \natexlab [1]{#1}%
\providecommand \enquote  [1]{``#1''}%
\providecommand \bibnamefont  [1]{#1}%
\providecommand \bibfnamefont [1]{#1}%
\providecommand \citenamefont [1]{#1}%
\providecommand \href@noop [0]{\@secondoftwo}%
\providecommand \href [0]{\begingroup \@sanitize@url \@href}%
\providecommand \@href[1]{\@@startlink{#1}\@@href}%
\providecommand \@@href[1]{\endgroup#1\@@endlink}%
\providecommand \@sanitize@url [0]{\catcode `\\12\catcode `\$12\catcode
  `\&12\catcode `\#12\catcode `\^12\catcode `\_12\catcode `\%12\relax}%
\providecommand \@@startlink[1]{}%
\providecommand \@@endlink[0]{}%
\providecommand \url  [0]{\begingroup\@sanitize@url \@url }%
\providecommand \@url [1]{\endgroup\@href {#1}{\urlprefix }}%
\providecommand \urlprefix  [0]{URL }%
\providecommand \Eprint [0]{\href }%
\providecommand \doibase [0]{http://dx.doi.org/}%
\providecommand \selectlanguage [0]{\@gobble}%
\providecommand \bibinfo  [0]{\@secondoftwo}%
\providecommand \bibfield  [0]{\@secondoftwo}%
\providecommand \translation [1]{[#1]}%
\providecommand \BibitemOpen [0]{}%
\providecommand \bibitemStop [0]{}%
\providecommand \bibitemNoStop [0]{.\EOS\space}%
\providecommand \EOS [0]{\spacefactor3000\relax}%
\providecommand \BibitemShut  [1]{\csname bibitem#1\endcsname}%
\let\auto@bib@innerbib\@empty
\bibitem [{\citenamefont {{Scoccimarro}}\ \emph {et~al.}(1999)\citenamefont
  {{Scoccimarro}}, \citenamefont {{Zaldarriaga}},\ and\ \citenamefont
  {{Hui}}}]{1999ApJ...527....1S}%
  \BibitemOpen
  \bibfield  {author} {\bibinfo {author} {\bibfnamefont {R.}~\bibnamefont
  {{Scoccimarro}}}, \bibinfo {author} {\bibfnamefont {M.}~\bibnamefont
  {{Zaldarriaga}}}, \ and\ \bibinfo {author} {\bibfnamefont {L.}~\bibnamefont
  {{Hui}}},\ }\href {\doibase 10.1086/308059} {\bibfield  {journal} {\bibinfo
  {journal} {\apj}\ }\textbf {\bibinfo {volume} {527}},\ \bibinfo {pages} {1}
  (\bibinfo {year} {1999})},\ \Eprint {http://arxiv.org/abs/astro-ph/9901099}
  {arXiv:astro-ph/9901099 [astro-ph]} \BibitemShut {NoStop}%
\bibitem [{\citenamefont {{Sefusatti}}\ \emph {et~al.}(2006)\citenamefont
  {{Sefusatti}}, \citenamefont {{Crocce}}, \citenamefont {{Pueblas}},\ and\
  \citenamefont {{Scoccimarro}}}]{2006PhRvD..74b3522S}%
  \BibitemOpen
  \bibfield  {author} {\bibinfo {author} {\bibfnamefont {E.}~\bibnamefont
  {{Sefusatti}}}, \bibinfo {author} {\bibfnamefont {M.}~\bibnamefont
  {{Crocce}}}, \bibinfo {author} {\bibfnamefont {S.}~\bibnamefont {{Pueblas}}},
  \ and\ \bibinfo {author} {\bibfnamefont {R.}~\bibnamefont {{Scoccimarro}}},\
  }\href {\doibase 10.1103/PhysRevD.74.023522} {\bibfield  {journal} {\bibinfo
  {journal} {\prd}\ }\textbf {\bibinfo {volume} {74}},\ \bibinfo {eid} {023522}
  (\bibinfo {year} {2006})},\ \Eprint {http://arxiv.org/abs/astro-ph/0604505}
  {arXiv:astro-ph/0604505 [astro-ph]} \BibitemShut {NoStop}%
\bibitem [{\citenamefont {{Gil-Mar{\'\i}n}}\ \emph {et~al.}(2017)\citenamefont
  {{Gil-Mar{\'\i}n}}, \citenamefont {{Percival}}, \citenamefont {{Verde}},
  \citenamefont {{Brownstein}}, \citenamefont {{Chuang}}, \citenamefont
  {{Kitaura}}, \citenamefont {{Rodr{\'\i}guez-Torres}},\ and\ \citenamefont
  {{Olmstead}}}]{2017MNRAS.465.1757G}%
  \BibitemOpen
  \bibfield  {author} {\bibinfo {author} {\bibfnamefont {H.}~\bibnamefont
  {{Gil-Mar{\'\i}n}}}, \bibinfo {author} {\bibfnamefont {W.~J.}\ \bibnamefont
  {{Percival}}}, \bibinfo {author} {\bibfnamefont {L.}~\bibnamefont {{Verde}}},
  \bibinfo {author} {\bibfnamefont {J.~R.}\ \bibnamefont {{Brownstein}}},
  \bibinfo {author} {\bibfnamefont {C.-H.}\ \bibnamefont {{Chuang}}}, \bibinfo
  {author} {\bibfnamefont {F.-S.}\ \bibnamefont {{Kitaura}}}, \bibinfo {author}
  {\bibfnamefont {S.~A.}\ \bibnamefont {{Rodr{\'\i}guez-Torres}}}, \ and\
  \bibinfo {author} {\bibfnamefont {M.~D.}\ \bibnamefont {{Olmstead}}},\ }\href
  {\doibase 10.1093/mnras/stw2679} {\bibfield  {journal} {\bibinfo  {journal}
  {\mnras}\ }\textbf {\bibinfo {volume} {465}},\ \bibinfo {pages} {1757}
  (\bibinfo {year} {2017})},\ \Eprint {http://arxiv.org/abs/1606.00439}
  {arXiv:1606.00439 [astro-ph.CO]} \BibitemShut {NoStop}%
\bibitem [{\citenamefont {{Pearson}}\ and\ \citenamefont
  {{Samushia}}(2018)}]{2018MNRAS.478.4500P}%
  \BibitemOpen
  \bibfield  {author} {\bibinfo {author} {\bibfnamefont {D.~W.}\ \bibnamefont
  {{Pearson}}}\ and\ \bibinfo {author} {\bibfnamefont {L.}~\bibnamefont
  {{Samushia}}},\ }\href {\doibase 10.1093/mnras/sty1266} {\bibfield  {journal}
  {\bibinfo  {journal} {\mnras}\ }\textbf {\bibinfo {volume} {478}},\ \bibinfo
  {pages} {4500} (\bibinfo {year} {2018})},\ \Eprint
  {http://arxiv.org/abs/1712.04970} {arXiv:1712.04970 [astro-ph.CO]}
  \BibitemShut {NoStop}%
\bibitem [{\citenamefont {{Scoccimarro}}\ \emph {et~al.}(2001)\citenamefont
  {{Scoccimarro}}, \citenamefont {{Feldman}}, \citenamefont {{Fry}},\ and\
  \citenamefont {{Frieman}}}]{2001ApJ...546..652S}%
  \BibitemOpen
  \bibfield  {author} {\bibinfo {author} {\bibfnamefont {R.}~\bibnamefont
  {{Scoccimarro}}}, \bibinfo {author} {\bibfnamefont {H.~A.}\ \bibnamefont
  {{Feldman}}}, \bibinfo {author} {\bibfnamefont {J.~N.}\ \bibnamefont
  {{Fry}}}, \ and\ \bibinfo {author} {\bibfnamefont {J.~A.}\ \bibnamefont
  {{Frieman}}},\ }\href {\doibase 10.1086/318284} {\bibfield  {journal}
  {\bibinfo  {journal} {\apj}\ }\textbf {\bibinfo {volume} {546}},\ \bibinfo
  {pages} {652} (\bibinfo {year} {2001})},\ \Eprint
  {http://arxiv.org/abs/astro-ph/0004087} {arXiv:astro-ph/0004087 [astro-ph]}
  \BibitemShut {NoStop}%
\bibitem [{\citenamefont {{Szapudi}}(2004)}]{2004ApJ...605L..89S}%
  \BibitemOpen
  \bibfield  {author} {\bibinfo {author} {\bibfnamefont {I.}~\bibnamefont
  {{Szapudi}}},\ }\href {\doibase 10.1086/420894} {\bibfield  {journal}
  {\bibinfo  {journal} {\apjl}\ }\textbf {\bibinfo {volume} {605}},\ \bibinfo
  {pages} {L89} (\bibinfo {year} {2004})},\ \Eprint
  {http://arxiv.org/abs/astro-ph/0404476} {arXiv:astro-ph/0404476 [astro-ph]}
  \BibitemShut {NoStop}%
\bibitem [{\citenamefont {{Fergusson}}\ \emph {et~al.}(2012)\citenamefont
  {{Fergusson}}, \citenamefont {{Regan}},\ and\ \citenamefont
  {{Shellard}}}]{2012PhRvD..86f3511F}%
  \BibitemOpen
  \bibfield  {author} {\bibinfo {author} {\bibfnamefont {J.~R.}\ \bibnamefont
  {{Fergusson}}}, \bibinfo {author} {\bibfnamefont {D.~M.}\ \bibnamefont
  {{Regan}}}, \ and\ \bibinfo {author} {\bibfnamefont {E.~P.~S.}\ \bibnamefont
  {{Shellard}}},\ }\href {\doibase 10.1103/PhysRevD.86.063511} {\bibfield
  {journal} {\bibinfo  {journal} {\prd}\ }\textbf {\bibinfo {volume} {86}},\
  \bibinfo {eid} {063511} (\bibinfo {year} {2012})},\ \Eprint
  {http://arxiv.org/abs/1008.1730} {arXiv:1008.1730 [astro-ph.CO]} \BibitemShut
  {NoStop}%
\bibitem [{\citenamefont {{Schmittfull}}\ \emph {et~al.}(2013)\citenamefont
  {{Schmittfull}}, \citenamefont {{Regan}},\ and\ \citenamefont
  {{Shellard}}}]{2013PhRvD..88f3512S}%
  \BibitemOpen
  \bibfield  {author} {\bibinfo {author} {\bibfnamefont {M.~M.}\ \bibnamefont
  {{Schmittfull}}}, \bibinfo {author} {\bibfnamefont {D.~M.}\ \bibnamefont
  {{Regan}}}, \ and\ \bibinfo {author} {\bibfnamefont {E.~P.~S.}\ \bibnamefont
  {{Shellard}}},\ }\href {\doibase 10.1103/PhysRevD.88.063512} {\bibfield
  {journal} {\bibinfo  {journal} {\prd}\ }\textbf {\bibinfo {volume} {88}},\
  \bibinfo {eid} {063512} (\bibinfo {year} {2013})},\ \Eprint
  {http://arxiv.org/abs/1207.5678} {arXiv:1207.5678 [astro-ph.CO]} \BibitemShut
  {NoStop}%
\bibitem [{\citenamefont {{Hung}}\ \emph {et~al.}(2019)\citenamefont {{Hung}},
  \citenamefont {{Manera}},\ and\ \citenamefont
  {{Shellard}}}]{2019arXiv190903248H}%
  \BibitemOpen
  \bibfield  {author} {\bibinfo {author} {\bibfnamefont {J.}~\bibnamefont
  {{Hung}}}, \bibinfo {author} {\bibfnamefont {M.}~\bibnamefont {{Manera}}}, \
  and\ \bibinfo {author} {\bibfnamefont {E.~P.~S.}\ \bibnamefont
  {{Shellard}}},\ }\href@noop {} {\bibfield  {journal} {\bibinfo  {journal}
  {arXiv e-prints}\ ,\ \bibinfo {eid} {arXiv:1909.03248}} (\bibinfo {year}
  {2019})},\ \Eprint {http://arxiv.org/abs/1909.03248} {arXiv:1909.03248
  [astro-ph.CO]} \BibitemShut {NoStop}%
\bibitem [{\citenamefont {{Philcox}}(2020)}]{2020arXiv200501739P}%
  \BibitemOpen
  \bibfield  {author} {\bibinfo {author} {\bibfnamefont {O.~H.~E.}\
  \bibnamefont {{Philcox}}},\ }\href@noop {} {\bibfield  {journal} {\bibinfo
  {journal} {arXiv e-prints}\ ,\ \bibinfo {eid} {arXiv:2005.01739}} (\bibinfo
  {year} {2020})},\ \Eprint {http://arxiv.org/abs/2005.01739} {arXiv:2005.01739
  [astro-ph.CO]} \BibitemShut {NoStop}%
\bibitem [{\citenamefont {{Angulo}}\ \emph
  {et~al.}(2015{\natexlab{a}})\citenamefont {{Angulo}}, \citenamefont
  {{Foreman}}, \citenamefont {{Schmittfull}},\ and\ \citenamefont
  {{Senatore}}}]{2015JCAP...10..039A}%
  \BibitemOpen
  \bibfield  {author} {\bibinfo {author} {\bibfnamefont {R.~E.}\ \bibnamefont
  {{Angulo}}}, \bibinfo {author} {\bibfnamefont {S.}~\bibnamefont {{Foreman}}},
  \bibinfo {author} {\bibfnamefont {M.}~\bibnamefont {{Schmittfull}}}, \ and\
  \bibinfo {author} {\bibfnamefont {L.}~\bibnamefont {{Senatore}}},\ }\href
  {\doibase 10.1088/1475-7516/2015/10/039} {\bibfield  {journal} {\bibinfo
  {journal} {\jcap}\ }\textbf {\bibinfo {volume} {2015}},\ \bibinfo {eid} {039}
  (\bibinfo {year} {2015}{\natexlab{a}})},\ \Eprint
  {http://arxiv.org/abs/1406.4143} {arXiv:1406.4143 [astro-ph.CO]} \BibitemShut
  {NoStop}%
\bibitem [{\citenamefont {{Baldauf}}\ \emph {et~al.}(2015)\citenamefont
  {{Baldauf}}, \citenamefont {{Mirbabayi}}, \citenamefont {{Simonovi{\'c}}},\
  and\ \citenamefont {{Zaldarriaga}}}]{2015PhRvD..92d3514B}%
  \BibitemOpen
  \bibfield  {author} {\bibinfo {author} {\bibfnamefont {T.}~\bibnamefont
  {{Baldauf}}}, \bibinfo {author} {\bibfnamefont {M.}~\bibnamefont
  {{Mirbabayi}}}, \bibinfo {author} {\bibfnamefont {M.}~\bibnamefont
  {{Simonovi{\'c}}}}, \ and\ \bibinfo {author} {\bibfnamefont {M.}~\bibnamefont
  {{Zaldarriaga}}},\ }\href {\doibase 10.1103/PhysRevD.92.043514} {\bibfield
  {journal} {\bibinfo  {journal} {\prd}\ }\textbf {\bibinfo {volume} {92}},\
  \bibinfo {eid} {043514} (\bibinfo {year} {2015})},\ \Eprint
  {http://arxiv.org/abs/1504.04366} {arXiv:1504.04366 [astro-ph.CO]}
  \BibitemShut {NoStop}%
\bibitem [{\citenamefont {{Lazanu}}\ and\ \citenamefont
  {{Liguori}}(2018)}]{2018JCAP...04..055L}%
  \BibitemOpen
  \bibfield  {author} {\bibinfo {author} {\bibfnamefont {A.}~\bibnamefont
  {{Lazanu}}}\ and\ \bibinfo {author} {\bibfnamefont {M.}~\bibnamefont
  {{Liguori}}},\ }\href {\doibase 10.1088/1475-7516/2018/04/055} {\bibfield
  {journal} {\bibinfo  {journal} {\jcap}\ }\textbf {\bibinfo {volume} {2018}},\
  \bibinfo {eid} {055} (\bibinfo {year} {2018})},\ \Eprint
  {http://arxiv.org/abs/1803.03184} {arXiv:1803.03184 [astro-ph.CO]}
  \BibitemShut {NoStop}%
\bibitem [{\citenamefont {{Eisenstein}}\ \emph {et~al.}(2007)\citenamefont
  {{Eisenstein}}, \citenamefont {{Seo}}, \citenamefont {{Sirko}},\ and\
  \citenamefont {{Spergel}}}]{2007ApJ...664..675E}%
  \BibitemOpen
  \bibfield  {author} {\bibinfo {author} {\bibfnamefont {D.~J.}\ \bibnamefont
  {{Eisenstein}}}, \bibinfo {author} {\bibfnamefont {H.-J.}\ \bibnamefont
  {{Seo}}}, \bibinfo {author} {\bibfnamefont {E.}~\bibnamefont {{Sirko}}}, \
  and\ \bibinfo {author} {\bibfnamefont {D.~N.}\ \bibnamefont {{Spergel}}},\
  }\href {\doibase 10.1086/518712} {\bibfield  {journal} {\bibinfo  {journal}
  {\apj}\ }\textbf {\bibinfo {volume} {664}},\ \bibinfo {pages} {675} (\bibinfo
  {year} {2007})},\ \Eprint {http://arxiv.org/abs/astro-ph/0604362}
  {arXiv:astro-ph/0604362 [astro-ph]} \BibitemShut {NoStop}%
\bibitem [{\citenamefont {{Weinberg}}(1992)}]{1992MNRAS.254..315W}%
  \BibitemOpen
  \bibfield  {author} {\bibinfo {author} {\bibfnamefont {D.~H.}\ \bibnamefont
  {{Weinberg}}},\ }\href {\doibase 10.1093/mnras/254.2.315} {\bibfield
  {journal} {\bibinfo  {journal} {\mnras}\ }\textbf {\bibinfo {volume} {254}},\
  \bibinfo {pages} {315} (\bibinfo {year} {1992})}\BibitemShut {NoStop}%
\bibitem [{\citenamefont {{Neyrinck}}\ \emph {et~al.}(2011)\citenamefont
  {{Neyrinck}}, \citenamefont {{Szapudi}},\ and\ \citenamefont
  {{Szalay}}}]{2011ApJ...731..116N}%
  \BibitemOpen
  \bibfield  {author} {\bibinfo {author} {\bibfnamefont {M.~C.}\ \bibnamefont
  {{Neyrinck}}}, \bibinfo {author} {\bibfnamefont {I.}~\bibnamefont
  {{Szapudi}}}, \ and\ \bibinfo {author} {\bibfnamefont {A.~S.}\ \bibnamefont
  {{Szalay}}},\ }\href {\doibase 10.1088/0004-637X/731/2/116} {\bibfield
  {journal} {\bibinfo  {journal} {\apj}\ }\textbf {\bibinfo {volume} {731}},\
  \bibinfo {eid} {116} (\bibinfo {year} {2011})},\ \Eprint
  {http://arxiv.org/abs/1009.5680} {arXiv:1009.5680 [astro-ph.CO]} \BibitemShut
  {NoStop}%
\bibitem [{\citenamefont {{Neyrinck}}(2011)}]{2011ApJ...742...91N}%
  \BibitemOpen
  \bibfield  {author} {\bibinfo {author} {\bibfnamefont {M.~C.}\ \bibnamefont
  {{Neyrinck}}},\ }\href {\doibase 10.1088/0004-637X/742/2/91} {\bibfield
  {journal} {\bibinfo  {journal} {\apj}\ }\textbf {\bibinfo {volume} {742}},\
  \bibinfo {eid} {91} (\bibinfo {year} {2011})},\ \Eprint
  {http://arxiv.org/abs/1105.2955} {arXiv:1105.2955 [astro-ph.CO]} \BibitemShut
  {NoStop}%
\bibitem [{\citenamefont {{Neyrinck}}\ \emph {et~al.}(2009)\citenamefont
  {{Neyrinck}}, \citenamefont {{Szapudi}},\ and\ \citenamefont
  {{Szalay}}}]{2009ApJ...698L..90N}%
  \BibitemOpen
  \bibfield  {author} {\bibinfo {author} {\bibfnamefont {M.~C.}\ \bibnamefont
  {{Neyrinck}}}, \bibinfo {author} {\bibfnamefont {I.}~\bibnamefont
  {{Szapudi}}}, \ and\ \bibinfo {author} {\bibfnamefont {A.~S.}\ \bibnamefont
  {{Szalay}}},\ }\href {\doibase 10.1088/0004-637X/698/2/L90} {\bibfield
  {journal} {\bibinfo  {journal} {\apjl}\ }\textbf {\bibinfo {volume} {698}},\
  \bibinfo {pages} {L90} (\bibinfo {year} {2009})},\ \Eprint
  {http://arxiv.org/abs/0903.4693} {arXiv:0903.4693 [astro-ph.CO]} \BibitemShut
  {NoStop}%
\bibitem [{\citenamefont {{Wang}}\ \emph {et~al.}(2011)\citenamefont {{Wang}},
  \citenamefont {{Neyrinck}}, \citenamefont {{Szapudi}}, \citenamefont
  {{Szalay}}, \citenamefont {{Chen}}, \citenamefont {{Lesgourgues}},
  \citenamefont {{Riotto}},\ and\ \citenamefont
  {{Sloth}}}]{2011ApJ...735...32W}%
  \BibitemOpen
  \bibfield  {author} {\bibinfo {author} {\bibfnamefont {X.}~\bibnamefont
  {{Wang}}}, \bibinfo {author} {\bibfnamefont {M.}~\bibnamefont {{Neyrinck}}},
  \bibinfo {author} {\bibfnamefont {I.}~\bibnamefont {{Szapudi}}}, \bibinfo
  {author} {\bibfnamefont {A.}~\bibnamefont {{Szalay}}}, \bibinfo {author}
  {\bibfnamefont {X.}~\bibnamefont {{Chen}}}, \bibinfo {author} {\bibfnamefont
  {J.}~\bibnamefont {{Lesgourgues}}}, \bibinfo {author} {\bibfnamefont
  {A.}~\bibnamefont {{Riotto}}}, \ and\ \bibinfo {author} {\bibfnamefont
  {M.}~\bibnamefont {{Sloth}}},\ }\href {\doibase 10.1088/0004-637X/735/1/32}
  {\bibfield  {journal} {\bibinfo  {journal} {\apj}\ }\textbf {\bibinfo
  {volume} {735}},\ \bibinfo {eid} {32} (\bibinfo {year} {2011})},\ \Eprint
  {http://arxiv.org/abs/1103.2166} {arXiv:1103.2166 [astro-ph.CO]} \BibitemShut
  {NoStop}%
\bibitem [{\citenamefont {{Peebles}}(1980)}]{1980lssu.book.....P}%
  \BibitemOpen
  \bibfield  {author} {\bibinfo {author} {\bibfnamefont {P.~J.~E.}\
  \bibnamefont {{Peebles}}},\ }\href@noop {} {\emph {\bibinfo {title} {{The
  large-scale structure of the universe}}}}\ (\bibinfo {year}
  {1980})\BibitemShut {NoStop}%
\bibitem [{\citenamefont {{Pisani}}\ \emph {et~al.}(2019)\citenamefont
  {{Pisani}}, \citenamefont {{Massara}}, \citenamefont {{Spergel}},
  \citenamefont {{Alonso}}, \citenamefont {{Baker}}, \citenamefont {{Cai}},
  \citenamefont {{Cautun}}, \citenamefont {{Davies}}, \citenamefont
  {{Demchenko}}, \citenamefont {{Dor{\'e}}}, \citenamefont {{Goulding}},
  \citenamefont {{Habouzit}}, \citenamefont {{Hamaus}}, \citenamefont
  {{Hawken}}, \citenamefont {{Hirata}}, \citenamefont {{Ho}}, \citenamefont
  {{Jain}}, \citenamefont {{Kreisch}}, \citenamefont {{Marulli}}, \citenamefont
  {{Padilla}}, \citenamefont {{Pollina}}, \citenamefont {{Sahl{\'e}n}},
  \citenamefont {{Sheth}}, \citenamefont {{Somerville}}, \citenamefont
  {{Szapudi}}, \citenamefont {{van de Weygaert}}, \citenamefont
  {{Villaescusa-Navarro}}, \citenamefont {{Wandelt}},\ and\ \citenamefont
  {{Wang}}}]{2019BAAS...51c..40P}%
  \BibitemOpen
  \bibfield  {author} {\bibinfo {author} {\bibfnamefont {A.}~\bibnamefont
  {{Pisani}}}, \bibinfo {author} {\bibfnamefont {E.}~\bibnamefont {{Massara}}},
  \bibinfo {author} {\bibfnamefont {D.~N.}\ \bibnamefont {{Spergel}}}, \bibinfo
  {author} {\bibfnamefont {D.}~\bibnamefont {{Alonso}}}, \bibinfo {author}
  {\bibfnamefont {T.}~\bibnamefont {{Baker}}}, \bibinfo {author} {\bibfnamefont
  {Y.-C.}\ \bibnamefont {{Cai}}}, \bibinfo {author} {\bibfnamefont
  {M.}~\bibnamefont {{Cautun}}}, \bibinfo {author} {\bibfnamefont
  {C.}~\bibnamefont {{Davies}}}, \bibinfo {author} {\bibfnamefont
  {V.}~\bibnamefont {{Demchenko}}}, \bibinfo {author} {\bibfnamefont
  {O.}~\bibnamefont {{Dor{\'e}}}}, \bibinfo {author} {\bibfnamefont
  {A.}~\bibnamefont {{Goulding}}}, \bibinfo {author} {\bibfnamefont
  {M.}~\bibnamefont {{Habouzit}}}, \bibinfo {author} {\bibfnamefont
  {N.}~\bibnamefont {{Hamaus}}}, \bibinfo {author} {\bibfnamefont
  {A.}~\bibnamefont {{Hawken}}}, \bibinfo {author} {\bibfnamefont {C.~M.}\
  \bibnamefont {{Hirata}}}, \bibinfo {author} {\bibfnamefont {S.}~\bibnamefont
  {{Ho}}}, \bibinfo {author} {\bibfnamefont {B.}~\bibnamefont {{Jain}}},
  \bibinfo {author} {\bibfnamefont {C.~D.}\ \bibnamefont {{Kreisch}}}, \bibinfo
  {author} {\bibfnamefont {F.}~\bibnamefont {{Marulli}}}, \bibinfo {author}
  {\bibfnamefont {N.}~\bibnamefont {{Padilla}}}, \bibinfo {author}
  {\bibfnamefont {G.}~\bibnamefont {{Pollina}}}, \bibinfo {author}
  {\bibfnamefont {M.}~\bibnamefont {{Sahl{\'e}n}}}, \bibinfo {author}
  {\bibfnamefont {R.~K.}\ \bibnamefont {{Sheth}}}, \bibinfo {author}
  {\bibfnamefont {R.}~\bibnamefont {{Somerville}}}, \bibinfo {author}
  {\bibfnamefont {I.}~\bibnamefont {{Szapudi}}}, \bibinfo {author}
  {\bibfnamefont {R.}~\bibnamefont {{van de Weygaert}}}, \bibinfo {author}
  {\bibfnamefont {F.}~\bibnamefont {{Villaescusa-Navarro}}}, \bibinfo {author}
  {\bibfnamefont {B.~D.}\ \bibnamefont {{Wandelt}}}, \ and\ \bibinfo {author}
  {\bibfnamefont {Y.}~\bibnamefont {{Wang}}},\ }\href@noop {} {\bibfield
  {journal} {\bibinfo  {journal} {\baas}\ }\textbf {\bibinfo {volume} {51}},\
  \bibinfo {eid} {40} (\bibinfo {year} {2019})},\ \Eprint
  {http://arxiv.org/abs/1903.05161} {arXiv:1903.05161 [astro-ph.CO]}
  \BibitemShut {NoStop}%
\bibitem [{\citenamefont {{Stoyan}}(1984)}]{doi:10.1002/mana.19841160115}%
  \BibitemOpen
  \bibfield  {author} {\bibinfo {author} {\bibfnamefont {D.}~\bibnamefont
  {{Stoyan}}},\ }\href {\doibase 10.1002/mana.19841160115} {\bibfield
  {journal} {\bibinfo  {journal} {Mathematische Nachrichten}\ }\textbf
  {\bibinfo {volume} {116}},\ \bibinfo {pages} {197} (\bibinfo {year}
  {1984})}\BibitemShut {NoStop}%
\bibitem [{\citenamefont {{Sheth}}(2005)}]{2005MNRAS.364..796S}%
  \BibitemOpen
  \bibfield  {author} {\bibinfo {author} {\bibfnamefont {R.~K.}\ \bibnamefont
  {{Sheth}}},\ }\href {\doibase 10.1111/j.1365-2966.2005.09609.x} {\bibfield
  {journal} {\bibinfo  {journal} {\mnras}\ }\textbf {\bibinfo {volume} {364}},\
  \bibinfo {pages} {796} (\bibinfo {year} {2005})},\ \Eprint
  {http://arxiv.org/abs/astro-ph/0511772} {arXiv:astro-ph/0511772 [astro-ph]}
  \BibitemShut {NoStop}%
\bibitem [{\citenamefont {{Sheth}}\ \emph {et~al.}(2005)\citenamefont
  {{Sheth}}, \citenamefont {{Connolly}},\ and\ \citenamefont
  {{Skibba}}}]{2005astro.ph.11773S}%
  \BibitemOpen
  \bibfield  {author} {\bibinfo {author} {\bibfnamefont {R.~K.}\ \bibnamefont
  {{Sheth}}}, \bibinfo {author} {\bibfnamefont {A.~J.}\ \bibnamefont
  {{Connolly}}}, \ and\ \bibinfo {author} {\bibfnamefont {R.}~\bibnamefont
  {{Skibba}}},\ }\href@noop {} {\bibfield  {journal} {\bibinfo  {journal}
  {arXiv e-prints}\ ,\ \bibinfo {eid} {astro-ph/0511773}} (\bibinfo {year}
  {2005})},\ \Eprint {http://arxiv.org/abs/astro-ph/0511773}
  {arXiv:astro-ph/0511773 [astro-ph]} \BibitemShut {NoStop}%
\bibitem [{\citenamefont {{Skibba}}\ \emph {et~al.}(2006)\citenamefont
  {{Skibba}}, \citenamefont {{Sheth}}, \citenamefont {{Connolly}},\ and\
  \citenamefont {{Scranton}}}]{2006MNRAS.369...68S}%
  \BibitemOpen
  \bibfield  {author} {\bibinfo {author} {\bibfnamefont {R.}~\bibnamefont
  {{Skibba}}}, \bibinfo {author} {\bibfnamefont {R.~K.}\ \bibnamefont
  {{Sheth}}}, \bibinfo {author} {\bibfnamefont {A.~J.}\ \bibnamefont
  {{Connolly}}}, \ and\ \bibinfo {author} {\bibfnamefont {R.}~\bibnamefont
  {{Scranton}}},\ }\href {\doibase 10.1111/j.1365-2966.2006.10196.x} {\bibfield
   {journal} {\bibinfo  {journal} {\mnras}\ }\textbf {\bibinfo {volume}
  {369}},\ \bibinfo {pages} {68} (\bibinfo {year} {2006})},\ \Eprint
  {http://arxiv.org/abs/astro-ph/0512463} {arXiv:astro-ph/0512463 [astro-ph]}
  \BibitemShut {NoStop}%
\bibitem [{\citenamefont {{Beisbart}}\ and\ \citenamefont
  {{Kerscher}}(2000)}]{2000ApJ...545....6B}%
  \BibitemOpen
  \bibfield  {author} {\bibinfo {author} {\bibfnamefont {C.}~\bibnamefont
  {{Beisbart}}}\ and\ \bibinfo {author} {\bibfnamefont {M.}~\bibnamefont
  {{Kerscher}}},\ }\href {\doibase 10.1086/317788} {\bibfield  {journal}
  {\bibinfo  {journal} {\apj}\ }\textbf {\bibinfo {volume} {545}},\ \bibinfo
  {pages} {6} (\bibinfo {year} {2000})},\ \Eprint
  {http://arxiv.org/abs/astro-ph/0003358} {arXiv:astro-ph/0003358 [astro-ph]}
  \BibitemShut {NoStop}%
\bibitem [{\citenamefont {{Gottl{\"o}ber}}\ \emph {et~al.}(2002)\citenamefont
  {{Gottl{\"o}ber}}, \citenamefont {{Kerscher}}, \citenamefont {{Kravtsov}},
  \citenamefont {{Faltenbacher}}, \citenamefont {{Klypin}},\ and\ \citenamefont
  {{M{\"u}ller}}}]{2002A&A...387..778G}%
  \BibitemOpen
  \bibfield  {author} {\bibinfo {author} {\bibfnamefont {S.}~\bibnamefont
  {{Gottl{\"o}ber}}}, \bibinfo {author} {\bibfnamefont {M.}~\bibnamefont
  {{Kerscher}}}, \bibinfo {author} {\bibfnamefont {A.~V.}\ \bibnamefont
  {{Kravtsov}}}, \bibinfo {author} {\bibfnamefont {A.}~\bibnamefont
  {{Faltenbacher}}}, \bibinfo {author} {\bibfnamefont {A.}~\bibnamefont
  {{Klypin}}}, \ and\ \bibinfo {author} {\bibfnamefont {V.}~\bibnamefont
  {{M{\"u}ller}}},\ }\href {\doibase 10.1051/0004-6361:20020339} {\bibfield
  {journal} {\bibinfo  {journal} {\aap}\ }\textbf {\bibinfo {volume} {387}},\
  \bibinfo {pages} {778} (\bibinfo {year} {2002})},\ \Eprint
  {http://arxiv.org/abs/astro-ph/0203148} {arXiv:astro-ph/0203148 [astro-ph]}
  \BibitemShut {NoStop}%
\bibitem [{\citenamefont {{White}}(2016)}]{2016JCAP...11..057W}%
  \BibitemOpen
  \bibfield  {author} {\bibinfo {author} {\bibfnamefont {M.}~\bibnamefont
  {{White}}},\ }\href {\doibase 10.1088/1475-7516/2016/11/057} {\bibfield
  {journal} {\bibinfo  {journal} {\jcap}\ }\textbf {\bibinfo {volume} {2016}},\
  \bibinfo {eid} {057} (\bibinfo {year} {2016})},\ \Eprint
  {http://arxiv.org/abs/1609.08632} {arXiv:1609.08632 [astro-ph.CO]}
  \BibitemShut {NoStop}%
\bibitem [{\citenamefont {{Valogiannis}}\ and\ \citenamefont
  {{Bean}}(2018)}]{2018PhRvD..97b3535V}%
  \BibitemOpen
  \bibfield  {author} {\bibinfo {author} {\bibfnamefont {G.}~\bibnamefont
  {{Valogiannis}}}\ and\ \bibinfo {author} {\bibfnamefont {R.}~\bibnamefont
  {{Bean}}},\ }\href {\doibase 10.1103/PhysRevD.97.023535} {\bibfield
  {journal} {\bibinfo  {journal} {\prd}\ }\textbf {\bibinfo {volume} {97}},\
  \bibinfo {eid} {023535} (\bibinfo {year} {2018})},\ \Eprint
  {http://arxiv.org/abs/1708.05652} {arXiv:1708.05652 [astro-ph.CO]}
  \BibitemShut {NoStop}%
\bibitem [{\citenamefont {{Armijo}}\ \emph {et~al.}(2018)\citenamefont
  {{Armijo}}, \citenamefont {{Cai}}, \citenamefont {{Padilla}}, \citenamefont
  {{Li}},\ and\ \citenamefont {{Peacock}}}]{2018MNRAS.478.3627A}%
  \BibitemOpen
  \bibfield  {author} {\bibinfo {author} {\bibfnamefont {J.}~\bibnamefont
  {{Armijo}}}, \bibinfo {author} {\bibfnamefont {Y.-C.}\ \bibnamefont {{Cai}}},
  \bibinfo {author} {\bibfnamefont {N.}~\bibnamefont {{Padilla}}}, \bibinfo
  {author} {\bibfnamefont {B.}~\bibnamefont {{Li}}}, \ and\ \bibinfo {author}
  {\bibfnamefont {J.~A.}\ \bibnamefont {{Peacock}}},\ }\href {\doibase
  10.1093/mnras/sty1335} {\bibfield  {journal} {\bibinfo  {journal} {\mnras}\
  }\textbf {\bibinfo {volume} {478}},\ \bibinfo {pages} {3627} (\bibinfo {year}
  {2018})},\ \Eprint {http://arxiv.org/abs/1801.08975} {arXiv:1801.08975
  [astro-ph.CO]} \BibitemShut {NoStop}%
\bibitem [{\citenamefont {{Hern{\'a}ndez-Aguayo}}\ \emph
  {et~al.}(2018)\citenamefont {{Hern{\'a}ndez-Aguayo}}, \citenamefont
  {{Baugh}},\ and\ \citenamefont {{Li}}}]{2018MNRAS.479.4824H}%
  \BibitemOpen
  \bibfield  {author} {\bibinfo {author} {\bibfnamefont {C.}~\bibnamefont
  {{Hern{\'a}ndez-Aguayo}}}, \bibinfo {author} {\bibfnamefont {C.~M.}\
  \bibnamefont {{Baugh}}}, \ and\ \bibinfo {author} {\bibfnamefont
  {B.}~\bibnamefont {{Li}}},\ }\href {\doibase 10.1093/mnras/sty1822}
  {\bibfield  {journal} {\bibinfo  {journal} {\mnras}\ }\textbf {\bibinfo
  {volume} {479}},\ \bibinfo {pages} {4824} (\bibinfo {year} {2018})},\ \Eprint
  {http://arxiv.org/abs/1801.08880} {arXiv:1801.08880 [astro-ph.CO]}
  \BibitemShut {NoStop}%
\bibitem [{\citenamefont {{Aviles}}\ \emph {et~al.}(2020)\citenamefont
  {{Aviles}}, \citenamefont {{Koyama}}, \citenamefont {{Cervantes-Cota}},
  \citenamefont {{Winther}},\ and\ \citenamefont {{Li}}}]{2020JCAP...01..006A}%
  \BibitemOpen
  \bibfield  {author} {\bibinfo {author} {\bibfnamefont {A.}~\bibnamefont
  {{Aviles}}}, \bibinfo {author} {\bibfnamefont {K.}~\bibnamefont {{Koyama}}},
  \bibinfo {author} {\bibfnamefont {J.~L.}\ \bibnamefont {{Cervantes-Cota}}},
  \bibinfo {author} {\bibfnamefont {H.~A.}\ \bibnamefont {{Winther}}}, \ and\
  \bibinfo {author} {\bibfnamefont {B.}~\bibnamefont {{Li}}},\ }\href {\doibase
  10.1088/1475-7516/2020/01/006} {\bibfield  {journal} {\bibinfo  {journal}
  {\jcap}\ }\textbf {\bibinfo {volume} {2020}},\ \bibinfo {eid} {006} (\bibinfo
  {year} {2020})},\ \Eprint {http://arxiv.org/abs/1911.06362} {arXiv:1911.06362
  [astro-ph.CO]} \BibitemShut {NoStop}%
\bibitem [{\citenamefont {{Massara}}\ \emph {et~al.}(2020)\citenamefont
  {{Massara}}, \citenamefont {{Villaescusa-Navarro}}, \citenamefont {{Ho}},
  \citenamefont {{Dalal}},\ and\ \citenamefont
  {{Spergel}}}]{2020arXiv200111024M}%
  \BibitemOpen
  \bibfield  {author} {\bibinfo {author} {\bibfnamefont {E.}~\bibnamefont
  {{Massara}}}, \bibinfo {author} {\bibfnamefont {F.}~\bibnamefont
  {{Villaescusa-Navarro}}}, \bibinfo {author} {\bibfnamefont {S.}~\bibnamefont
  {{Ho}}}, \bibinfo {author} {\bibfnamefont {N.}~\bibnamefont {{Dalal}}}, \
  and\ \bibinfo {author} {\bibfnamefont {D.~N.}\ \bibnamefont {{Spergel}}},\
  }\href@noop {} {\bibfield  {journal} {\bibinfo  {journal} {arXiv e-prints}\
  ,\ \bibinfo {eid} {arXiv:2001.11024}} (\bibinfo {year} {2020})},\ \Eprint
  {http://arxiv.org/abs/2001.11024} {arXiv:2001.11024 [astro-ph.CO]}
  \BibitemShut {NoStop}%
\bibitem [{\citenamefont {{Baumann}}\ \emph {et~al.}(2012)\citenamefont
  {{Baumann}}, \citenamefont {{Nicolis}}, \citenamefont {{Senatore}},\ and\
  \citenamefont {{Zaldarriaga}}}]{2012JCAP...07..051B}%
  \BibitemOpen
  \bibfield  {author} {\bibinfo {author} {\bibfnamefont {D.}~\bibnamefont
  {{Baumann}}}, \bibinfo {author} {\bibfnamefont {A.}~\bibnamefont
  {{Nicolis}}}, \bibinfo {author} {\bibfnamefont {L.}~\bibnamefont
  {{Senatore}}}, \ and\ \bibinfo {author} {\bibfnamefont {M.}~\bibnamefont
  {{Zaldarriaga}}},\ }\href {\doibase 10.1088/1475-7516/2012/07/051} {\bibfield
   {journal} {\bibinfo  {journal} {\jcap}\ }\textbf {\bibinfo {volume}
  {2012}},\ \bibinfo {eid} {051} (\bibinfo {year} {2012})},\ \Eprint
  {http://arxiv.org/abs/1004.2488} {arXiv:1004.2488 [astro-ph.CO]} \BibitemShut
  {NoStop}%
\bibitem [{\citenamefont {{Carrasco}}\ \emph {et~al.}(2012)\citenamefont
  {{Carrasco}}, \citenamefont {{Hertzberg}},\ and\ \citenamefont
  {{Senatore}}}]{2012JHEP...09..082C}%
  \BibitemOpen
  \bibfield  {author} {\bibinfo {author} {\bibfnamefont {J.~J.~M.}\
  \bibnamefont {{Carrasco}}}, \bibinfo {author} {\bibfnamefont {M.~P.}\
  \bibnamefont {{Hertzberg}}}, \ and\ \bibinfo {author} {\bibfnamefont
  {L.}~\bibnamefont {{Senatore}}},\ }\href {\doibase 10.1007/JHEP09(2012)082}
  {\bibfield  {journal} {\bibinfo  {journal} {Journal of High Energy Physics}\
  }\textbf {\bibinfo {volume} {2012}},\ \bibinfo {eid} {82} (\bibinfo {year}
  {2012})},\ \Eprint {http://arxiv.org/abs/1206.2926} {arXiv:1206.2926
  [astro-ph.CO]} \BibitemShut {NoStop}%
\bibitem [{\citenamefont {{Senatore}}\ and\ \citenamefont
  {{Zaldarriaga}}(2014)}]{2014arXiv1409.1225S}%
  \BibitemOpen
  \bibfield  {author} {\bibinfo {author} {\bibfnamefont {L.}~\bibnamefont
  {{Senatore}}}\ and\ \bibinfo {author} {\bibfnamefont {M.}~\bibnamefont
  {{Zaldarriaga}}},\ }\href@noop {} {\bibfield  {journal} {\bibinfo  {journal}
  {arXiv e-prints}\ ,\ \bibinfo {eid} {arXiv:1409.1225}} (\bibinfo {year}
  {2014})},\ \Eprint {http://arxiv.org/abs/1409.1225} {arXiv:1409.1225
  [astro-ph.CO]} \BibitemShut {NoStop}%
\bibitem [{\citenamefont {{Senatore}}(2015)}]{2015JCAP...11..007S}%
  \BibitemOpen
  \bibfield  {author} {\bibinfo {author} {\bibfnamefont {L.}~\bibnamefont
  {{Senatore}}},\ }\href {\doibase 10.1088/1475-7516/2015/11/007} {\bibfield
  {journal} {\bibinfo  {journal} {\jcap}\ }\textbf {\bibinfo {volume} {2015}},\
  \bibinfo {eid} {007} (\bibinfo {year} {2015})},\ \Eprint
  {http://arxiv.org/abs/1406.7843} {arXiv:1406.7843 [astro-ph.CO]} \BibitemShut
  {NoStop}%
\bibitem [{\citenamefont {{Angulo}}\ \emph
  {et~al.}(2015{\natexlab{b}})\citenamefont {{Angulo}}, \citenamefont
  {{Fasiello}}, \citenamefont {{Senatore}},\ and\ \citenamefont
  {{Vlah}}}]{2015JCAP...09..029A}%
  \BibitemOpen
  \bibfield  {author} {\bibinfo {author} {\bibfnamefont {R.}~\bibnamefont
  {{Angulo}}}, \bibinfo {author} {\bibfnamefont {M.}~\bibnamefont
  {{Fasiello}}}, \bibinfo {author} {\bibfnamefont {L.}~\bibnamefont
  {{Senatore}}}, \ and\ \bibinfo {author} {\bibfnamefont {Z.}~\bibnamefont
  {{Vlah}}},\ }\href {\doibase 10.1088/1475-7516/2015/09/029} {\bibfield
  {journal} {\bibinfo  {journal} {\jcap}\ }\textbf {\bibinfo {volume} {2015}},\
  \bibinfo {eid} {029} (\bibinfo {year} {2015}{\natexlab{b}})},\ \Eprint
  {http://arxiv.org/abs/1503.08826} {arXiv:1503.08826 [astro-ph.CO]}
  \BibitemShut {NoStop}%
\bibitem [{\citenamefont {{Perko}}\ \emph {et~al.}(2016)\citenamefont
  {{Perko}}, \citenamefont {{Senatore}}, \citenamefont {{Jennings}},\ and\
  \citenamefont {{Wechsler}}}]{2016arXiv161009321P}%
  \BibitemOpen
  \bibfield  {author} {\bibinfo {author} {\bibfnamefont {A.}~\bibnamefont
  {{Perko}}}, \bibinfo {author} {\bibfnamefont {L.}~\bibnamefont {{Senatore}}},
  \bibinfo {author} {\bibfnamefont {E.}~\bibnamefont {{Jennings}}}, \ and\
  \bibinfo {author} {\bibfnamefont {R.~H.}\ \bibnamefont {{Wechsler}}},\
  }\href@noop {} {\bibfield  {journal} {\bibinfo  {journal} {arXiv e-prints}\
  ,\ \bibinfo {eid} {arXiv:1610.09321}} (\bibinfo {year} {2016})},\ \Eprint
  {http://arxiv.org/abs/1610.09321} {arXiv:1610.09321 [astro-ph.CO]}
  \BibitemShut {NoStop}%
\bibitem [{\citenamefont {{Bernardeau}}\ \emph {et~al.}(2002)\citenamefont
  {{Bernardeau}}, \citenamefont {{Colombi}}, \citenamefont {{Gazta{\~n}aga}},\
  and\ \citenamefont {{Scoccimarro}}}]{2002PhR...367....1B}%
  \BibitemOpen
  \bibfield  {author} {\bibinfo {author} {\bibfnamefont {F.}~\bibnamefont
  {{Bernardeau}}}, \bibinfo {author} {\bibfnamefont {S.}~\bibnamefont
  {{Colombi}}}, \bibinfo {author} {\bibfnamefont {E.}~\bibnamefont
  {{Gazta{\~n}aga}}}, \ and\ \bibinfo {author} {\bibfnamefont {R.}~\bibnamefont
  {{Scoccimarro}}},\ }\href {\doibase 10.1016/S0370-1573(02)00135-7} {\bibfield
   {journal} {\bibinfo  {journal} {\physrep}\ }\textbf {\bibinfo {volume}
  {367}},\ \bibinfo {pages} {1} (\bibinfo {year} {2002})},\ \Eprint
  {http://arxiv.org/abs/astro-ph/0112551} {arXiv:astro-ph/0112551 [astro-ph]}
  \BibitemShut {NoStop}%
\bibitem [{\citenamefont {{Saito}}\ \emph {et~al.}(2009)\citenamefont
  {{Saito}}, \citenamefont {{Takada}},\ and\ \citenamefont
  {{Taruya}}}]{2009PhRvD..80h3528S}%
  \BibitemOpen
  \bibfield  {author} {\bibinfo {author} {\bibfnamefont {S.}~\bibnamefont
  {{Saito}}}, \bibinfo {author} {\bibfnamefont {M.}~\bibnamefont {{Takada}}}, \
  and\ \bibinfo {author} {\bibfnamefont {A.}~\bibnamefont {{Taruya}}},\ }\href
  {\doibase 10.1103/PhysRevD.80.083528} {\bibfield  {journal} {\bibinfo
  {journal} {\prd}\ }\textbf {\bibinfo {volume} {80}},\ \bibinfo {eid} {083528}
  (\bibinfo {year} {2009})},\ \Eprint {http://arxiv.org/abs/0907.2922}
  {arXiv:0907.2922 [astro-ph.CO]} \BibitemShut {NoStop}%
\bibitem [{\citenamefont {{Blas}}\ \emph
  {et~al.}(2014{\natexlab{a}})\citenamefont {{Blas}}, \citenamefont {{Garny}},
  \citenamefont {{Konstandin}},\ and\ \citenamefont
  {{Lesgourgues}}}]{2014JCAP...11..039B}%
  \BibitemOpen
  \bibfield  {author} {\bibinfo {author} {\bibfnamefont {D.}~\bibnamefont
  {{Blas}}}, \bibinfo {author} {\bibfnamefont {M.}~\bibnamefont {{Garny}}},
  \bibinfo {author} {\bibfnamefont {T.}~\bibnamefont {{Konstandin}}}, \ and\
  \bibinfo {author} {\bibfnamefont {J.}~\bibnamefont {{Lesgourgues}}},\ }\href
  {\doibase 10.1088/1475-7516/2014/11/039} {\bibfield  {journal} {\bibinfo
  {journal} {\jcap}\ }\textbf {\bibinfo {volume} {2014}},\ \bibinfo {eid} {039}
  (\bibinfo {year} {2014}{\natexlab{a}})},\ \Eprint
  {http://arxiv.org/abs/1408.2995} {arXiv:1408.2995 [astro-ph.CO]} \BibitemShut
  {NoStop}%
\bibitem [{\citenamefont {{F{\"u}hrer}}\ and\ \citenamefont
  {{Wong}}(2015)}]{2015JCAP...03..046F}%
  \BibitemOpen
  \bibfield  {author} {\bibinfo {author} {\bibfnamefont {F.}~\bibnamefont
  {{F{\"u}hrer}}}\ and\ \bibinfo {author} {\bibfnamefont {Y.~Y.~Y.}\
  \bibnamefont {{Wong}}},\ }\href {\doibase 10.1088/1475-7516/2015/03/046}
  {\bibfield  {journal} {\bibinfo  {journal} {\jcap}\ }\textbf {\bibinfo
  {volume} {2015}},\ \bibinfo {eid} {046} (\bibinfo {year} {2015})},\ \Eprint
  {http://arxiv.org/abs/1412.2764} {arXiv:1412.2764 [astro-ph.CO]} \BibitemShut
  {NoStop}%
\bibitem [{\citenamefont {{Senatore}}\ and\ \citenamefont
  {{Zaldarriaga}}(2017)}]{2017arXiv170704698S}%
  \BibitemOpen
  \bibfield  {author} {\bibinfo {author} {\bibfnamefont {L.}~\bibnamefont
  {{Senatore}}}\ and\ \bibinfo {author} {\bibfnamefont {M.}~\bibnamefont
  {{Zaldarriaga}}},\ }\href@noop {} {\bibfield  {journal} {\bibinfo  {journal}
  {arXiv e-prints}\ ,\ \bibinfo {eid} {arXiv:1707.04698}} (\bibinfo {year}
  {2017})},\ \Eprint {http://arxiv.org/abs/1707.04698} {arXiv:1707.04698
  [astro-ph.CO]} \BibitemShut {NoStop}%
\bibitem [{\citenamefont {{Saito}}\ \emph {et~al.}(2008)\citenamefont
  {{Saito}}, \citenamefont {{Takada}},\ and\ \citenamefont
  {{Taruya}}}]{2008PhRvL.100s1301S}%
  \BibitemOpen
  \bibfield  {author} {\bibinfo {author} {\bibfnamefont {S.}~\bibnamefont
  {{Saito}}}, \bibinfo {author} {\bibfnamefont {M.}~\bibnamefont {{Takada}}}, \
  and\ \bibinfo {author} {\bibfnamefont {A.}~\bibnamefont {{Taruya}}},\ }\href
  {\doibase 10.1103/PhysRevLett.100.191301} {\bibfield  {journal} {\bibinfo
  {journal} {\prl}\ }\textbf {\bibinfo {volume} {100}},\ \bibinfo {eid}
  {191301} (\bibinfo {year} {2008})},\ \Eprint {http://arxiv.org/abs/0801.0607}
  {arXiv:0801.0607 [astro-ph]} \BibitemShut {NoStop}%
\bibitem [{\citenamefont {{Chudaykin}}\ and\ \citenamefont
  {{Ivanov}}(2019)}]{2019JCAP...11..034C}%
  \BibitemOpen
  \bibfield  {author} {\bibinfo {author} {\bibfnamefont {A.}~\bibnamefont
  {{Chudaykin}}}\ and\ \bibinfo {author} {\bibfnamefont {M.~M.}\ \bibnamefont
  {{Ivanov}}},\ }\href {\doibase 10.1088/1475-7516/2019/11/034} {\bibfield
  {journal} {\bibinfo  {journal} {\jcap}\ }\textbf {\bibinfo {volume} {2019}},\
  \bibinfo {eid} {034} (\bibinfo {year} {2019})},\ \Eprint
  {http://arxiv.org/abs/1907.06666} {arXiv:1907.06666 [astro-ph.CO]}
  \BibitemShut {NoStop}%
\bibitem [{\citenamefont {{Lewis}}\ and\ \citenamefont
  {{Challinor}}(2011)}]{2011ascl.soft02026L}%
  \BibitemOpen
  \bibfield  {author} {\bibinfo {author} {\bibfnamefont {A.}~\bibnamefont
  {{Lewis}}}\ and\ \bibinfo {author} {\bibfnamefont {A.}~\bibnamefont
  {{Challinor}}},\ }\href@noop {} {\enquote {\bibinfo {title} {{CAMB: Code for
  Anisotropies in the Microwave Background}},}\ } (\bibinfo {year} {2011}),\
  \Eprint {http://arxiv.org/abs/1102.026} {ascl:1102.026} \BibitemShut
  {NoStop}%
\bibitem [{\citenamefont {{Blas}}\ \emph {et~al.}(2011)\citenamefont {{Blas}},
  \citenamefont {{Lesgourgues}},\ and\ \citenamefont
  {{Tram}}}]{2011JCAP...07..034B}%
  \BibitemOpen
  \bibfield  {author} {\bibinfo {author} {\bibfnamefont {D.}~\bibnamefont
  {{Blas}}}, \bibinfo {author} {\bibfnamefont {J.}~\bibnamefont
  {{Lesgourgues}}}, \ and\ \bibinfo {author} {\bibfnamefont {T.}~\bibnamefont
  {{Tram}}},\ }\href {\doibase 10.1088/1475-7516/2011/07/034} {\bibfield
  {journal} {\bibinfo  {journal} {\jcap}\ }\textbf {\bibinfo {volume} {2011}},\
  \bibinfo {eid} {034} (\bibinfo {year} {2011})},\ \Eprint
  {http://arxiv.org/abs/1104.2933} {arXiv:1104.2933 [astro-ph.CO]} \BibitemShut
  {NoStop}%
\bibitem [{\citenamefont {{Hamilton}}(2000)}]{2000MNRAS.312..257H}%
  \BibitemOpen
  \bibfield  {author} {\bibinfo {author} {\bibfnamefont {A.~J.~S.}\
  \bibnamefont {{Hamilton}}},\ }\href {\doibase
  10.1046/j.1365-8711.2000.03071.x} {\bibfield  {journal} {\bibinfo  {journal}
  {\mnras}\ }\textbf {\bibinfo {volume} {312}},\ \bibinfo {pages} {257}
  (\bibinfo {year} {2000})},\ \Eprint {http://arxiv.org/abs/astro-ph/9905191}
  {arXiv:astro-ph/9905191 [astro-ph]} \BibitemShut {NoStop}%
\bibitem [{\citenamefont {{Schmittfull}}\ \emph {et~al.}(2016)\citenamefont
  {{Schmittfull}}, \citenamefont {{Vlah}},\ and\ \citenamefont
  {{McDonald}}}]{2016PhRvD..93j3528S}%
  \BibitemOpen
  \bibfield  {author} {\bibinfo {author} {\bibfnamefont {M.}~\bibnamefont
  {{Schmittfull}}}, \bibinfo {author} {\bibfnamefont {Z.}~\bibnamefont
  {{Vlah}}}, \ and\ \bibinfo {author} {\bibfnamefont {P.}~\bibnamefont
  {{McDonald}}},\ }\href {\doibase 10.1103/PhysRevD.93.103528} {\bibfield
  {journal} {\bibinfo  {journal} {\prd}\ }\textbf {\bibinfo {volume} {93}},\
  \bibinfo {eid} {103528} (\bibinfo {year} {2016})},\ \Eprint
  {http://arxiv.org/abs/1603.04405} {arXiv:1603.04405 [astro-ph.CO]}
  \BibitemShut {NoStop}%
\bibitem [{\citenamefont {{McEwen}}\ \emph {et~al.}(2016)\citenamefont
  {{McEwen}}, \citenamefont {{Fang}}, \citenamefont {{Hirata}},\ and\
  \citenamefont {{Blazek}}}]{2016JCAP...09..015M}%
  \BibitemOpen
  \bibfield  {author} {\bibinfo {author} {\bibfnamefont {J.~E.}\ \bibnamefont
  {{McEwen}}}, \bibinfo {author} {\bibfnamefont {X.}~\bibnamefont {{Fang}}},
  \bibinfo {author} {\bibfnamefont {C.~M.}\ \bibnamefont {{Hirata}}}, \ and\
  \bibinfo {author} {\bibfnamefont {J.~A.}\ \bibnamefont {{Blazek}}},\ }\href
  {\doibase 10.1088/1475-7516/2016/09/015} {\bibfield  {journal} {\bibinfo
  {journal} {\jcap}\ }\textbf {\bibinfo {volume} {2016}},\ \bibinfo {eid} {015}
  (\bibinfo {year} {2016})},\ \Eprint {http://arxiv.org/abs/1603.04826}
  {arXiv:1603.04826 [astro-ph.CO]} \BibitemShut {NoStop}%
\bibitem [{\citenamefont {{Villaescusa-Navarro}}\ \emph
  {et~al.}(2019)\citenamefont {{Villaescusa-Navarro}}, \citenamefont {{Hahn}},
  \citenamefont {{Massara}}, \citenamefont {{Banerjee}}, \citenamefont
  {{Delgado}}, \citenamefont {{Kodi Ramanah}}, \citenamefont {{Charnock}},
  \citenamefont {{Giusarma}}, \citenamefont {{Li}}, \citenamefont {{Allys}},
  \citenamefont {{Brochard}}, \citenamefont {{Chiang}}, \citenamefont {{He}},
  \citenamefont {{Pisani}}, \citenamefont {{Obuljen}}, \citenamefont {{Feng}},
  \citenamefont {{Castorina}}, \citenamefont {{Contardo}}, \citenamefont
  {{Kreisch}}, \citenamefont {{Nicola}}, \citenamefont {{Scoccimarro}},
  \citenamefont {{Verde}}, \citenamefont {{Viel}}, \citenamefont {{Ho}},
  \citenamefont {{Mallat}}, \citenamefont {{Wand elt}},\ and\ \citenamefont
  {{Spergel}}}]{2019arXiv190905273V}%
  \BibitemOpen
  \bibfield  {author} {\bibinfo {author} {\bibfnamefont {F.}~\bibnamefont
  {{Villaescusa-Navarro}}}, \bibinfo {author} {\bibfnamefont {C.}~\bibnamefont
  {{Hahn}}}, \bibinfo {author} {\bibfnamefont {E.}~\bibnamefont {{Massara}}},
  \bibinfo {author} {\bibfnamefont {A.}~\bibnamefont {{Banerjee}}}, \bibinfo
  {author} {\bibfnamefont {A.~M.}\ \bibnamefont {{Delgado}}}, \bibinfo {author}
  {\bibfnamefont {D.}~\bibnamefont {{Kodi Ramanah}}}, \bibinfo {author}
  {\bibfnamefont {T.}~\bibnamefont {{Charnock}}}, \bibinfo {author}
  {\bibfnamefont {E.}~\bibnamefont {{Giusarma}}}, \bibinfo {author}
  {\bibfnamefont {Y.}~\bibnamefont {{Li}}}, \bibinfo {author} {\bibfnamefont
  {E.}~\bibnamefont {{Allys}}}, \bibinfo {author} {\bibfnamefont
  {A.}~\bibnamefont {{Brochard}}}, \bibinfo {author} {\bibfnamefont {C.-T.}\
  \bibnamefont {{Chiang}}}, \bibinfo {author} {\bibfnamefont {S.}~\bibnamefont
  {{He}}}, \bibinfo {author} {\bibfnamefont {A.}~\bibnamefont {{Pisani}}},
  \bibinfo {author} {\bibfnamefont {A.}~\bibnamefont {{Obuljen}}}, \bibinfo
  {author} {\bibfnamefont {Y.}~\bibnamefont {{Feng}}}, \bibinfo {author}
  {\bibfnamefont {E.}~\bibnamefont {{Castorina}}}, \bibinfo {author}
  {\bibfnamefont {G.}~\bibnamefont {{Contardo}}}, \bibinfo {author}
  {\bibfnamefont {C.~D.}\ \bibnamefont {{Kreisch}}}, \bibinfo {author}
  {\bibfnamefont {A.}~\bibnamefont {{Nicola}}}, \bibinfo {author}
  {\bibfnamefont {R.}~\bibnamefont {{Scoccimarro}}}, \bibinfo {author}
  {\bibfnamefont {L.}~\bibnamefont {{Verde}}}, \bibinfo {author} {\bibfnamefont
  {M.}~\bibnamefont {{Viel}}}, \bibinfo {author} {\bibfnamefont
  {S.}~\bibnamefont {{Ho}}}, \bibinfo {author} {\bibfnamefont {S.}~\bibnamefont
  {{Mallat}}}, \bibinfo {author} {\bibfnamefont {B.}~\bibnamefont {{Wand
  elt}}}, \ and\ \bibinfo {author} {\bibfnamefont {D.~N.}\ \bibnamefont
  {{Spergel}}},\ }\href@noop {} {\bibfield  {journal} {\bibinfo  {journal}
  {arXiv e-prints}\ ,\ \bibinfo {eid} {arXiv:1909.05273}} (\bibinfo {year}
  {2019})},\ \Eprint {http://arxiv.org/abs/1909.05273} {arXiv:1909.05273
  [astro-ph.CO]} \BibitemShut {NoStop}%
\bibitem [{\citenamefont {{Sirko}}(2005)}]{2005ApJ...634..728S}%
  \BibitemOpen
  \bibfield  {author} {\bibinfo {author} {\bibfnamefont {E.}~\bibnamefont
  {{Sirko}}},\ }\href {\doibase 10.1086/497090} {\bibfield  {journal} {\bibinfo
   {journal} {\apj}\ }\textbf {\bibinfo {volume} {634}},\ \bibinfo {pages}
  {728} (\bibinfo {year} {2005})},\ \Eprint
  {http://arxiv.org/abs/astro-ph/0503106} {arXiv:astro-ph/0503106 [astro-ph]}
  \BibitemShut {NoStop}%
\bibitem [{\citenamefont {{Senatore}}\ and\ \citenamefont
  {{Zaldarriaga}}(2015)}]{2015JCAP...02..013S}%
  \BibitemOpen
  \bibfield  {author} {\bibinfo {author} {\bibfnamefont {L.}~\bibnamefont
  {{Senatore}}}\ and\ \bibinfo {author} {\bibfnamefont {M.}~\bibnamefont
  {{Zaldarriaga}}},\ }\href {\doibase 10.1088/1475-7516/2015/02/013} {\bibfield
   {journal} {\bibinfo  {journal} {\jcap}\ }\textbf {\bibinfo {volume}
  {2015}},\ \bibinfo {eid} {013} (\bibinfo {year} {2015})},\ \Eprint
  {http://arxiv.org/abs/1404.5954} {arXiv:1404.5954 [astro-ph.CO]} \BibitemShut
  {NoStop}%
\bibitem [{\citenamefont {{Philcox}}\ \emph {et~al.}(2020)\citenamefont
  {{Philcox}}, \citenamefont {{Spergel}},\ and\ \citenamefont
  {{Villaescusa-Navarro}}}]{2020PhRvD.101l3520P}%
  \BibitemOpen
  \bibfield  {author} {\bibinfo {author} {\bibfnamefont {O.~H.~E.}\
  \bibnamefont {{Philcox}}}, \bibinfo {author} {\bibfnamefont {D.~N.}\
  \bibnamefont {{Spergel}}}, \ and\ \bibinfo {author} {\bibfnamefont
  {F.}~\bibnamefont {{Villaescusa-Navarro}}},\ }\href {\doibase
  10.1103/PhysRevD.101.123520} {\bibfield  {journal} {\bibinfo  {journal}
  {\prd}\ }\textbf {\bibinfo {volume} {101}},\ \bibinfo {eid} {123520}
  (\bibinfo {year} {2020})},\ \Eprint {http://arxiv.org/abs/2004.09515}
  {arXiv:2004.09515 [astro-ph.CO]} \BibitemShut {NoStop}%
\bibitem [{\citenamefont {{Konstandin}}\ \emph {et~al.}(2019)\citenamefont
  {{Konstandin}}, \citenamefont {{Porto}},\ and\ \citenamefont
  {{Rubira}}}]{2019JCAP...11..027K}%
  \BibitemOpen
  \bibfield  {author} {\bibinfo {author} {\bibfnamefont {T.}~\bibnamefont
  {{Konstandin}}}, \bibinfo {author} {\bibfnamefont {R.~A.}\ \bibnamefont
  {{Porto}}}, \ and\ \bibinfo {author} {\bibfnamefont {H.}~\bibnamefont
  {{Rubira}}},\ }\href {\doibase 10.1088/1475-7516/2019/11/027} {\bibfield
  {journal} {\bibinfo  {journal} {\jcap}\ }\textbf {\bibinfo {volume} {2019}},\
  \bibinfo {eid} {027} (\bibinfo {year} {2019})},\ \Eprint
  {http://arxiv.org/abs/1906.00997} {arXiv:1906.00997 [astro-ph.CO]}
  \BibitemShut {NoStop}%
\bibitem [{\citenamefont {{McDonald}}\ and\ \citenamefont
  {{Roy}}(2009)}]{2009JCAP...08..020M}%
  \BibitemOpen
  \bibfield  {author} {\bibinfo {author} {\bibfnamefont {P.}~\bibnamefont
  {{McDonald}}}\ and\ \bibinfo {author} {\bibfnamefont {A.}~\bibnamefont
  {{Roy}}},\ }\href {\doibase 10.1088/1475-7516/2009/08/020} {\bibfield
  {journal} {\bibinfo  {journal} {\jcap}\ }\textbf {\bibinfo {volume} {2009}},\
  \bibinfo {eid} {020} (\bibinfo {year} {2009})},\ \Eprint
  {http://arxiv.org/abs/0902.0991} {arXiv:0902.0991 [astro-ph.CO]} \BibitemShut
  {NoStop}%
\bibitem [{\citenamefont {{Assassi}}\ \emph {et~al.}(2014)\citenamefont
  {{Assassi}}, \citenamefont {{Baumann}}, \citenamefont {{Green}},\ and\
  \citenamefont {{Zaldarriaga}}}]{2014JCAP...08..056A}%
  \BibitemOpen
  \bibfield  {author} {\bibinfo {author} {\bibfnamefont {V.}~\bibnamefont
  {{Assassi}}}, \bibinfo {author} {\bibfnamefont {D.}~\bibnamefont
  {{Baumann}}}, \bibinfo {author} {\bibfnamefont {D.}~\bibnamefont {{Green}}},
  \ and\ \bibinfo {author} {\bibfnamefont {M.}~\bibnamefont {{Zaldarriaga}}},\
  }\href {\doibase 10.1088/1475-7516/2014/08/056} {\bibfield  {journal}
  {\bibinfo  {journal} {\jcap}\ }\textbf {\bibinfo {volume} {2014}},\ \bibinfo
  {eid} {056} (\bibinfo {year} {2014})},\ \Eprint
  {http://arxiv.org/abs/1402.5916} {arXiv:1402.5916 [astro-ph.CO]} \BibitemShut
  {NoStop}%
\bibitem [{\citenamefont {{Schmittfull}}\ \emph {et~al.}(2017)\citenamefont
  {{Schmittfull}}, \citenamefont {{Baldauf}},\ and\ \citenamefont
  {{Zaldarriaga}}}]{2017PhRvD..96b3505S}%
  \BibitemOpen
  \bibfield  {author} {\bibinfo {author} {\bibfnamefont {M.}~\bibnamefont
  {{Schmittfull}}}, \bibinfo {author} {\bibfnamefont {T.}~\bibnamefont
  {{Baldauf}}}, \ and\ \bibinfo {author} {\bibfnamefont {M.}~\bibnamefont
  {{Zaldarriaga}}},\ }\href {\doibase 10.1103/PhysRevD.96.023505} {\bibfield
  {journal} {\bibinfo  {journal} {\prd}\ }\textbf {\bibinfo {volume} {96}},\
  \bibinfo {eid} {023505} (\bibinfo {year} {2017})},\ \Eprint
  {http://arxiv.org/abs/1704.06634} {arXiv:1704.06634 [astro-ph.CO]}
  \BibitemShut {NoStop}%
\bibitem [{\citenamefont {{Chiang}}\ \emph {et~al.}(2014)\citenamefont
  {{Chiang}}, \citenamefont {{Wagner}}, \citenamefont {{Schmidt}},\ and\
  \citenamefont {{Komatsu}}}]{2014JCAP...05..048C}%
  \BibitemOpen
  \bibfield  {author} {\bibinfo {author} {\bibfnamefont {C.-T.}\ \bibnamefont
  {{Chiang}}}, \bibinfo {author} {\bibfnamefont {C.}~\bibnamefont {{Wagner}}},
  \bibinfo {author} {\bibfnamefont {F.}~\bibnamefont {{Schmidt}}}, \ and\
  \bibinfo {author} {\bibfnamefont {E.}~\bibnamefont {{Komatsu}}},\ }\href
  {\doibase 10.1088/1475-7516/2014/05/048} {\bibfield  {journal} {\bibinfo
  {journal} {\jcap}\ }\textbf {\bibinfo {volume} {2014}},\ \bibinfo {eid} {048}
  (\bibinfo {year} {2014})},\ \Eprint {http://arxiv.org/abs/1403.3411}
  {arXiv:1403.3411 [astro-ph.CO]} \BibitemShut {NoStop}%
\bibitem [{\citenamefont {{McDonald}}(2006)}]{2006PhRvD..74j3512M}%
  \BibitemOpen
  \bibfield  {author} {\bibinfo {author} {\bibfnamefont {P.}~\bibnamefont
  {{McDonald}}},\ }\href {\doibase 10.1103/PhysRevD.74.103512} {\bibfield
  {journal} {\bibinfo  {journal} {\prd}\ }\textbf {\bibinfo {volume} {74}},\
  \bibinfo {eid} {103512} (\bibinfo {year} {2006})},\ \Eprint
  {http://arxiv.org/abs/astro-ph/0609413} {arXiv:astro-ph/0609413 [astro-ph]}
  \BibitemShut {NoStop}%
\bibitem [{\citenamefont {{Simonovi{\'c}}}\ \emph {et~al.}(2018)\citenamefont
  {{Simonovi{\'c}}}, \citenamefont {{Baldauf}}, \citenamefont {{Zaldarriaga}},
  \citenamefont {{Carrasco}},\ and\ \citenamefont
  {{Kollmeier}}}]{2018JCAP...04..030S}%
  \BibitemOpen
  \bibfield  {author} {\bibinfo {author} {\bibfnamefont {M.}~\bibnamefont
  {{Simonovi{\'c}}}}, \bibinfo {author} {\bibfnamefont {T.}~\bibnamefont
  {{Baldauf}}}, \bibinfo {author} {\bibfnamefont {M.}~\bibnamefont
  {{Zaldarriaga}}}, \bibinfo {author} {\bibfnamefont {J.~J.}\ \bibnamefont
  {{Carrasco}}}, \ and\ \bibinfo {author} {\bibfnamefont {J.~A.}\ \bibnamefont
  {{Kollmeier}}},\ }\href {\doibase 10.1088/1475-7516/2018/04/030} {\bibfield
  {journal} {\bibinfo  {journal} {\jcap}\ }\textbf {\bibinfo {volume} {2018}},\
  \bibinfo {eid} {030} (\bibinfo {year} {2018})},\ \Eprint
  {http://arxiv.org/abs/1708.08130} {arXiv:1708.08130 [astro-ph.CO]}
  \BibitemShut {NoStop}%
\bibitem [{\citenamefont {{Scoccimarro}}\ and\ \citenamefont
  {{Frieman}}(1996)}]{1996ApJS..105...37S}%
  \BibitemOpen
  \bibfield  {author} {\bibinfo {author} {\bibfnamefont {R.}~\bibnamefont
  {{Scoccimarro}}}\ and\ \bibinfo {author} {\bibfnamefont {J.}~\bibnamefont
  {{Frieman}}},\ }\href {\doibase 10.1086/192306} {\bibfield  {journal}
  {\bibinfo  {journal} {\apjs}\ }\textbf {\bibinfo {volume} {105}},\ \bibinfo
  {pages} {37} (\bibinfo {year} {1996})},\ \Eprint
  {http://arxiv.org/abs/astro-ph/9509047} {arXiv:astro-ph/9509047 [astro-ph]}
  \BibitemShut {NoStop}%
\bibitem [{\citenamefont {{Jain}}\ and\ \citenamefont
  {{Bertschinger}}(1996)}]{1996ApJ...456...43J}%
  \BibitemOpen
  \bibfield  {author} {\bibinfo {author} {\bibfnamefont {B.}~\bibnamefont
  {{Jain}}}\ and\ \bibinfo {author} {\bibfnamefont {E.}~\bibnamefont
  {{Bertschinger}}},\ }\href {\doibase 10.1086/176625} {\bibfield  {journal}
  {\bibinfo  {journal} {\apj}\ }\textbf {\bibinfo {volume} {456}},\ \bibinfo
  {pages} {43} (\bibinfo {year} {1996})},\ \Eprint
  {http://arxiv.org/abs/astro-ph/9503025} {arXiv:astro-ph/9503025 [astro-ph]}
  \BibitemShut {NoStop}%
\bibitem [{\citenamefont {{Blas}}\ \emph {et~al.}(2013)\citenamefont {{Blas}},
  \citenamefont {{Garny}},\ and\ \citenamefont
  {{Konstandin}}}]{2013JCAP...09..024B}%
  \BibitemOpen
  \bibfield  {author} {\bibinfo {author} {\bibfnamefont {D.}~\bibnamefont
  {{Blas}}}, \bibinfo {author} {\bibfnamefont {M.}~\bibnamefont {{Garny}}}, \
  and\ \bibinfo {author} {\bibfnamefont {T.}~\bibnamefont {{Konstandin}}},\
  }\href {\doibase 10.1088/1475-7516/2013/09/024} {\bibfield  {journal}
  {\bibinfo  {journal} {\jcap}\ }\textbf {\bibinfo {volume} {2013}},\ \bibinfo
  {eid} {024} (\bibinfo {year} {2013})},\ \Eprint
  {http://arxiv.org/abs/1304.1546} {arXiv:1304.1546 [astro-ph.CO]} \BibitemShut
  {NoStop}%
\bibitem [{\citenamefont {{Carrasco}}\ \emph {et~al.}(2014)\citenamefont
  {{Carrasco}}, \citenamefont {{Foreman}}, \citenamefont {{Green}},\ and\
  \citenamefont {{Senatore}}}]{2014JCAP...07..056C}%
  \BibitemOpen
  \bibfield  {author} {\bibinfo {author} {\bibfnamefont {J.~J.~M.}\
  \bibnamefont {{Carrasco}}}, \bibinfo {author} {\bibfnamefont
  {S.}~\bibnamefont {{Foreman}}}, \bibinfo {author} {\bibfnamefont
  {D.}~\bibnamefont {{Green}}}, \ and\ \bibinfo {author} {\bibfnamefont
  {L.}~\bibnamefont {{Senatore}}},\ }\href {\doibase
  10.1088/1475-7516/2014/07/056} {\bibfield  {journal} {\bibinfo  {journal}
  {\jcap}\ }\textbf {\bibinfo {volume} {2014}},\ \bibinfo {eid} {056} (\bibinfo
  {year} {2014})},\ \Eprint {http://arxiv.org/abs/1304.4946} {arXiv:1304.4946
  [astro-ph.CO]} \BibitemShut {NoStop}%
\bibitem [{\citenamefont {{Blas}}\ \emph
  {et~al.}(2014{\natexlab{b}})\citenamefont {{Blas}}, \citenamefont {{Garny}},\
  and\ \citenamefont {{Konstandin}}}]{2014JCAP...01..010B}%
  \BibitemOpen
  \bibfield  {author} {\bibinfo {author} {\bibfnamefont {D.}~\bibnamefont
  {{Blas}}}, \bibinfo {author} {\bibfnamefont {M.}~\bibnamefont {{Garny}}}, \
  and\ \bibinfo {author} {\bibfnamefont {T.}~\bibnamefont {{Konstandin}}},\
  }\href {\doibase 10.1088/1475-7516/2014/01/010} {\bibfield  {journal}
  {\bibinfo  {journal} {\jcap}\ }\textbf {\bibinfo {volume} {2014}},\ \bibinfo
  {eid} {010} (\bibinfo {year} {2014}{\natexlab{b}})},\ \Eprint
  {http://arxiv.org/abs/1309.3308} {arXiv:1309.3308 [astro-ph.CO]} \BibitemShut
  {NoStop}%
\end{thebibliography}%

\end{document}